\documentclass[aps,prb, twocolumn,floatfix,notitlepage, superscriptaddress,longbibliography]{revtex4-2}
\usepackage{pgf}
\usepackage[utf8]{inputenc}
\usepackage{amsmath}
\usepackage{braket}
\usepackage{amssymb}
\usepackage{amsmath}
\usepackage{stackrel}
\usepackage{graphicx}
\usepackage{overpic}
\usepackage{epstopdf}
\usepackage[pdftex,bookmarks,colorlinks]{hyperref}
\usepackage{hyperref}
\usepackage{amsmath}
\usepackage{amssymb}
\usepackage{braket}

\newcommand{\Tr}[1]{\operatorname{Tr} #1}

\newcommand{\nep}{\textrm{e}}

\definecolor{caribbeangreen}{rgb}{0.0, 0.8, 0.6}

\newcommand{\ud}{\mathrm{d}}

\begin{document}
\title{Weak ergodicity breaking in Josephson-junction arrays}
%
\author{Angelo Russomanno}
\affiliation{Scuola Superiore Meridionale, Università di Napoli Federico II
Largo San Marcellino 10, I-80138 Napoli, Italy}
\affiliation{Dipartimento di Fisica, Universit\`a di Napoli ``Federico II'', Monte S. Angelo, I-80126 Napoli, Italy}
\author{Michele Fava}
\affiliation{Rudolf Peierls Centre for Theoretical Physics, Clarendon Laboratory, University of Oxford, Oxford OX1 3PU, United Kingdom}
\author{Rosario Fazio}
\affiliation{Abdus Salam ICTP, Strada Costiera 11, I-34151 Trieste, Italy}
\affiliation{Dipartimento di Fisica, Universit\`a di Napoli ``Federico II'', Monte S. Angelo, I-80126 Napoli, Italy}
\affiliation{Scuola Superiore Meridionale, Università di Napoli Federico II
Largo San Marcellino 10, I-80138 Napoli, Italy}
%
%
%
\begin{abstract}
{We study the quantum dynamics of Josephson junction arrays. We find isolated groups of low-entanglement eigenstates, {that persist even when the Josephson interaction is strong enough to destroy the overall organization of the spectrum in multiplets, and a perturbative description is no longer possible}. These eigenstates lie in the inner part of the spectrum, far from the spectral edge, and {provide} a weak ergodicity breaking{, reminiscent} of the quantum scars. Due to the presence of these eigenstates, initializing with a charge-density-wave state, the system does not thermalize and the charge-density-wave order persists for long times.  {Considering global ergodicity probes, we find that the system tends towards more ergodicity for increasing system size: The parameter range where the bulk of the eigenstates look nonergodic shrinks for increasing system size.} We study two geometries, a one-dimensional chain and a two-leg ladder. In the latter case, adding a magnetic flux makes the system more ergodic.} 
\end{abstract}
\maketitle
\section{Introduction}
Quantum thermalization has been the focus of {quite a lot of scientific activity} in recent years (see~\cite{silva_rmp,kafri_2016} for a review). In the vast majority of the existing numerical and analytical studies, quantum thermalization relies on the properties of the eigenstates, which must be locally equivalent to thermal density matrices, the so-called eigenstate thermalization (ETH)~\cite{Rigol_Nat,Prosen_PRL98,Sred_PRE94,Deutsch_PRA91}. In particular, ergodicity and thermalization {come in association with} eigenstates delocalized in the Hilbert space~\cite{kafri_2016,silva_rmp}. When this does not happen -- for instance in many-body localization (MBL)~\cite{abanin_rmp} -- the dynamics is so constrained that local observables can never thermalize. 
Preparing the system in a {state out of equilibrium, the system is frozen and never reaches thermalization}, as it has been experimentally verified~\cite{Bloch2019}.
MBL usually occurs in disordered systems, but it has been demonstrated to occur also in some clean systems~\cite{karpov2020disorderfree,Adam_prb,Adam_prl19,Adam_prl171}, even in a clean chain of Josephson junctions at high energies~\cite{altshuler}. 

There are some peculiar clean systems (like the PXP model)~\cite{PhysRevB.99.161101,PhysRevB.98.155134} which show some features of {non ergodicity similar to} MBL, although the bulk of the eigenstates is delocalized in the Hilbert space and locally thermal. In these cases, 
there are a minority of eigenstates which are nonthermal, the so-called many-body quantum scars~\cite{2021_serbyn_nat}. {These scars are a nontrivial ergodicity breaking phenomenon, being their energy far from the ground state and in the inner part of the spectrum, surrounded by thermal eigenstates. They} have a significant effect on the dynamics: They hinder thermalization, and this happens only when the system is prepared in some special initial states. These states are factorized and simple to prepare, and have a large overlap with the nonthermal eigenstates. This dynamical phenomenon is known as weak ergodicity breaking~\cite{2021_serbyn_nat}. {Many-body quantum scars are a purely quantum phenomenon radically different from the single-particle ones, which appear around isolated periodic trajectories of a fully-chaotic (ergodic and mixing) classical Hamiltonian system~\cite{lichtenberg1983regular,Berry_regirr78:proceeding}, when one quantizes it~\cite{Heller_Les_Houches,scars_turner}.}

{Questions of quantum thermalization have been addressed also in the context of Josephson junction arrays~\cite{houck,FAZIO2001235}. 
It has been theoretically predicted that a linear Josephson-junction chain shows MBL in the regime of high energies~\cite{altshuler}, {a Bose-glass phase has been experimentally observed in a disordered Josephson-junction~\cite{PhysRevLett.119.167701},} and in a similar model -- the Bose-Hubbard chain -- nonergodic behavior~\cite{russomanno2020nonergodic,kollath,kollath1,PhysRevA.90.033606,biroli2010effect} and Bose-glass dynamics~\cite{Carleo} have been numerically observed.

{In this work w}e focus on Josephson-junction arrays in a regime {of energies smaller than Ref.}~\cite{altshuler} and find a weak-ergodicity breaking phenomenon reminiscent of quantum scars. We find indeed nonthermal eigenstates of the
Hamiltonian, that have small entanglement entropy and lie in
the inner part of the spectrum, far from the spectral edge. {The latter is an important point: lying far from the spectral edges these eigenstates cannot be confused with a finite-size effect.} These eigenstates are the superposition of few degenerate
charge configurations dressed by the interaction, are organized
in small isolated groups, and -- although they are a minority --
we argue that they are infinite in number, even at finite size.

{A part of these special eigenstates are organized in doublets. These doublets have small entanglement entropy equal to $\simeq\log 2$. Another set of low-entanglement eigenstates is organized in 6-tuplets, and these states have entanglement entropy $\log 6$.  Both sets of states persist for values of the Josephson coupling $E_J$ up to order of the charging energy scale $E_C$. Being $L$ the system size, $E_J\ll E_C/L$ defines the regime where perturbation theory is valid, the spectrum is organized in multiplets, and one expects to find low-entanglement eigenstates in the inner part of the spectrum. Our special eigenstates can be found already in this regime using perturbation theory, but they remarkably survive much beyond it.}

{These low-entanglement eigenstates hinder thermalization. The $\log 2$ states affect the dynamics when the system is initialized with a period-2 charge density wave. There is an interval of Josephson couplings, where the initial state has a large overlap with one of the $\log 2$ eigenstates. In correspondence with that, the dynamics provides a charge-density-wave order persisting at long times. A similar persistence occurs when the system is prepared in a charge-density-wave state with period 3. In a thermalizing case, in contrast, the charge-density-wave order would melt after a transient. As in the case of quantum scars, we see absence of thermalization when we initialize with simple factorized states. We have therefore a strong effect on the dynamics of a  minority of low-entanglement eigenstates with large overlap with the initial state, similarly to quantum scars.}

{For values of $E_J$ beyond the perturbative regime, we apply tDMRG and probe the long-time persistence of the charge-density wave order, and the related existence of the nonthermal eigenstates, up to $L=60$ for the $\log 2$ states and $L=90$ for the $\log 6$ ones (see Sec.~\ref{dindyno:sec}). So, the nonthermal eigenstates (or at least their effect on the dynamics) persists for quite large system sizes, although no extrapolation to the thermodynamic limit is possible.}

Looking at global probes of ergodicity (involving averages over the bulk of the eigenstates and the eigenvalues of the Hamiltonian), one can see a diffuse tendency of the system towards more ergodicity for increasing system size. The system is definitely not ergodic at small Josephson coupling, but this parameter range shrinks as the system size is increased.
{We can access to limited system sizes and truncations of the Hilbert space, and we cannot easily extrapolate to larger values. Therefore, our results are not contrast with the prediction of many-body localization for this model at high energies and small coupling~\cite{altshuler}.} 
In particular, for the energies we have access to, we see that a dynamical behavior suggesting MBL (the persistence of the charge density wave) is actually due to the effect of a isolated eigenstates, similarly to what happens in the quantum-scar phenomenon.

We consider global probes of ergodicity both for a one-dimensional chain and a ladder geometry. {In this work, we have a model with short-range interactions and consider the properties of local observables (or objects that have been shown to share similar eigenstate-thermalization properties, like the half-system entanglement entropy). In this context eigenstate thermalization -- the Hamiltonian eigenstates being locally thermal -- and quantum chaos -- the Hamiltonian behaving in some respects as a random matrix -- occur together~\cite{kafri_2016}, and they go below the common name of ``ergodicity''. So, it is equivalent to look at global probes of quantum chaos or global probes of eigenstate thermalization (in both cases averages over the spectrum or the bulk eigenvalues), and we talk about global probes of ergodicity. There are cases of long-range systems where quantum chaos and eigenstate thermalization have no clear relation~\cite{michele2021,gradenigo}, or cases of integrable systems where a nonlocal observable thermalizes~\cite{gradenigo}, but they go beyond the scope of our work.}

{The global probe of eigenstate thermalization we choose is the smoothness quantifier of the eigenstate-expectations of a local observable. The idea is that if there is eigenstate thermalization the expectations of local observables behave as the microcanonical ones, that are smooth in energy~\cite{kafri_2016,PhysRevB.82.174411,michele2021}. For quantum chaos, we consider the inverse-participation ratio~\cite{thouless} (IPR) of the Hamiltonian eigenstates in the basis of the charge eigenstates (a measure of delocalization in this basis), and apply to it the smoothness quantifier. If there is quantum chaos, the Hamiltonian eigenstates behave as random states delocalized in the basis of the charge eigenstates over a energy window provided by $E_J$, and eigenstates with nearby energies look similar, giving rise to a smooth IPR~\cite{Torres_Herrera_2015,Santos_2012,Santos_2010,Gubin_2012}. We probe also quantum chaos looking at the average level-spacing ratio, which measures how much the spectrum of the Hamiltonian is similar to a random matrix~\cite{PhysRevB.82.174411,Haake,kafri_2016}. }

{All the global probes of ergodicity provide results in agreement and show a regime of small $E_J$ where there is no quantum chaos and no eigenstate thermalization. This regime shrinks for increasing system size, so the range of parameters where the system behaves ergodically becomes larger for increasing system size. This is true both for the chain geometry and the ladder one. In the latter case, we can add a finite magnetic flux. 
Looking at the global indicators of ergodicity,} we see that the finite magnetic flux makes the system more ergodic. {Together with that, the eigenstates of the Hamiltonian with finite flux become more delocalized in the basis of the charge eigenstates. A somewhat similar} delocalizing effect of the magnetic flux occurs in Anderson localization~\cite{rammer,rama} and MBL~\cite{PhysRevB.101.134203}. There, one has localization in physical space and the system is disordered and non-interacting. Our results show a similar effect in a many-body clean system, and here the localization phenomenon occurs in the many-body charge eigenstates basis. 

The paper is organized as follows. In Sec.~\ref{mod:sec} we present the two models (the Josephson chain and the Josephson ladder) we use in this paper.  
In Sec.~\ref{locini:sec} we show the existence of small-entanglement nonthermal eigenstates, which are similar to charge-density wave ordered states, and are a minority of the Hilbert space. In Sec~\ref{pert:app} we apply perturbation theory for $E_J\ll {E_C}$ and show that the small-entanglement nonthermal eigenstates we discuss, can already be seen in this approximation. In Sec.~\ref{dyno:sec} we show the effect on the dynamics of these eigenstates: Preparing the system in a charge density wave state, the order persists, for small enough Josephson coupling. In Sec.~\ref{ergo:sec} we discuss the global properties of ergodicity of the system.  
 In Sec.~\ref{conc:sec} we draw our conclusions. In Appendix~\ref{fluc:app} we discuss the energy fluctuations of the charge eigenstates (which is important for our discussion of global ergodicity probes) and in Appendix~\ref{effect:app} we provide a small discussion of the independence of our results on the chosen Hilbert-space truncation.
\section{Models} \label{mod:sec}
The {first} model we consider is a linear array of Josephson junctions represented in the scheme of Fig.~\ref{schema:fig}(a). Its Hamiltonian is
\begin{equation}\label{Ham0:eqn}
  \hat{H}=\sum_{i=1}^L\frac{1}{2}E_C\,\hat{q}_i^2-\sum_{i=1}^{L}E_J\cos\left(\hat{\theta}_i-\hat{\theta}_{i+1}\right)\,,
\end{equation}
{where $L$ is the length of the array. The system is a sequence of superconducting island [circles in Fig.~\ref{schema:fig}(a)] separated by junctions [crosses].} The first term of the Hamiltonian is the charging energy related to ground capacitance of an island (we neglect the {junction-capacitances}~\cite{altshuler}, {so that the charge interaction is onsite only and the screening length~\cite{Cole,PhysRevB.49.16773}  is vanishing}). The second term is the contribution due to the Josephson coupling between the islands [see the scheme in Fig.~\ref{schema:fig}(a)]. The canonical variables are the superconducting phases of the islands $\hat{\theta}_i$ and the corresponding number of Cooper pairs $\hat{q}_i$; In an appropriate experimental regime they behave in a quantum way~\cite{Leggett,clarke} and obey the commutation relation $\left[\hat{\theta}_i,\hat{q}_j\right]=i\delta_{i,j}$.
We will assume periodic boundary conditions, $\hat{q}_{L+1}\equiv\hat{q}_1$, $\hat{\theta}_{L+1}\equiv\hat{\theta}_1$.
\begin{figure}[h!]
  \begin{center}
    \begin{tabular}{c}
      \begin{overpic}[width=80mm]{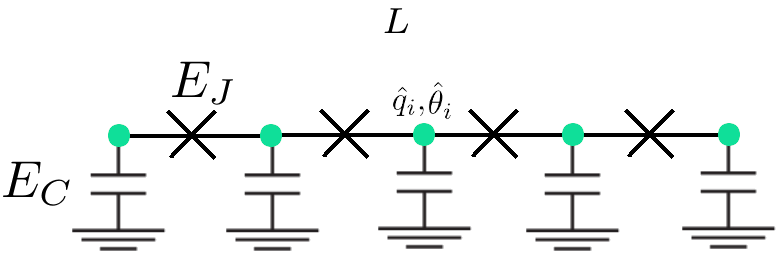}\put(0,22){(a)}\end{overpic}\\
      \\
      \hspace{0.4cm}\begin{overpic}[width=78mm]{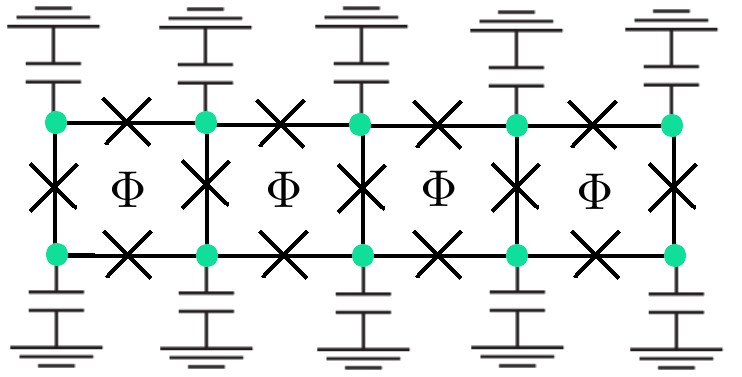}\put(-6,22){(b)}\end{overpic}\\
    \end{tabular}
  \end{center}
\caption{(Panel a) Scheme of the one-dimensional chain of Josephson junctions Chain model.. {The crosses are the junctions and the circles are the superconducting islands, and also the ground capacitances are shown.} (Panel b) Scheme of the Josephson-junctions ladder with magnetic flux $\Phi$. We assume that the ground capacitances dominates over the capacitances of the junctions, and neglect the latter.}
\label{schema:fig}
\end{figure}

This is a charge-conserving Hamiltonian. The total charge operator is defined as $\hat{Q}=\sum_i\hat{q}_i${, and is conserved}~\cite{note_Q}.
The charge operator $\hat{q}_i$ is the number of Cooper pairs, so is quantized with integer eigenvalues. This fact and the commutation relation imply $\nep^{i\hat{\theta}_j}\ket{q_j}=\ket{q_j+1}$. Using this relation, we can write the Hamiltonian in the basis of eigenstates of the charge operators $\{\hat{q_i}\}$~\cite{notabile}.
In any numerical implementation, a truncation of the Hilbert space is needed, forcing all the $q_i$ to run from $-M$ to $M$, for some integer $M$. We construct the truncated Hilbert space with the basis $\ket{q_1,\,q_2,\,\ldots,\,q_L}$ with $|q_j|\leq M\;\forall j$. 

We consider also a ladder configuration where we can put a magnetic flux $\Phi$ [see the scheme in Fig.~\ref{schema:fig}(b)]. Calling $\Phi_0$ the flux quantum, we find the Hamiltonian to be
\begin{align}\label{ham_fl:eqn}
  \hat{H}&=\sum_{i=1}^L\sum_{k=1}^2\frac{1}{2}E_C\,\hat{q}_{i,\,k}^2-\sum_{i=1}^{L}E_J\cos\left(\hat{\theta}_{i,\,1}-\hat{\theta}_{i,\,2}\right)\nonumber\\
    &-\sum_{i=1}^{L}E_J\cos\left(\hat{\theta}_{i,\,1}-\hat{\theta}_{i+1,\,1}\right)\nonumber\\
    &-\sum_{i=1}^{L}E_J\cos\left(\hat{\theta}_{i,\,2}-\hat{\theta}_{i+1,\,2}-2\pi\Phi/\Phi_0\right)\,,
\end{align}
where the phases $\hat{\theta}_{i,\,k}$ are now gauge invariant~\cite{tinkham}. We take periodic boundary conditions along the horizontal direction $i$, and open boundary conditions along the vertical direction $k$. Thanks to the usual commutation rules $\left[\hat{\theta}_{i,\,k},\hat{q}_{j,\,q}\right]=i\delta_{i,j}\delta_{k,q}$, the construction of the Hilbert space and the expansion in eigenstates of $\hat{q}_{i,\,k}$ are strictly analogous to the one considered for Eq.~\eqref{Ham0:eqn}~\cite{notabile}. Also the resulting structure is very similar, with the important difference of some phase factors $\exp(\pm 2\pi i \Phi/\Phi_0)$ in some of the Josephson terms. In all the paper, we choose $\Phi/\Phi_0=1/\phi$, where $\phi$ is the golden ratio. We do that in order to avoid any possible effect of commensuration with the ladder~\cite{PhysRevB.14.2239}. {We term the $\Phi=0$ case as ``vanishing flux'' and the $\Phi=\Phi_0/\phi$ case as ``finite flux''.} 

%

{We exploit the symmetries of the Hamiltonian in order to reduce our analysis to a Hilbert subspace. The Hamiltonian conserves the charge, and we always restrict to the subspace with total charge $Q=0$. Moreover, there is translation invariance along the horizontal direction, and we further restrict to the subspace with vanishing horizontal momentum. The chain model has reflection symmetry, and the ladder model with $\Phi=0$ has reflection symmetries along both the horizontal and the vertical axes. We further restrict to the subspace even under these symmetries. We call the resulting operator the ``fully even Hamiltonian'' and the corresponding subspace the ``fully even subspace''. Of course, when considering a dynamics with an initial state with less symmetries than the Hamiltonian, we will restrict to a larger Hilbert subspace.}

When the flux $\Phi$ is nonvanishing, the symmetries of the problem are reduced. In the $\Phi=0$ case, the symmetries are the translations along the ladder, the swap of the two rows of the ladder, the reflection along the row of the ladder and the parity $q\to -q$.
If $\Phi\neq 0$ the swap symmetry disappears, and when the inversion and the parity act, must also the flux $\Phi$ change sign in order to keep the Hamiltonian unchanged. This property is related to the fact  that the magnetic flux is an axial vector and changes sign under reflections. So, the inversion symmetry and the parity disappear as symmetries. The new symmetry which results is the composition of inversion and parity. Applying the flux gives rise therefore to a symmetry breaking.
 

We numerically study the dynamics using exact diagonalization when possible, the Krylov method implemented in the {\sc Expokit} package~\cite{EXPOKIT}, or the tDMRG~\cite{Schollwock_rev} algorithm implemented through the ITensor Library~\cite{ITensor,notaD}. 
{(With the first two methods we can restrict to the fully even subspace.)} Throughout the paper we set $E_C=1$.

%
%
\section{Weak ergodicity breaking}\label{locini:sec}
%
%
%
%
In order to probe if an eigenstate of the Hamiltonian $\ket{\phi_\alpha}$ is {nonthermal}, we probe 
%
%
the entanglement entropy. 
{In order to define it, we} divide the system in two partitions 
$A$ and $B$ in the following way. {Both in the ladder and the chain model, assuming $L$ even, we cut the system perpendicularly to the horizontal direction of Fig.~\ref{schema:fig} in two equal parts. We get in this way $A$ and $B$ subsystems {with the same size and shape}, and $L/2$ long.~\cite{note_s}} So the full Hilbert space 
(not the symmetrized one) has a tensor product structure $\mathcal{H}=\mathcal{H}_A\otimes\mathcal{H}_B$. We define the half-chain entanglement entropy for each eigenstate as
\begin{equation} \label{entropy:eqn}
  S_{L/2}^{(\alpha)}=-\Tr_A[\hat{\rho}_A^{(\alpha)}\log\hat{\rho}_A^{(\alpha)}]\quad{\rm with}\quad \hat{\rho}_A^{(\alpha)}=\Tr_B[\ket{\phi_\alpha}\bra{\phi_\alpha}]\,,
\end{equation}
where $\Tr_B$ is the partial trace over $\mathcal{H}_B$. 

{It is important to emphasize that, if an eigenstate $\ket{\phi_\alpha}$ is locally thermal and obeys ETH, its half-system entanglement entropy {is $S_{L/2}^{(\alpha)}= \frac{L}{2}s_{\rm m.c.}(E_\alpha)+O(1)$, where  $s_{\rm m.c.}(E_\alpha)$ is the microcanonical entropy density at energy $E_\alpha$}~\cite{russomanno2020nonergodic,PhysRevB.93.134201,Huang_NPB19}. Therefore, if an eigenstate has an entanglement entropy order 1, does not for sure obey ETH and is therefore nonthermal. }

Violations of ETH at the spectral boundaries {are not very surprising (for instance, for noncritical Hamiltonians, the ground state and eigenstates with vanishing excitation-energy density obey an area-law entanglement~\cite{saito,RevModPhys.82.277,Hastings_2007}). Moreover, many of the violations of ETH at the spectral boundaries are believed to be a finite-size effect. The point is that near the boundaries the microcanonical entropy is much smaller than in the inner part of the spectrum, then an ETH eigenstate is a random superposition of less factorized eigenstates. So, the expectation over an ETH eigenstate near the spectral boundary has larger fluctuations around the microcanonical value. As a consequence, at the small system sizes which exact diagonalization can attain, states at the center of the spectrum show a very good ETH behavior, while states at the boundaries show fluctuations around ETH (see for instance~\cite{Santos_2010,Torres_Herrera_2015,PhysRevA.90.033606,russomanno2020nonergodic}). So, finding isolated small-entanglement eigenstates far from the edges of the spectrum, surrounded by eigenstates with large entanglement -- as we are going to do -- is a more interesting and unexpected phenomenon.}

{Before focusing on the entanglement entropy, let us define some special charge eigenstates important for this analysis as
\begin{align}\label{stat:eqn}
  \ket{\psi_2(q)}&=\ket{-q,\,q,\,-q,\,q,\,\ldots}\nonumber\\
  \ket{\psi_3(q,-q,0)}&=\ket{q,\,-q,\,0,\,q,\,-q,\,0,\,\ldots}\,.
\end{align}
They are both charge-density wave states~\cite{notec}, the first one with periodicity 2, the second with periodicity 3. For the latter, we consider also the related states $\ket{\psi_3(\Pi(q,-q,0))}$, where a permutation $\Pi$ is applied to the periodically-repeating domain $(q,-q,0)$. Notice that in both cases we restrict to the Hilbert space sector with vanishing total charge $Q=0$.} 
%

Let us start considering the entanglement entropy of the eigenstates in the chain model [Fig.~\ref{IPR1:fig}(a)]. We see two series of low-entanglement-entropy states, one near $S_{L/2}\sim\log(2)$ and another near $S_{L/2}\sim\log(6)$. {We see that both series of eigenstates persist also for values of $E_J$ -- for instance $E_J=0.8$ -- which lie out of the regime {where the spectrum is overall organized in multiplets}. [In this regime, defined by $E_J\ll E_C/L$, perturbation theory can be applied -- see Sec.~\ref{pert:app}].} The $\log 2$ eigenstates shown in Fig.~\ref{IPR1:fig} are even under the symmetries of the Hamiltonian, and correspond to states similar to
\begin{equation}\label{pipp:eqn}
  \ket{\psi_+(q)}\equiv\frac{1}{\sqrt{2}}\left[\ket{\psi_2(q)}+\ket{\psi_2(-q)}\right]\,,
\end{equation}
with $\ket{\psi_2(q)}$, $\ket{\psi_2(-q)}$ dressed by the interaction with other
charge configurations.
{We can directly check this. From one side, the $\log 2$ eigenstates occur at energies near to $E^{(0)}(q)=\braket{\psi_+(q)|\hat{H}|\psi_+(q)}=Lq^2/2$. From the other -- as we are going to show in Sec.~\ref{dindyno:sec} for $q=2$ -- the eigenstate having maximum overlap with $\ket{\psi_+(2)}$, is right the one having entanglement entropy $\simeq\log 2$ and energy $\simeq E^{(0)}(2)$. 

Actually, there is not only the even $\log 2$ eigenstate, but also the odd one, that in the limit $E_J\ll E_C/L$ has the form $\ket{\psi_-(q)}=\frac{1}{\sqrt{2}}\left[\ket{\psi_2(q)}-\ket{\psi_2(-q)}\right]$. The latter is outside the fully even subspace and we cannot see it in our numerics. At second order in $E_J$, the Hamiltonian is degenerate in the 2 states $\ket{\psi_2(q)}$, $\ket{\psi_2(-q)}$ (see Sec.~\ref{pert:app}). At further orders, the two states are coupled and the eigenstates become the quasidegenerate $\ket{\psi_\pm(q)}$.}



{The $E_J=0$-Hamiltonian is also degenerate in the 6 states $\ket{\psi_3(\Pi(-q,q,0))}$, where $\Pi$ is a generic permutation of the three arguments. Applying the interaction $E_J\ll E_C$, the eigenstates become superpositions of these six states. Here we restrict to the fully even subspace, and we can see only the superposition of the 6 states with equal weights and equal phases. This fact provides the value $\log(6)$ of the entanglement entropy. Considering the full Hilbert space, we can construct 6 possible superpositions (each one with different phases), so each point at $\log(6)$ corresponds to a sextuplet. What is remarkable, is that this spectral structure persists also when $E_J$ is not infinitesimal, but has a value order 1.}

Some remarks are in order. First, here the truncation $M$ is finite, but we can see low-entanglement eigenstates far from the upper edge of the energy spectrum. So we can believe that these eigenstates persist also in the limit $M\to\infty$. It is also important to notice that the states of the form Eq.~\eqref{stat:eqn} are a countable infinity when $M\to\infty$. The nonthermal eigenstates are a dressed superposition of a finite number of them, so they are also infinite, at any finite size $L$. This fact is in contrast with the usual many-body quantum scars, whose number grows with the system size~\cite{scars_turner,2021_serbyn_nat}. {Let us emphasize again that we focus our attention on nontrivial low-entanglement states, {\em i.e.} the ones surviving beyond the perturbative regime, and lying far from the spectral edges (so clearly distinct from a finite-size effect or an artifact of the truncation~\cite{notaccola}.)}

{Second,} we see a dependence of the structure of nonthermal eigenstates on the divisors of $L$. 2 is a divisor of both $6$ and $8$, so we see $\log(2)$ states in both cases [Fig.~\ref{IPR1:fig}(a,b)], but we find $\log(6)$ eigenstates only in the former [Fig.~\ref{IPR1:fig}(a)]. In this case, 3 is a divisor of 6, and the period-3 charge density wave states $\ket{\psi_3}$ [Eq.~\eqref{stat:eqn}] needed for the $\log(6)$ eigenstates can be accommodated. {For the case $L=8$ we see also eigenstates with entanglement entropy $\log 4$. In the perturbative regime, they come from the superposition of 4 states, the charge-density-wave state with periodicity 4 $\ket{\psi_4(q)}=\ket{q,\,0,\,-q,\,0,\,q,\,0,\,-q,\,0,\,\ldots}$, and the other distinct 3 obtained by translating it. Also the presence of these states depends on the divisors of $L$ ($L$ must be a multiple of 4), but we can see that they are fragile and disappear when one moves away from the perturbative regime.}


{Moving to the case of the ladder without magnetic flux, we see that the situation is very similar to the chain-model case. Still restricting our numerics to the fully even subspace, also here there are the $\log 2$ nonthermal eigenstates, as we see in the $S_{L/2}^{(\alpha)}$ versus $E_\alpha$ plots in Fig.~\ref{IPR:fig}. These eigenstates appear around the energies $\mathcal{E}^{(0)}(q)=Lq^2$, and we mark with a blue circle the one for $q=2$ and coupling values $E_J=0.05,\,0.2$ in Fig.~\ref{IPR:fig}. This even eigenstate has an odd quasidegenerate partner, that lies outside the subspace where we perform the exact diagonalization. As we discuss in better detail in Sec.~\ref{dynlad:sec}, these eigenstates have a large overlap respectively with the even and odd superpositions of the charge-density wave states $\ket{\psi_2(\pm q)}$ [Eq.~\eqref{staf:eqn}]. 

{These eigenstates can be found for small $E_J$ using perturbation theory (App.~\ref{pert:app}). Going beyond the perturbative regime, although the overall multiplet structure of the spectrum disappears, these small entanglement eigenstates can still be recognized, but they are dressed by the interaction and the entanglement entropy increases ($E_J=0.8$ -- orange circle in Fig.~\ref{IPR:fig}). Such a strong effect was not present in the linear chain for the same value of $E_J$ (Fig.~\ref{IPR1:fig}), so we see that the effect of the Josephson coupling in the ladder is stronger than in the chain.}

We do not explicitly discuss the case with finite flux $\Phi$, but we expect that the small-entanglement eigenstates are still there, because they can be found in the perturbative regime (App.~\ref{pert:app}). Outside this regime, they might survive in a parameter range different than the case $\Phi=0$. As we are going to see in Sec.~\ref{dynlad:sec}, results of the dynamics suggest that this parameter range is smaller.

{ In the next section} we are going to see how these {low-entanglement} eigenstates {appear already in the perturbative regime.}

%
%
\begin{figure}
  \begin{center}
  \begin{tabular}{c}
     \begin{overpic}[width=80mm]{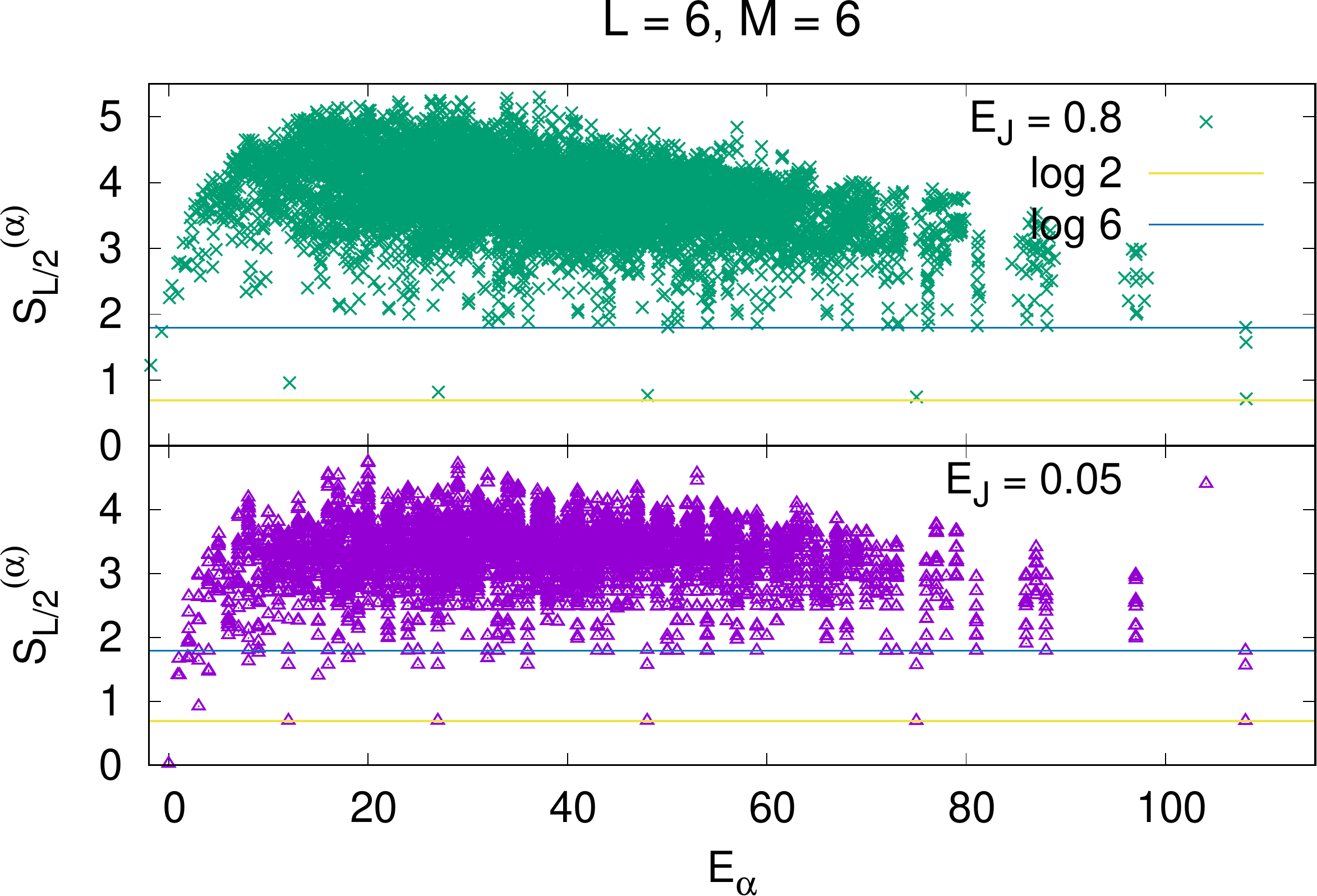}\put(10,66){(a)}\end{overpic}\\
     \\
     \begin{overpic}[width=80mm]{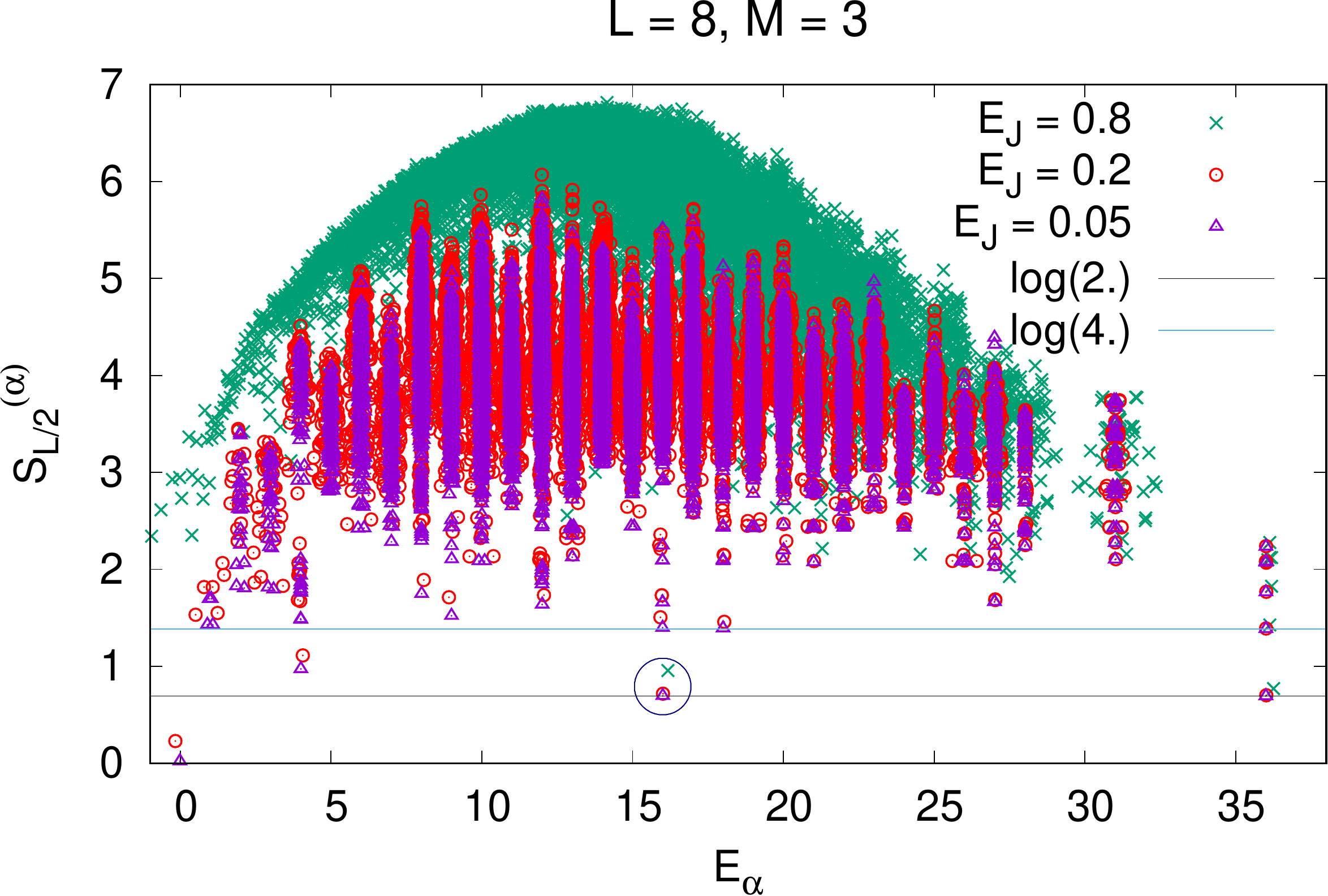}\put(10,66){(b)}\end{overpic}
   \end{tabular}
 \end{center}
\caption{Chain model. (Panel a) Half-system entanglement entropy $S_{L/2}^{(\alpha)}$ versus $E_\alpha$ for $L=M=6$ and different $E_J$. The horizontal lines mark the ergodicity-breaking eigenstates with entanglement entropy $\log(2)$ and $\log(6)$. These eigenstates persist also for $E_J=0.8$, well beyond the perturbative regime. (Panel b) The same for $L=8$, $M=3$. Inside the circle there is the even $\log 2$ eigenstate for $q=2$, which is physical and not a finite size effect, being far from the spectral edges. We can see also the $\log 4$ eigenstates, coming from superpositions of charge-density-wave states with period 4, {and persisting only} in the small-$E_J$ perturbative regime, {where the spectrum is globally organized in multiplets.}} 
\label{IPR1:fig}
\end{figure}
\begin{figure}
  \begin{center}
  \begin{tabular}{c}
     \begin{overpic}[width=80mm]{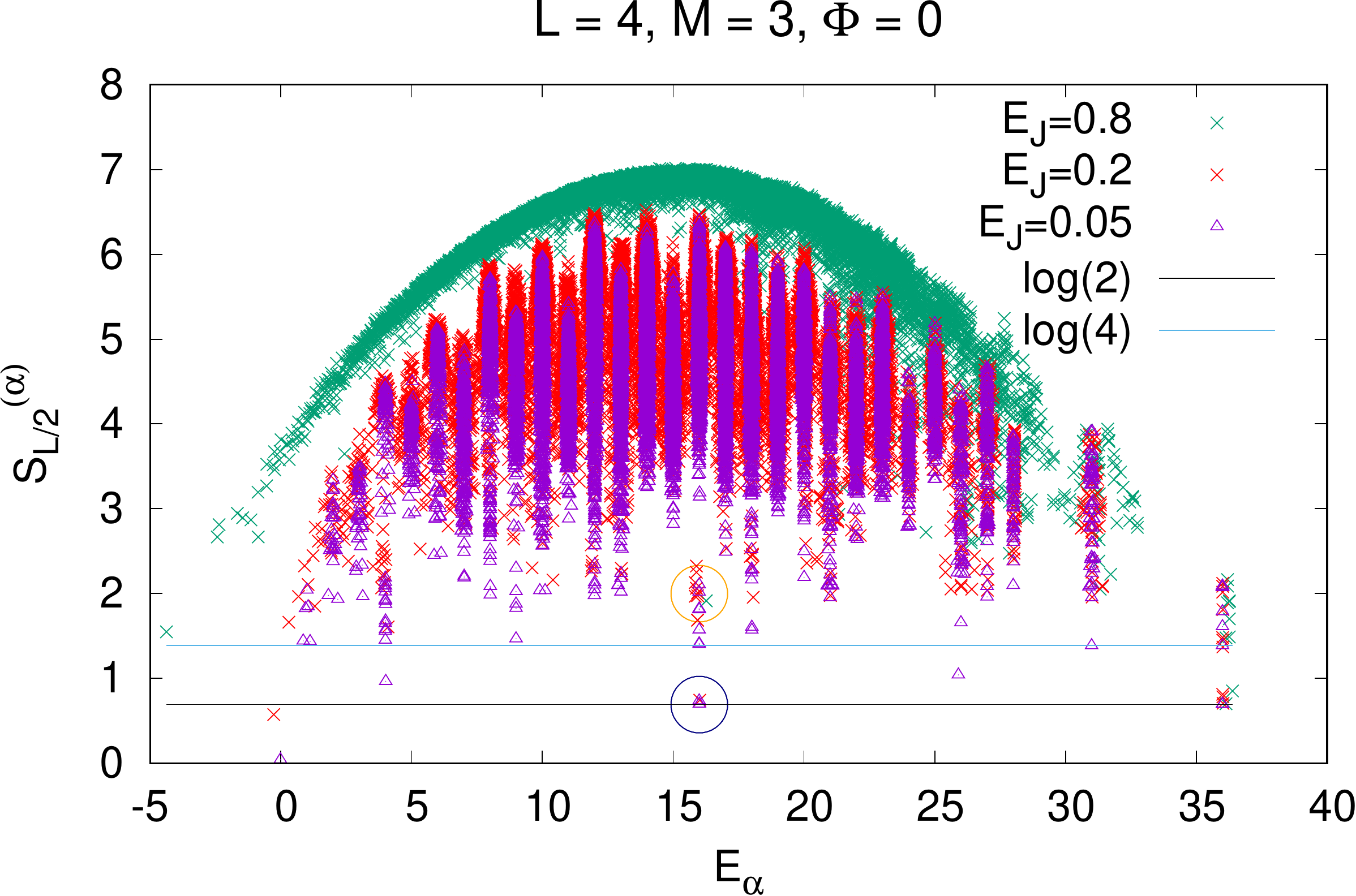}\put(20,32){}\end{overpic}\\
%
    \end{tabular}
  \end{center}
\caption{Ladder model with vanishing flux $\Phi=0$. Half-system entanglement entropy $S_{L/2}^{(\alpha)}$ versus $E_\alpha$ for $L=4$, $M=3$ and different $E_J$. Notice the even $\log 2$ eigenstates for $q=2$ (blue circle). {For $E_J=0.8$ the state can still be recognized, but the entanglement entropy is larger (orange circle)}. For $E_J\ll 1$ there are also some $\log 4$ states which disappear when $E_J$ is larger ($E_J=0.8$), the spectrum is no more organized in multiplets, and perturbation theory cannot be applied.}
\label{IPR:fig}
\end{figure}
\section{{Perturbation theory}}\label{pert:app}

{
In this section we show that the low-entanglement eigenstates discussed above can be understood through a perturbative analysis, when $E_J\ll E_C/L$.}
{
In order to apply perturbation theory, let us write our Hamiltonians Eq.~\eqref{Ham0:eqn} and~\eqref{ham_fl:eqn} as 
\begin{equation}\label{brano:eqn}
\hat{H}=\hat{H}_0+E_J\hat{V}\,,
\end{equation}
where $\hat{H}_0$ is the charging part and $E_J\hat{V}$ the Josephson part.
The spectrum of $\hat{H}_0$ splits the Hilbert space into degenerate multiplets. When a perturbatively small $E_J$ is considered, the degeneracies within multiples are lifted. In the perturbative regime we expect the energy dispersion acquired by each multiplet to be proportional to $E_J L$, while the separation between distinct multiplets is proportional to $E_C$. We expect perturbation theory to yield correct results as long as the multiplets are well-separated, i.e. $E_J\ll E_C/L$.}

{
In this regime, within each multiplet we can identify a lowest-energy state and an highest-energy state, which can be respectively described as the ground state and the highest-energy state of the effective Hamiltonian obtained from perturbation theory and projected to the multiplet in question.
Crucially, if the effective Hamiltonian projected to a multiplet is approximately local and has a gap above its ground state (or below the highest-energy state), then the ground state (or highest-energy state) within the multiplet has following properties. (i) The state changes adiabatically with $E_J$. (ii) The state does not obey ETH, e.g. its entanglement entropy will obey an area law. These are standard properties for a ground state (or an highest excited state), but in this context they are not trivial as the ground state (or the highest-excited state) within some multiplet is embedded in the middle of the spectrum of $\hat{H}$.}

{
	In the rest of this section we exemplify the discussion by considering the state $\ket{\psi_+(q)}$ defined in Eq.~\eqref{pipp:eqn} and showing that, in the perturbative regime, it is adiabatically connected to the highest-energy state of the multiplet with energy $E^{(0)}\simeq L\frac{E_C}{2}q^2$. The latter state is the version of $\ket{\psi_+(q)}$ dressed by the
interaction and provides the even log 2 eigenstate. We remark that this multiplet is far from the ground state of $\hat{H_0}$ and also $\hat{H}_0$ does not have an highest-excited state, since its eigenvalues are not bounded from above. Therefore we would naively expect the states within this multiplet to obey some form of ETH.
}


%
%

%
 The eigenstates of $\hat{H}_0$ are the charge configurations $\ket{q_1,\,q_2,\,\ldots,\,q_L}$, simultaneous eigenstates of the local charge operators $\hat{q}_j$. {At $E_J=0$ the spectrum is organized in degenerate multiplets, and each multiplet {contains} the set of all possible site permutations of some charge configuration (all them have the same charging energy $E_{\{q_j\}}^{(0)}=\frac{E_C}{2}\sum_jq_j^2$).} Applying second order perturbation theory, the states in each multiplet mix with each other, and inside each multiplet, one can write an effective Hamiltonian~\cite{note_mik}
{
\begin{align}
  \hat{H}^{(2)} &= \hat{H}_0+E_J\sum_{\{q_j\},\,\{q_j'\}}\hat{\Pi}_{\{q_j'\}}\hat{V}\hat{\Pi}_{\{q_j\}}\nonumber\\
  &+E_J^2\sum_{\{q_j'\},\,\{q_j\},\,\{p_j\}}\frac{\hat{\Pi}_{\{q_j'\}}\hat{V}\hat{\Pi}_{\{p_j\}}\hat{V}\hat{\Pi}_{\{q_j\}}}{E_{q}-E_{\{p_j\}}}\,,
\end{align}
where we have defined the projectors as $\hat{\Pi}_{\{q_j\}}=\ket{\{q_j\}}\bra{\{q_j\}}$, and the charge configurations $\ket{\{q_j\}}$, $\ket{\{q_j'\}}$ belong to the considered multiplet and have degenerate charging energy $E_q$, while the configurations $\ket{\{p_j\}}$ belong to the other multiplets and have charging energies $E_{\{p_j\}}$.}

Let us now focus on the chain model charge configurations $\ket{q_1,\,q_2,\,\ldots,\,q_L}$ such that $|q_j-q_{j+1}|\geq 3$ $\forall\,j\in1,\,\ldots,\,L$. {{Neglecting non-local terms (see Ref.~\onlinecite{note_mik} for further discussion),} one sees that $\hat{V}$ at first order does not mix these charge-configuration states with other ones inside the corresponding multiplet. One finds that these charge-configurations states} are eigenstates of the effective Hamiltonian $\hat{H}^{(2)}$ with eigenvalue
\begin{equation} \label{bruno:eqn}
  E_{\{q_j\}}^{(2)}=E_{\{q_j\}}^{(0)}+E_J^2\sum_{j=1}^L\frac{2}{(q_j-q_{j+1})^2-1}\,.
\end{equation}
For the ladder model, the analysis is the same, but one has to consider states such that $|q_j-q_{i}|\geq 3$ $\forall$ $j$ and $i$ nearest neighbors and the sum in restrict the sum in Eq.~\eqref{bruno:eqn} runs on all the nearest-neighbor pairs of charges.

We can consider some cases. If we consider the multiplet where half of the sites have charge $+q$, and half have charge $-q$, we find that $E_{\{q_j\}}^{(2)}$ is maximized by two classical configurations, the $\ket{\psi_2(\pm q)}$ defined in Eq.~\eqref{stat:eqn} for the chain model and in Eq.~\eqref{staf:eqn} for the ladder model. The two $\ket{\psi_2(\pm q)}$ are degenerate at order $E_J^2$; Because we restrict to subspaces fully even under the symmetries of the Hamiltonian, we get $\ket{\psi_+(q)}=\frac{1}{\sqrt{2}}(\ket{\psi_2(q)}+\ket{\psi_2(-q)})$, whose entanglement entropy is $\log 2$ in Figs.~\ref{IPR1:fig} and~\ref{IPR:fig}.
{To any order in perturbation theory these will change smoothly with $E_J$, since the mixing with the other states in the multiplet is protected by the finite gap (proportional to $E_J^2/E_C$) between $\ket{\psi_2(\pm q)}$ and the rest of the states in the multiplet.
Note, however, that this analysis is inconclusive beyond the perturbative regime, where the gap with states from other multiplets will become exponentially small in $L$.
However, as we have shown numerically,} these $\log 2$ eigenstates survive well beyond the perturbative regime. We emphasize that Eq.~\eqref{bruno:eqn} does not depend on the phases, so in the ladder case is valid also for a {nonvanishing} flux $\Phi\neq 0$.

We can also choose to restrict to multiplets where the configurations are such that half of the sites have charge 0, one fourth charge $q$ and one fourth charge $-q$. Considering the linear chain, we see that there are exactly 4 distinct configurations maximizing Eq.~\eqref{bruno:eqn}. All them are charge density waves with period four, one is $\ket{q,\,0,\,-q,\,0,\,\ldots}$ (where $\ldots$ mean repetition with periodicity), and the other 3 are obtained by means of translations. Inside the fully even subspace, we see the equal-weight superpositions of these four, which have entanglement entropy $\log 4$, as we see in Fig.~\ref{IPR1:fig}. These states are more fragile and do not survive beyond the perturbative regime.

{Remarkably, the $\log 2$ and $\log 6$ eigenstates persist beyond the perturbative regime, and have important effects on the dynamics, forbidding thermalization and leading to the persistence of an initial charge-density wave order, as we clarify in the next section.}
\section{Dynamics}\label{dyno:sec}
{Both the states considered in Eq.~\eqref{stat:eqn} are simple product states which have a charge-density wave order. If we initialize the dynamics in one of these states, because of the large overlap with the nonthermal eigenstates introduced above, there is no thermalization. In particular, the charge-density wave order does not melt, but persists for long times.
\subsection{Chain model}\label{dindyno:sec}
  In order to probe this persistence, we have to consider some observables. If we initialize with a period-2 charge density wave -- for instance $\ket{\psi_2(q)}$ in Eq.~\eqref{stat:eqn} -- we can consider the normalized imbalance
\begin{equation} \label{iotta:eqn}
  \hat{\mathcal{I}}\equiv\frac{1}{qL}\sum_{i=1}^L(-1)^i\hat{q}_i
\end{equation}
(the full imbalance is $q\hat{\mathcal{I}}$), and the expectation of its powers, $\mathcal{I}_{(\alpha)}(t)=\braket{\psi(t)|\hat{\mathcal{I}}^\alpha|\psi(t)}$ {(that's to say its moments)}. The imbalance is nonvanishing if there is a period-2 charge density wave, and is a standard object used to probe localization and nonergodicity phenomena~\cite{Bloch2019,russomanno2020nonergodic,Carleo}.}

{If we chose to initialize the system in one of the $\ket{\psi_2(\pm q)}$, and studied the expectation of the imbalance $\mathcal{I}_1(t)$, we would have got Rabi oscillations of this quantity around 0 [see Fig.~\ref{comparison_M:fig}(a,b) and~\cite{russomanno2020nonergodic} for the detailed discussion if this phenomenon in the Bose-Hubbard model]. The states $\ket{\psi_2(\pm q)}$ are resonant at second order in $E_J$ (see Appendix~\ref{pert:app}), and they are coupled at further orders in $E_J$, so that there are the Rabi oscillations between the two (~\cite{russomanno2020nonergodic} and Appendix~\ref{effect:app}). The result is that the long time-average of $\mathcal{I}_1(t)$ is always vanishing, and one cannot use it to do statements on the persistence of the charge-density order.}

{That's the reason why we consider the dynamics of the operator $\hat{\mathcal{I}}^2$. This operator is degenerate on $\ket{\psi_2(\pm q)}$, so one cannot see Rabi oscillations in the dynamics of its expectation $\mathcal{I}_2(t)$, but convergence to a regime where there are oscillations around a finite value. We can probe this value through a time average; When this value is order one and does not scale to 0 with system size, we can speak about persistence of the charge density order. Moreover, due to the degeneracy of $\hat{\mathcal{I}}^2$ in $\ket{\psi_2(\pm q)}$, we can initialize with the even superposition of the two $\ket{\psi_+(q)}$ [see Eq.~\eqref{pipp:eqn}], and get the same dynamics for $\mathcal{I}_2(t)$.}

{So we focus on the long-time average of the second moment of the imbalance $\mathcal{I}_{(2)}=\frac{1}{T}\int_0^T\mathcal{I}_{(2)}(t)\ud t$, choosing $T$ big enough so that convergence is reached. We take $q=2$ and show $\mathcal{I}_{(2)}$ versus $E_J$ in Fig.~\ref{I2:fig}(a). We see a clear decrease of $\mathcal{I}_{(2)}$ with $L$ for $E_J\gtrsim 0.75$. For smaller values of $E_J$, $\mathcal{I}_{(2)}$ is almost independent of $E_J$. For $L\leq 8$ we can do exact diagonalization, and the limit $T\to\infty$ can be performed exactly, so that the infinite-time average is $\mathcal{I}_2=\sum_\alpha |\braket{\phi_\alpha|\psi_+(2)}|^2\braket{\phi_\alpha|\hat{\mathcal{I}}^2|\phi_\alpha}$, where $\ket{\phi_\alpha}$ are the eigenstates of the Hamiltonian. For $L=10$ we perform the evolution using the Krylov technique with dimension of the Krylov subspace $m=100$ (for smaller $L$ this value provides results indistinguishable from exact diagonalization).} 


As we have discussed in Sec.~\ref{locini:sec} and~\ref{pert:app}, the even $q=2$ nonthermal $\log 2$ eigenstate should have large overlap with the initial state $\ket{\psi_+(2)}$, and so play an important role in the dynamics.
{We confirm this prediction evaluating} the maximum over the eigenstates $\ket{\phi_\alpha}$ of the square overlap $|\braket{\psi_+(2)|\phi_\alpha}|^2$ [upper panel of Fig.~\ref{I2:fig}(b)]. We see that the maximum square overlap is near to 1 in the interval of $E_J$ where $\mathcal{I}_{(2)}$ is almost independent of $L$ and near 1, {so the dynamics is mainly affected by a single eigenstate}. The energy of the maximum-overlap eigenstate $\ket{\phi_{\overline{\alpha}}}$ -- let us call it $E_{\rm max}$ -- is very {near} to the energy $E_i\equiv E^{(0)}(2)=\braket{\psi_+(2)|\hat{H}|\psi_+(2)}=2L$ of the initial state [lower panel of Fig.~\ref{I2:fig}(b)]. We have directly checked that $\ket{\phi_{\overline{\alpha}}}$ has entanglement entropy $\simeq\log 2$. [an example for the case $E_J=0.05$ is shown in the blue circle of the inset in Fig.~\ref{I2:fig}(b)]. }

{So, we conclude that, when the charge density wave order persists, the dynamics is deeply affected by the eigenstate $\ket{\phi_{\overline{\alpha}}}$, which is {the even} $q=2$ nonthermal $\log 2$ eigenstates observed in Fig.~\ref{IPR1:fig} (inside the blue circle). 
In particular, it forbids the thermal melting of an initial charge-density wave order. The same situation occurs in the many-body quantum scar phenomenon. }
%
\begin{figure}
  \begin{center}
  \begin{tabular}{c}
     \begin{overpic}[width=80mm]{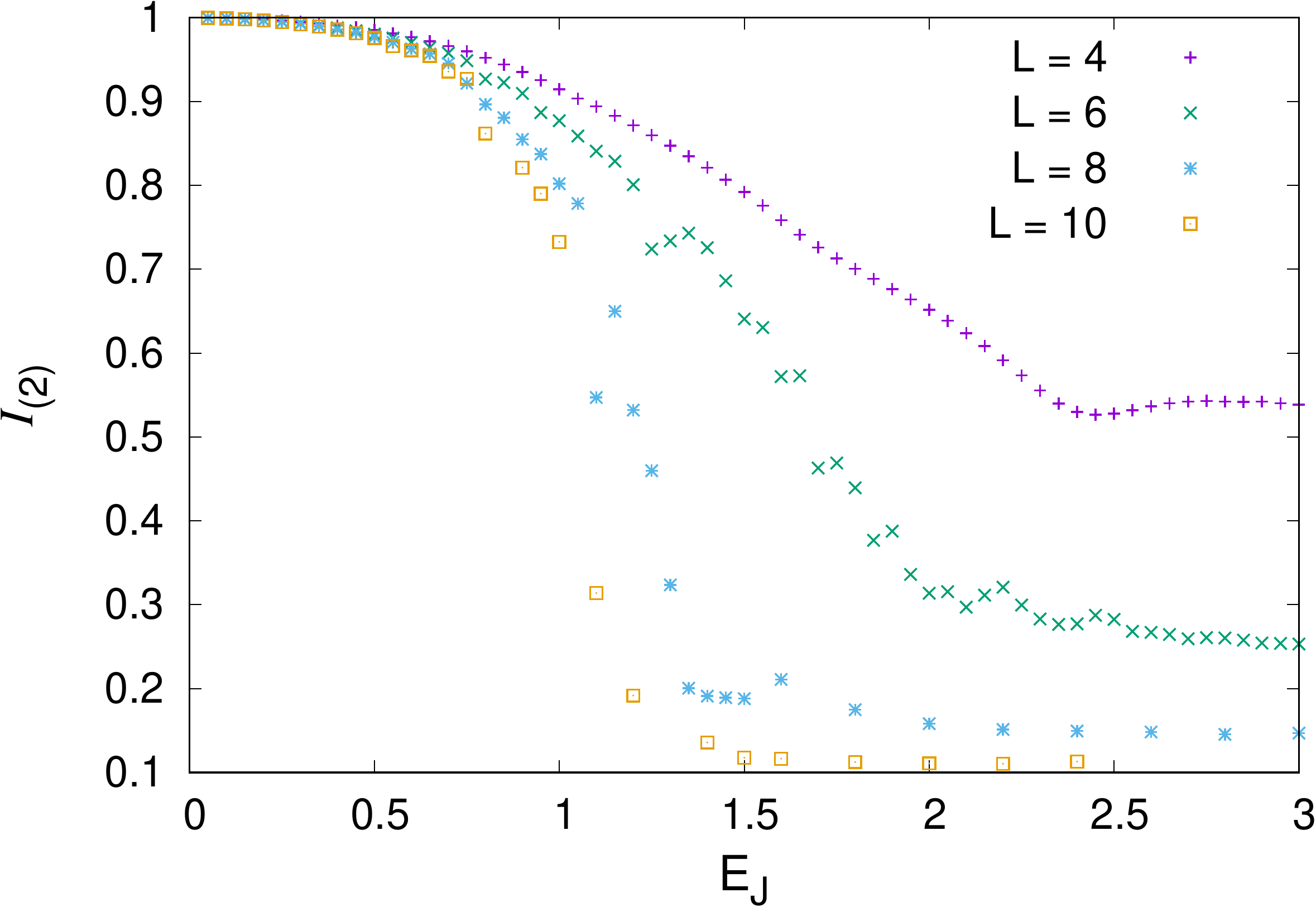}\put(15,71){(a)}\end{overpic}\\\vspace{3mm}\\
     \begin{overpic}[width=80mm]{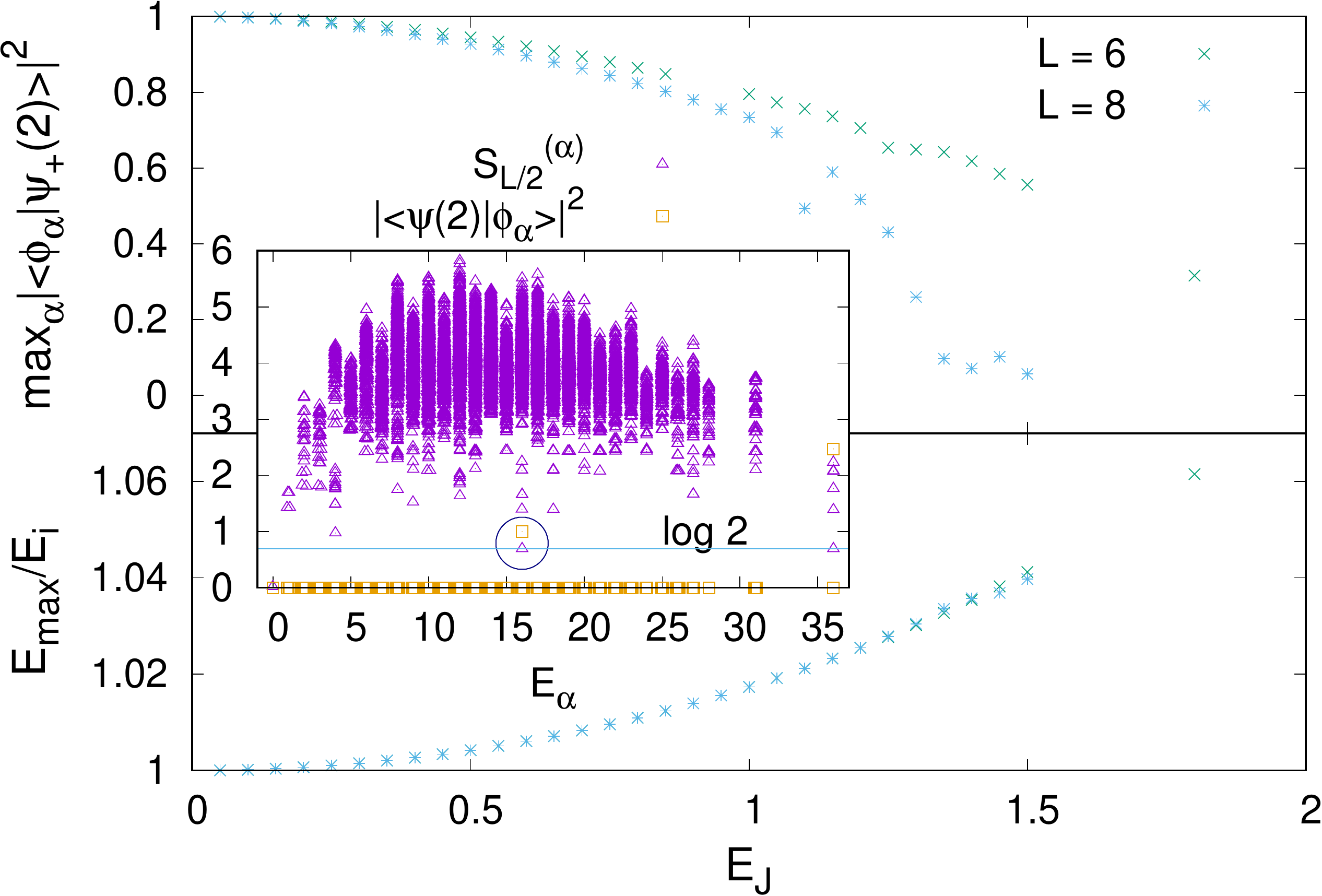}\put(15,70){(b)}\end{overpic}
   \end{tabular}
 \end{center}
\caption{Chain model.  Numerical parameters, $q=2$, $M=3$. (Panel a) $\mathcal{I}_{(2)}$ versus $E_J$. Exact diagonalization for $L\leq 8$, Krylov technique with dimension of the Krylov subspace $m=100$ for $L=10$. Initialization with $\ket{\psi_+(2)}$ for $L=4,\,6,\,8$; $\ket{\psi_2(2)}$ for $L=10$ (these two initial states provide the same dynamics for $\mathcal{I}_{(2)}$, as explained in the text). (Panel b, upper) Maximum over the eigenstates $\ket{\phi_\alpha}$ of the square overlap $|\braket{\phi_\alpha|\psi_+(2)}|^2$ versus $E_J$. (Panel b, lower) Energy $E_{\rm max}$ of the eigenstate $\ket{\phi_{\overline{\alpha}}}$ realizing this maximum divided by $E_i$ -- the energy of $\ket{\psi_+(2)}$ --, versus $E_J$. (Panel b, inset) On the same plot, for $E_J=0.05$, the entanglement entropy $S_{L/2}^{(\alpha)}$ and the square overlap $|\braket{\phi_\alpha|\psi_+(2)}|^2$ versus $E_\alpha$. The maximum-overlap eigenstate $\ket{\phi_{\overline{\alpha}}}$ has entanglement entropy $\simeq\log 2$ (inside the blue circle).} 
\label{I2:fig}
\end{figure}

The same persistence of $\mathcal{I}_{(2)}(t)$ at long times ---marking the corresponding persistence of the charge-density-wave order--- occurs for larger values of $q$. We can see that phenomenon in Fig.~\ref{I2t:fig}, where we show $\mathcal{I}_{(2)}(t)$ {versus $E_J t$}, initializing with $\ket{\psi_2(q)}$ for $q=4$. {In these figures, we plot for clarity only the envelope of the time traces, defined as the set of local maxima and local minima. We do that because, due to the very dense oscillations of $\mathcal{I}_{(2)}(t)$, the plots of the full curves would hide part of the others behind it [see Fig.~\ref{comparison_M:fig} for an example]. We emphasize that we fix $E_J$, and perform the plots versus $E_J t$ instead of $t$ because in this way we express time in units of $1/E_J$. This is a sensible thing to do, because $E_J$ is the scale of energy of the term in the Hamiltonian inducing the dynamics. With $E_J=0$, indeed, the initial state would be an Hamiltonian eigenstate and there would be no dynamics.} We consider different values of $E_J\leq 0.5$, and different values of $L$. These results have been obtained using the tDMRG~\cite{Schollwock_rev,notaD} algorithm (implemented through the ITensor Library~\cite{ITensor}). {We see that both the upper and lower branches of the envelope seem to tend towards some asymptotic curves.} In all these cases, $\mathcal{I}_{(2)}(t)$ stays near to 1, also for large system sizes{, suggesting that the low-entanglement states providing this persistence in time of the charge-density wave order still exist.} 
\begin{figure}
  \begin{center}
  \begin{tabular}{c}
     \hspace{-0.1cm}\begin{overpic}[width=90mm]{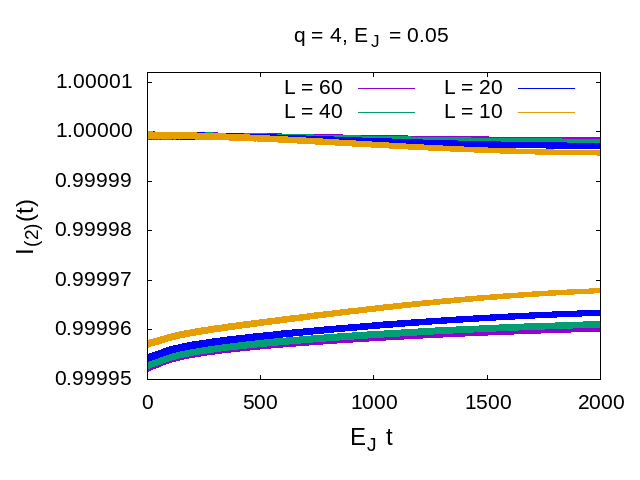}\put(22,68){}\end{overpic}\\
     \hspace{-0.1cm}\begin{overpic}[width=85mm]{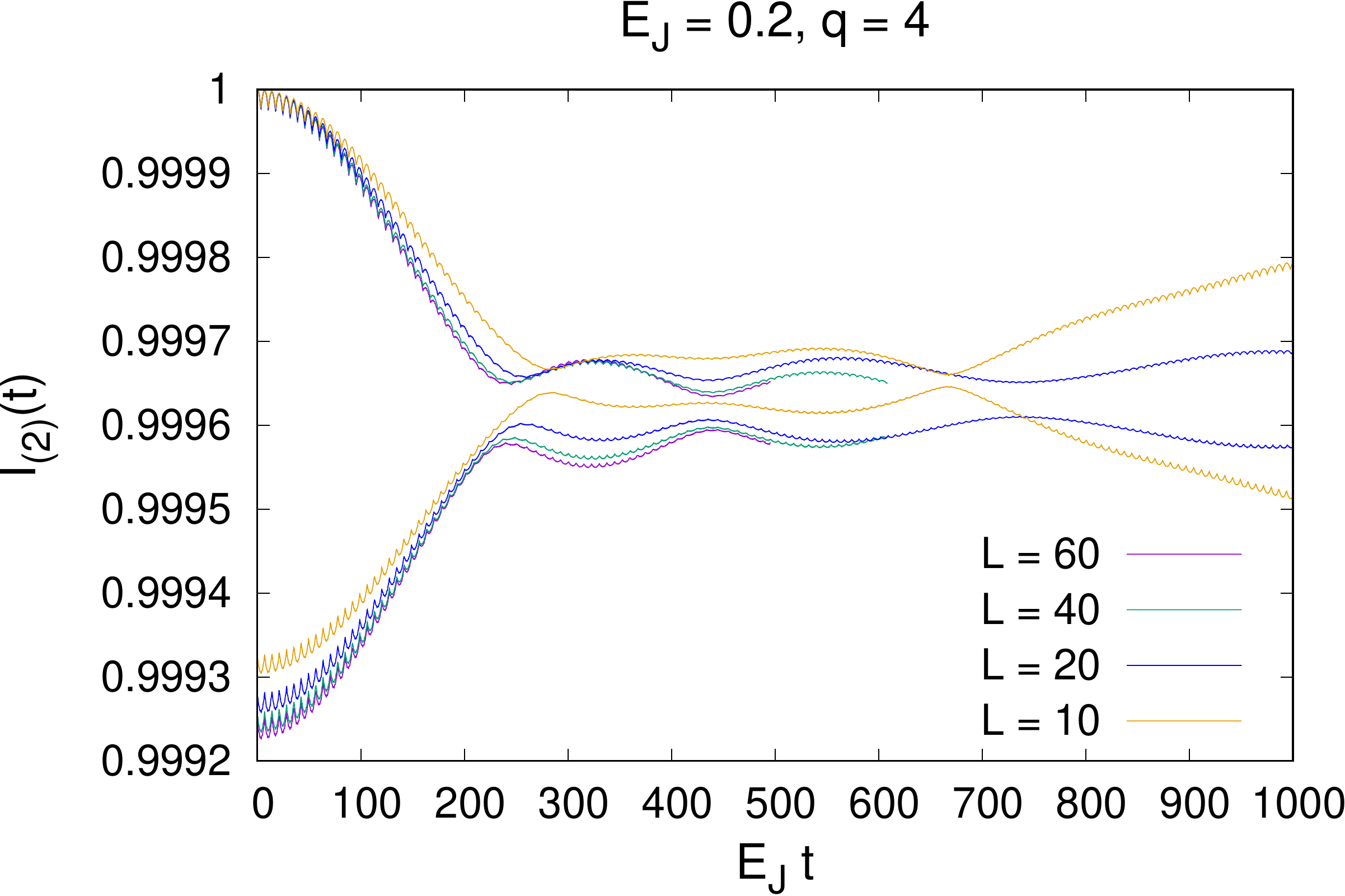}\put(20,67){}\end{overpic}\\
     \hspace{-0.1cm}\begin{overpic}[width=90mm]{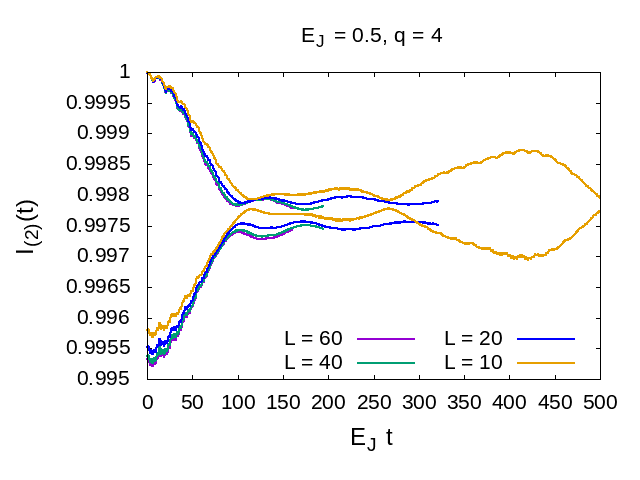}\put(22,68){}\end{overpic}
   \end{tabular}
 \end{center}
\caption{Chain model. $\mathcal{I}_{(2)}(t)$ versus $E_J t$ for different values of $E_J$ and $L$ {(plot of the envelopes of the curves)}. Numerical parameters, $q=4$, $M=5$, results obtained with tDMRG, initialization with $\ket{\psi_2(q)}$ [see Eq.~\eqref{stat:eqn}].}
\label{I2t:fig}
\end{figure}

For the initialization with period 3 in Eq.~\eqref{stat:eqn}, we probe the persistence of the charge density wave using a different observable
\begin{equation}\label{J3:eqn}
  \hat{J}_3 = \hat{K}^\dagger \hat{K},\quad \text{with}\quad \hat{K}\equiv\frac{1}{qL}\sum_{j=1}^L\nep^{2\pi i j /3}\hat{n}_j
\end{equation}

Also initializing with a charge density wave with period 3 ---the state $\ket{\psi_3(q,-q,0)}$ Eq.~\eqref{stat:eqn}--- one can see the persistence of the charge-density-wave order, if one considers the expectation of the operator $\hat{\mathcal{J}}_3$ [Eq.~\eqref{J3:eqn}]. We define this expectation $J_3(t)\equiv\braket{\psi(t)|\hat{\mathcal{J}}_3|\psi(t)}$, and to evaluate it, we perform the evolution using the tDMRG algorithm~\cite{notaD}. We plot $J_3(t)$ versus {$E_J t$} in Fig.~\ref{J3t:fig}. {Also here we plot the envelopes of the curves, to avoid that curves hide each other.} For $E_J\leq 0.2$, we see that $J_3(t)$ does not move very much from the initial value $\sim 1/3$, marking the persistence of the initial period-3 pattern, also for $L=90$ and $t$ order 1000.
\begin{figure}
  \begin{center}
  \begin{tabular}{c}
     \hspace{-0.1cm}\begin{overpic}[width=90mm]{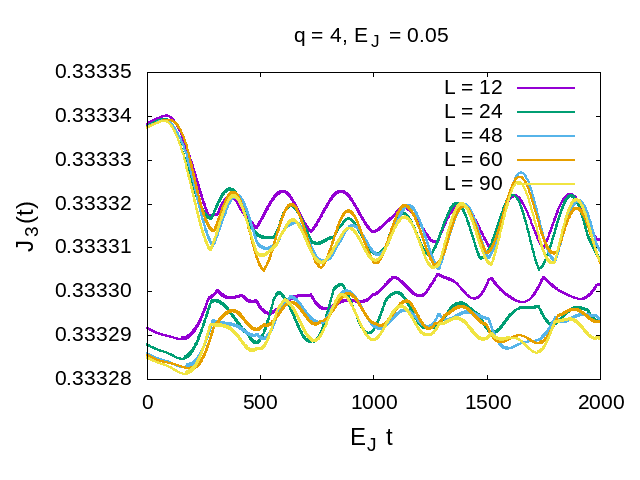}\put(40,59){}\end{overpic}\\
     \hspace{-0.1cm}\begin{overpic}[width=90mm]{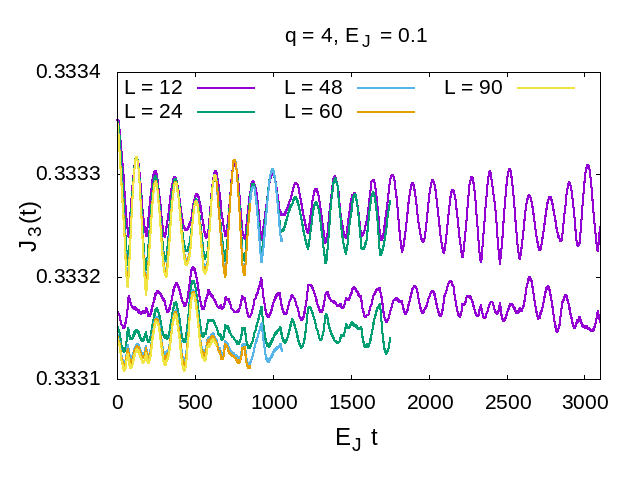}\put(70,59){}\end{overpic}\\
     \hspace{-0.1cm}\begin{overpic}[width=85mm]{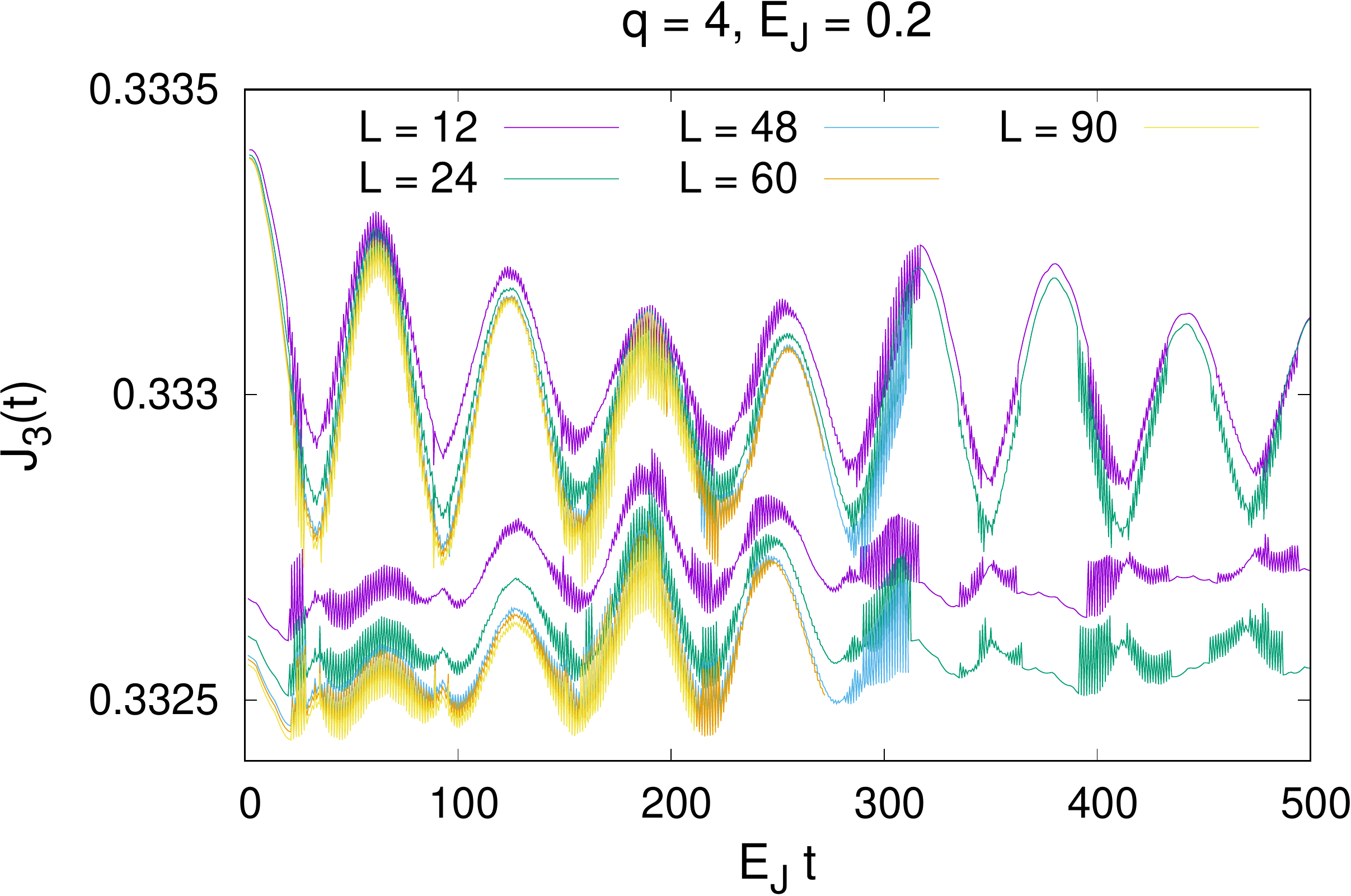}\put(80,55){}\end{overpic}\\
%
   \end{tabular}
 \end{center}
\caption{Chain model. $\mathcal{J}_{3}(t)$ versus $E_J t$ for different values of $E_J$ and $L$  {(plot of the envelopes of the curves)}. Numerical parameters, $q=4$, $M=5$, results obtained with tDMRG, initialization with $\ket{\psi_3(q,-q,0)}$ [see Eq.~\eqref{stat:eqn}].} 
\label{J3t:fig}
\end{figure}
\subsection{Ladder model}\label{dynlad:sec}
{We find similar results for the ladder model. In this case, we have to redefine the operators Eq.~\eqref{iotta:eqn} as
\begin{equation} \label{iottalad:eqn}
  \hat{\mathcal{I}}\equiv\frac{1}{2qL}\sum_{i=1}^L\sum_{k=1}^2(-1)^{i+k}\hat{q}_{i,\,k}\,,
\end{equation}
and the period-2 charge-density-wave states as 
\begin{equation}\label{staf:eqn}
  \ket{\psi_2(q)}=\left|\begin{array}{ccccc}-q,&\phantom{-}q,&-q,&\phantom{-}q,&\ldots\\\phantom{-}q,&-q,&\phantom{-}q,&-q,&\ldots\end{array}\right\rangle\,.
\end{equation}
Initializing the dynamics with $\ket{\psi_+(q=2)}$, defined in Eq.~\eqref{pipp:eqn} as the even superposition of $\ket{\psi_2(\pm q)}$, we see that $\mathcal{I}_2$ is almost independent of $L$ for $E_J\lesssim 0.7$ in the case with no flux [Fig.~\ref{I2lad:fig}(a)], for $E_J$ larger there is a stronger decrease with $L$. In the inset of Fig.~\ref{I2lad:fig}(a) we show the maximum over the eigenstates $\ket{\phi_\alpha}$ of the square overlap $|\braket{\phi_\alpha|\psi_+(q=2)}|^2$. We see that the overlap is near 1 and almost independent of $E_J$ for $E_J\lesssim 0.7$. So, in the interval where $\mathcal{I}_2$ is almost independent of $L$, the maximum overlap with $\ket{\psi_+(q=2)}$ is large, so there is an eigenstate -- call it $\ket{\phi_{\overline{\alpha}}}$ -- very similar to the initial state. As in the case of the chain, we have checked that $\ket{\phi_{\overline{\alpha}}}$ is one of the eigenstates with $\log 2$ entanglement entropy (the one inside the blue circle in Fig.~\ref{IPR:fig}). So this isolated nonthermal eigenstate has an important role in the dynamics, and forbids the thermal melting of the initial charge-density-wave order.}

{In the case with {finite flux} $\Phi=\Phi_0/\phi$, things are different. There is independence of $\mathcal{I}_2$ on $L$ for $E_J\lesssim 0.4$, and a faster decrease with $L$ for $E_J>0.8$ [Fig.~\ref{I2lad:fig}(b)]. So, at least for the considered values of $L$ ($L\leq 4$), one sees that the magnetic flux favours smaller values of $\mathcal{I}_2$, and of the maximum overlap [inset of Fig.~\ref{I2lad:fig}(b)]. So, in presence of the magnetic flux, the eigenstate with maximum overlap is slightly more different from $\ket{\psi_+(q=2)}$, and so deviates more from the ideal $\log 2$ eigenstate. This result suggests that the magnetic flux moves the system slightly more towards ergodicity. This surmise is confirmed analyzing quantum chaos and ergodicity looking at the bulk of the eigenstates and eigenvalues of the Hamiltonian, as we are going to do in the next section.}
\begin{figure}
  \begin{center}
  \begin{tabular}{c}
     \begin{overpic}[width=80mm]{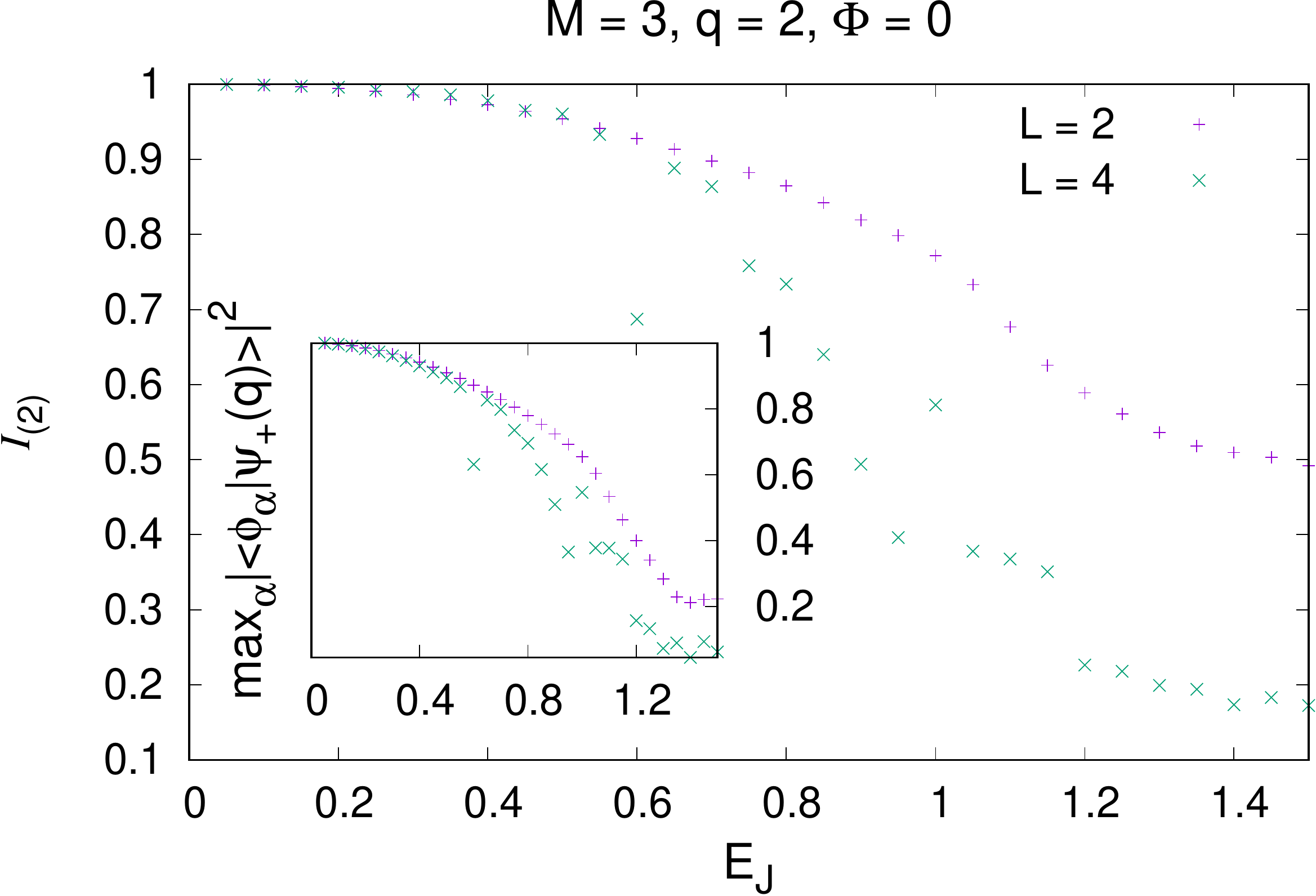}\put(15,66){(a)}\end{overpic}\\\vspace{1mm}\\
     \begin{overpic}[width=80mm]{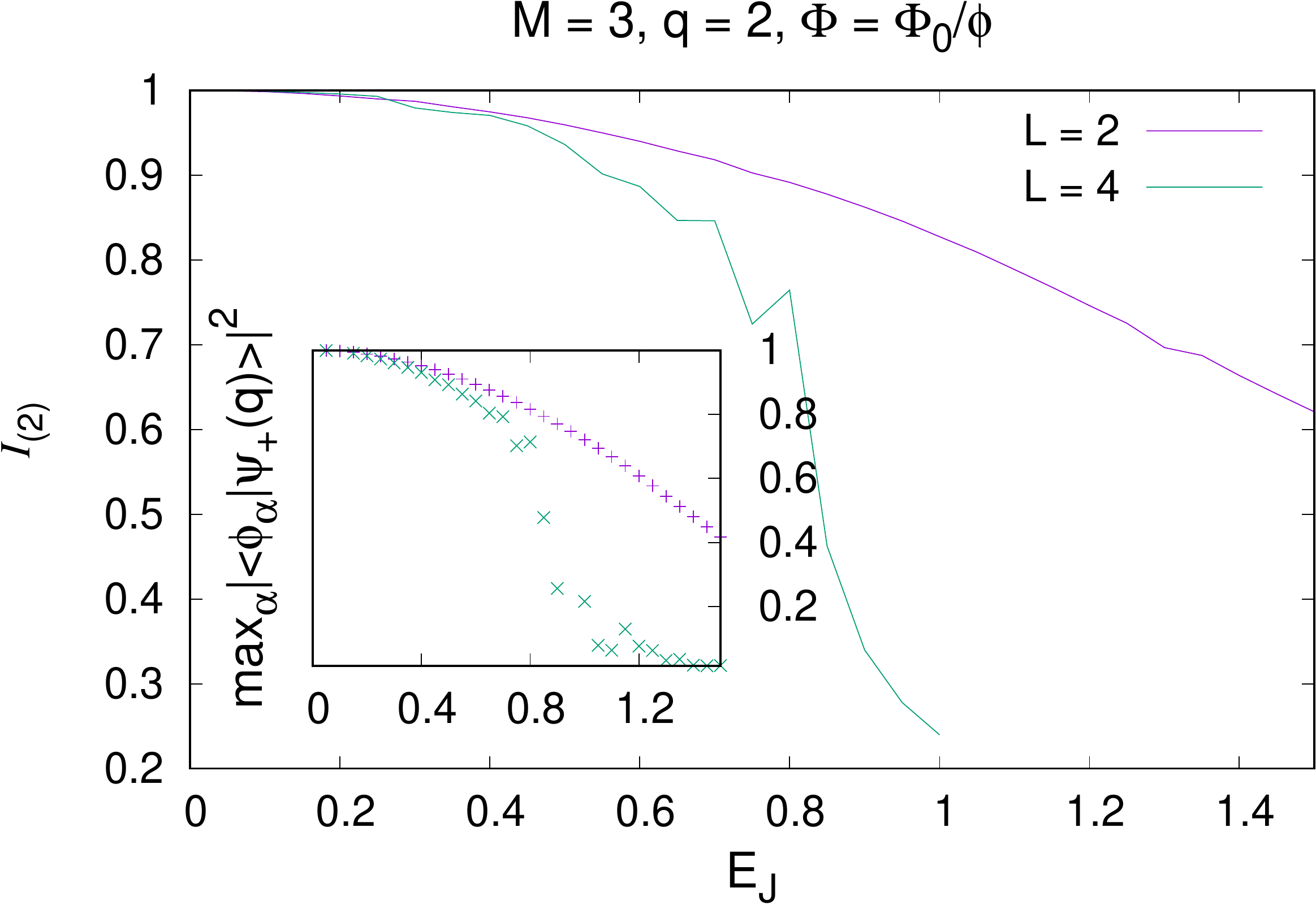}\put(15,66){(b)}\end{overpic}
   \end{tabular}
 \end{center}
\caption{Ladder model. Numerical parameters, $q=2$, $M=3$. (Main panels) $\mathcal{I}_{(2)}$ versus $E_J$ for $\Phi=0$ (Panel a) and {finite flux} $\Phi=\Phi_0/\phi$ (Panel b). (Insets) Maximum over the eigenstates $\ket{\phi_\alpha}$ of the square overlap $|\braket{\phi_\alpha|\psi_+(q=2)}|^2$, versus $E_J$, for $\Phi=0$ (Panel a) and {finite flux} $\Phi=\Phi_0/\phi$ (Panel b). Exact diagonalization everywhere, but in the $L=4$ curve of the main Panel b, where we have used Krylov technique with dimension of the Krylov subspace $m=50$.} 
\label{I2lad:fig}
\end{figure}
\section{Global probes of ergodicity}\label{ergo:sec}
{We exploit the fact that in short range systems, as the one we are discussing, quantum chaos implies locally thermal eigenstates~\cite{kafri_2016}. So, quantum chaos and ETH are strictly related (they go below the same name of ``quantum ergodicity''). That's why, in order to probe ergodicity, in the rest of this section we will introduce and use measures of quantum chaos and ETH, that apply to the bulk of the eigenstates and the eigenvalues of the Hamiltonian. These are integrated quantities, it is easy to compare their behavior, and we will see that they are in agreement with each other. They provide consistently a picture of absence of quantum chaos at small $E_J$. The interval of $E_J$ where this happens decreases when $L$ increases, so for bigger sizes there is a larger interval of parameters where the system is quantum ergodic. In the case of the ladder, a magnetic flux makes the system more ergodic.}
\subsection{Chain}
\subsubsection{{Smoothness of local observables}}
We start considering a direct probe of ETH. Given a local observable $\hat{O}$, and its expectations over the Hamiltonian eigenstates $O_\alpha\equiv\braket{\phi_\alpha|\hat{O}|\phi_\alpha}$, we define a smoothness quantifier~\cite{PhysRevB.82.174411,michele2021}
\begin{equation}\label{momo:eqn}
  \mathcal{M}(O)=\braket{|O_{\alpha+1}-O_\alpha|}\,,
\end{equation}
with $E_\alpha$ in increasing order, and the average is over $1/3$rd of the eigenstates (the central  ones) in order to avoid finite-size effects associated to the boundary of the spectrum. When the system obeys ETH (it is ``ergodic''), the curves $O_\alpha$ versus $E_\alpha$ tend to be smooth for increasing system size, {because they approach the curve $O_{\rm m.c.}(E)$ provided by the microcanonical ensemble.} Therefore $\mathcal{M}(O)$ should be small and decrease with the system size. When this does not happen, there is no ETH and the eigenstates are not locally thermal. We say in this case that the system shows a regular-like (or non-ergodic) behavior.

We consider as local observable the charging energy $\hat{E}_K=\frac{E_C}{2}\sum_j\hat{q}_j^2$. We see very small (and slightly decreasing) values of ${E_K}_\alpha$ at small $E_J$ [Fig.~\ref{r_vs_EJ_M2:fig}(a) for $M=2$ and Fig.~\ref{r_vs_EJ_M2:fig}(d) for $M=3$]. The reason is that at small $E_J$ the charging energy is almost degenerate in the multiplets [Fig.~\ref{EK:fig}]. Nevertheless, one can see a strong and clear decrease with $L$ only for $E_J\gtrsim 0.4$. As crossover point between regular-like and ergodic behavior can we take the maximum in $E_J$. We see that this maximum decreases with $L$ and its position shifts towards left, suggesting that the range of $E_J$ corresponding to ergodic behavior increases with $L$.
\subsubsection{{Smoothness of the inverse participation ratio}}\label{ergolo:sec}
The second quantity we consider is a probe of eigenstate quantum chaos. When there is quantum chaos the Hamiltonian of a system behaves in many respects as a random matrix, as it has been checked for systems with a chaotic classical limit~\cite{Haake,PhysRevLett.110.084101,Berry_Les_Houches}. In particular, when a perturbation inducing quantum chaos is applied to an integrable system, the new eigenstates should appear as random states in the basis of the unperturbed eigenstates. This implies two things. 

The first one is that the new eigenstates appear very delocalized in the basis of the unperturbed ones. One can check this fact by considering each of the new eigenstates, looking at the square overlaps of it and the unperturbed eigenstates, and probing if this overlap distribution is broad. A standard probe of broadness of this distribution (or equivalently delocalization) is the Shannon entropy~\cite{kafri_2016} (called also structural entropy in this context~\cite{Santos_2010}), and another is the inverse participation ratio~\cite{thouless} we discuss below.

The second thing is that when two new eigenstates are nearby in energy they look very similar. Each new eigenstate is the random superposition of unperturbed eigenstates taken in an energy window $\Delta E$ provided by the perturbation. If the new eigenstates are nearer in energy than $\Delta E$, we expect them to be very similar to each other. So, if we consider a probe of delocalization of the new eigenstate, and plot it versus the energy of the eigenstate, it should appear as a smooth curve if quantum chaos is present. This conjecture has been numerically checked to be true in many cases in the context of spin and fermionic chains~\cite{Torres_Herrera_2015,Santos_2012,Santos_2010,Gubin_2012}.

We can apply these methods also in our case, because we are exactly in the situation sketched above. We have a unperturbed Hamiltonian (the charging energy term in Eqs.~\eqref{Ham0:eqn} and~\eqref{ham_fl:eqn}) which is integrable in the standard way, because it has an extensive number of local integrals of motion~\cite{Berry_PRS77,Berry_Les_Houches,lichtenberg1983regular,arnol2013mathematical} (they are all the onsite charges $\hat{q}_j$). Its unperturbed eigenstates are the simultaneous eigenstates of the $\hat{q}_i$, $\ket{\{q_j\}}$. We apply to this integrable situation a perturbation provided by the Josephson terms in Eqs.~\eqref{Ham0:eqn} and~\eqref{ham_fl:eqn} and we want to see if they provide chaos. To do so, we look at the eigenstates $\ket{\phi_\alpha}$ of the new Hamiltonian resulting from the application of this perturbation, and study their delocalization in the basis of the unperturbed eigenstates. For each $\alpha$, we look at the broadness properties of the square-overlap distribution $|\braket{\{q_j\}|\phi_\alpha}|^2$. We choose to consider as broadness probe the inverse participation ratio
\begin{equation}\label{IPR:eq}
  {\rm IPR}_\alpha=\sum_{\{q_j\}}|\braket{\phi_\alpha|\{q_j\}}|^4\,.
\end{equation}
(Here the sum extends over all the unperturbed eigenstates consistent with the Hilbert space truncation.) The smaller is this IPR, the broader is the square overlap distribution, and the more delocalized is the eigenstate $\ket{\phi_\alpha}$ in the unperturbed-eigenstates basis.

As we have observed above, each $\ket{\phi_\alpha}$ is a superposition of unperturbed eigenstates coming from an energy window $\Delta E$. In order to estimate it, we do in the following way.
As we show in Appendix~\ref{fluc:app}, for each $\ket{\{q_j\}}$, the broadness in $E_\alpha$ of the $|\braket{\phi_\alpha|\{q_j\}}|^2$ distribution is {of} order $E_J\sqrt{L}$. We can take this quantity also as an estimate of $\Delta E$. If there is quantum chaos, nearby eigenstates whose energies $E_\alpha$ differ by less than $\Delta E$ should look very similar, and should have similar delocalization properties. Therefore, in presence of quantum chaos, delocalization probes like the ${\rm IPR}_\alpha$ should be smooth in $E_\alpha$. This fact can be quantified, by considering the scalar product of the overlap distributions of nearby eigenstates~\cite{Santos_2012}, or the fluctuations of delocalization probes at the center of the spectrum~\cite{Santos_2010}, but we choose a different (although equivalent) way. We simply apply our smoothness quantifier Eq.~\eqref{momo:eqn} to the quantity $\log{\rm IPR}_\alpha$. (We take the logarithm, because ${H_2}_\alpha=-\log{\rm IPR}_\alpha$ is the second R\'enyi entropy of the square overlap distribution, a quantity similar to the Shannon entropy).

We clearly see that $\mathcal{M}(\log{\rm IPR})$ increases with $L$ for small $E_J$ and decreases for large $E_J$ [Fig.~\ref{r_vs_EJ_M2:fig}(a) for $M=2$ and Fig.~\ref{r_vs_EJ_M2:fig}(d) for $M=3$]. So, the system becomes more regular-like with increasing $L$ at small $E_J$ and tends to ergodicity at large $E_J$. The intersection between different curves moves left for increasing $L$, so the regular-like region seems to shrink for increasing system size, in agreement with the results for $\mathcal{M}(E_K)$.
\subsubsection{Average level spacing ratio}
Another standard probe of quantum chaos is the average level spacing ratio $r$, defined as
\begin{equation} \label{rorro:eqn}
  r=\frac{1}{\mathcal{N}_S-2}\sum_{\alpha=1}^{\mathcal{N}_S-2}\frac{\min(E_{\alpha+2}-E_{\alpha+1},E_{\alpha+1}-E_\alpha)}{\max(E_{\alpha+2}-E_{\alpha+1},E_{\alpha+1}-E_\alpha)}\,,
\end{equation} 
where $\mathcal{N}_S$ is the dimension of the Hilbert subspace fully even under all the symmetries of the Hamiltonian, and the eigenenergies are in increasing order. The chain, the ladder without magnetic flux, and the ladder with magnetic flux have different symmetries, so different values of $\mathcal{N}_S$ and different possibilities of diagonalizing the Hamiltonian for large system size. That's why in different models the maximum $L$ is different.

When there is quantum chaos, the Hamiltonian behaves as a random matrix from the point of view of the spectrum and $r$ acquires a special value. If there is a basis where the Hamiltonian is real, this special value corresponds to the Wigner-Dyson value $r_{\rm WD}\simeq 0.5295$~\cite{Haake,PhysRevLett.110.084101}. This is true for both Hamiltonians -- Eq.~\eqref{Ham0:eqn} and Eq.~\eqref{ham_fl:eqn} --, because they are both real in at least one basis (the basis of the eigenstates of the phase). 
If $r$ is smaller than $r_{\rm WD}$ the behavior is different from a random matrix and there is no full quantum chaos. {A special value in this regime is the Poisson value $r_P\simeq 0.386$, that is proven to be generically attained by a classical integrable system when it is quantized~\cite{Berry_PRS77}, and has been frequently observed in nonthermalizing many-body quantum systems showing space localization~\cite{Schulz_2019,PhysRevB.82.174411,abanin_rmp}.}

The behavior of $r$ confirms the generalized tendency towards ergodicity for increasing system size which we have observed looking at the other probes. We see that $r$ increases with $L$ towards the Wigner-Dyson value [Fig.~\ref{r_vs_EJ_M2:fig}(c,f)].  In the case $M=3$ there is saturation of $r$ for $E_J\lesssim 0.3$, when moving from $L=8$ to $L=9$ [Fig.~\ref{r_vs_EJ_M2:fig}(f)]. So, in summary, the region of regular-like behavior shrinks with increasing $L$, but we have no way to extrapolate this behavior to values of $L$ larger than those provided by our numerical capabilities.
\begin{figure*}
  \begin{center} 
    \begin{tabular}{ccc}
     \hspace{-0.5cm}\begin{overpic}[width=60mm]{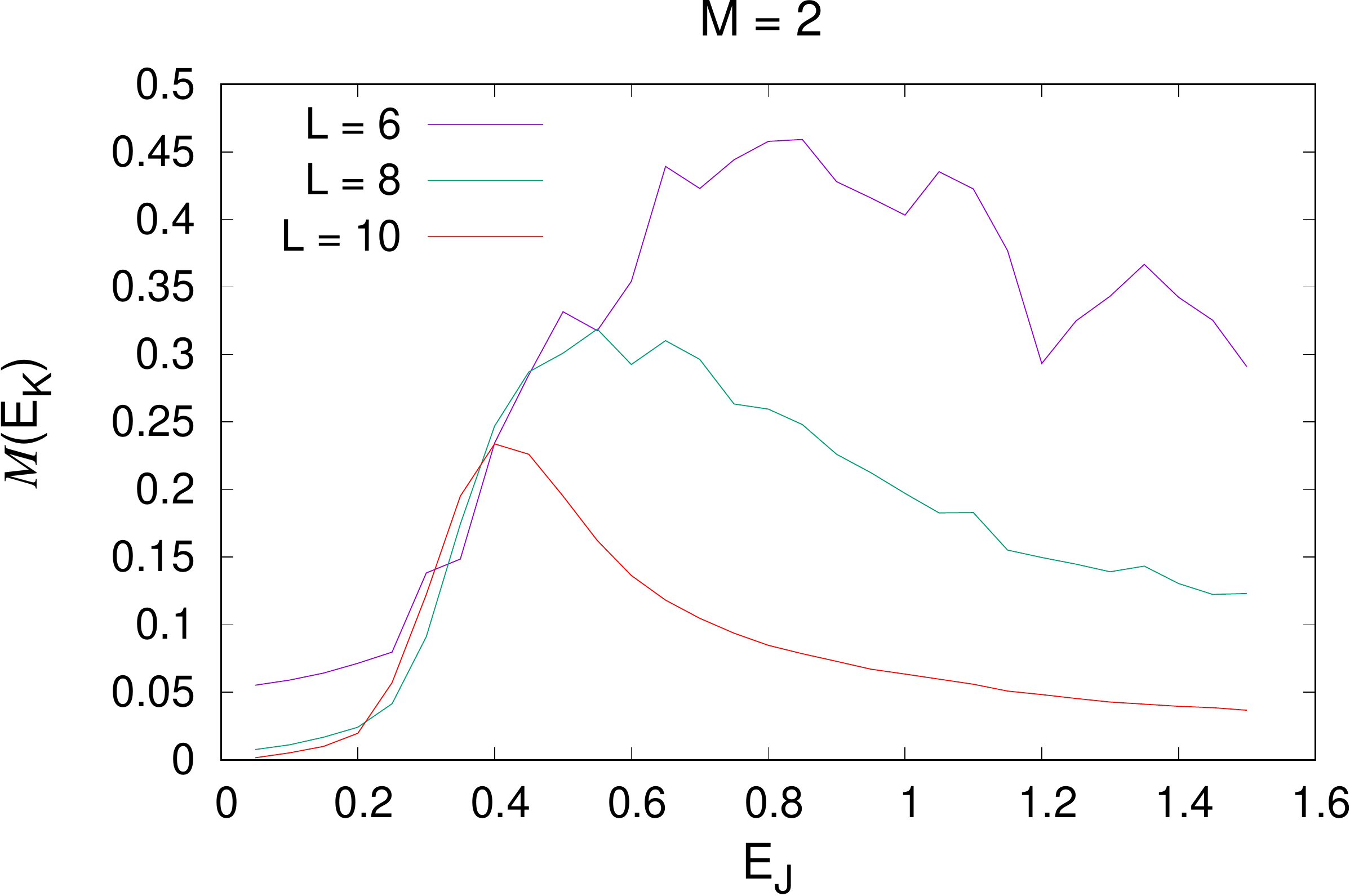}\put(15,64){(a)}\end{overpic}&
     \begin{overpic}[width=60mm]{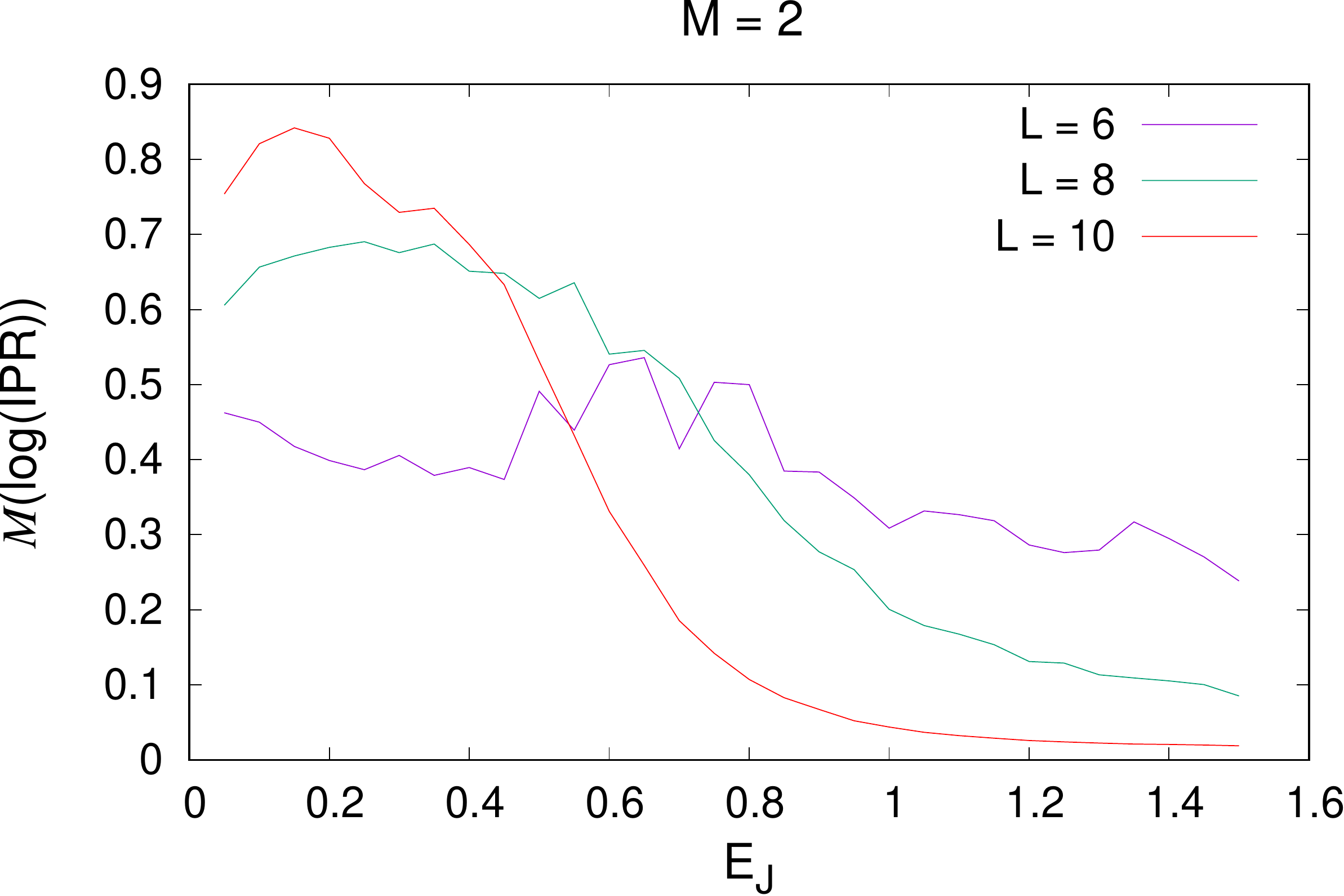}\put(15,64){(b)}\end{overpic}&
          \begin{overpic}[width=60mm]{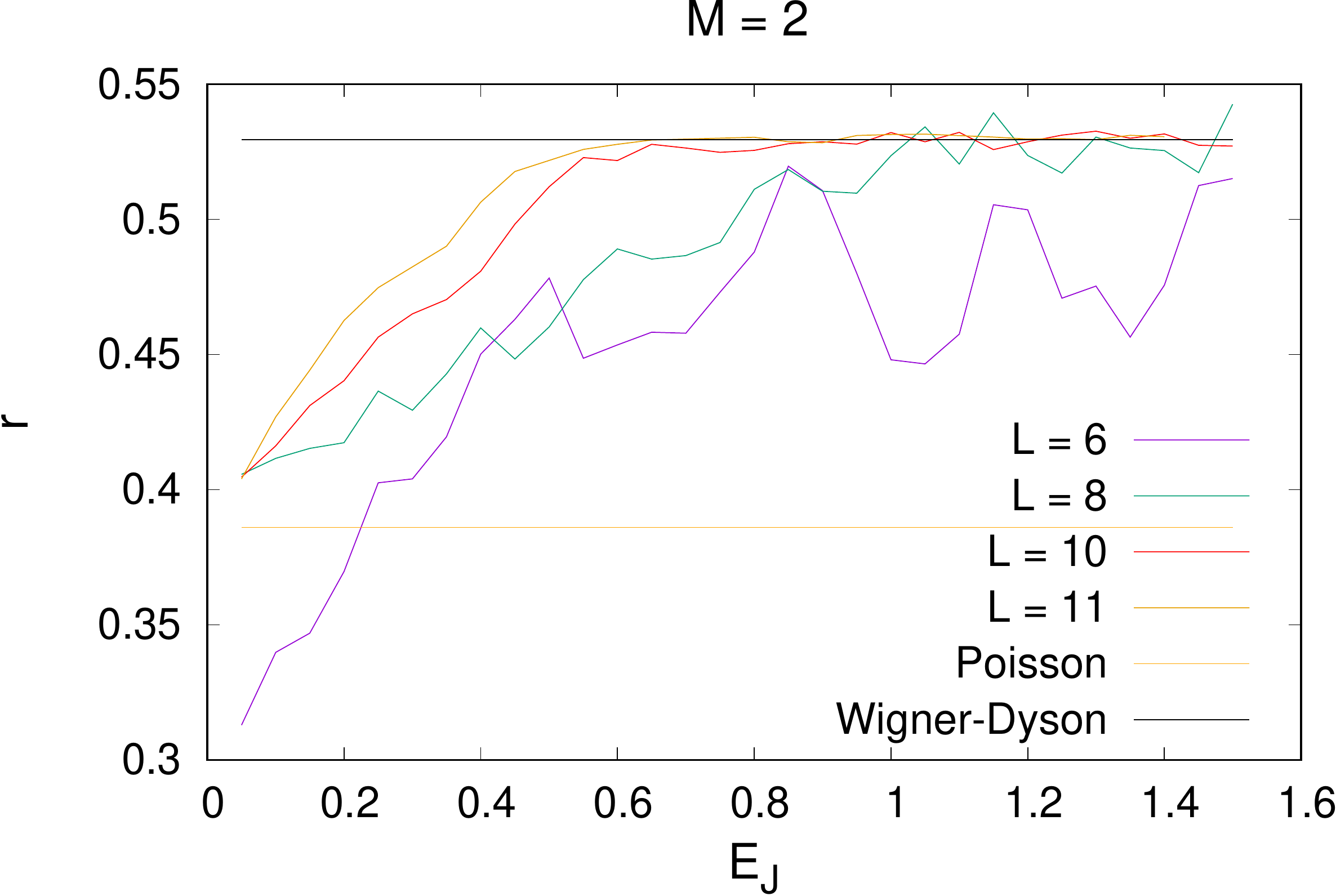}\put(15,64){(c)}\end{overpic}\\
%
     \hspace{-0.5cm}\begin{overpic}[width=60mm]{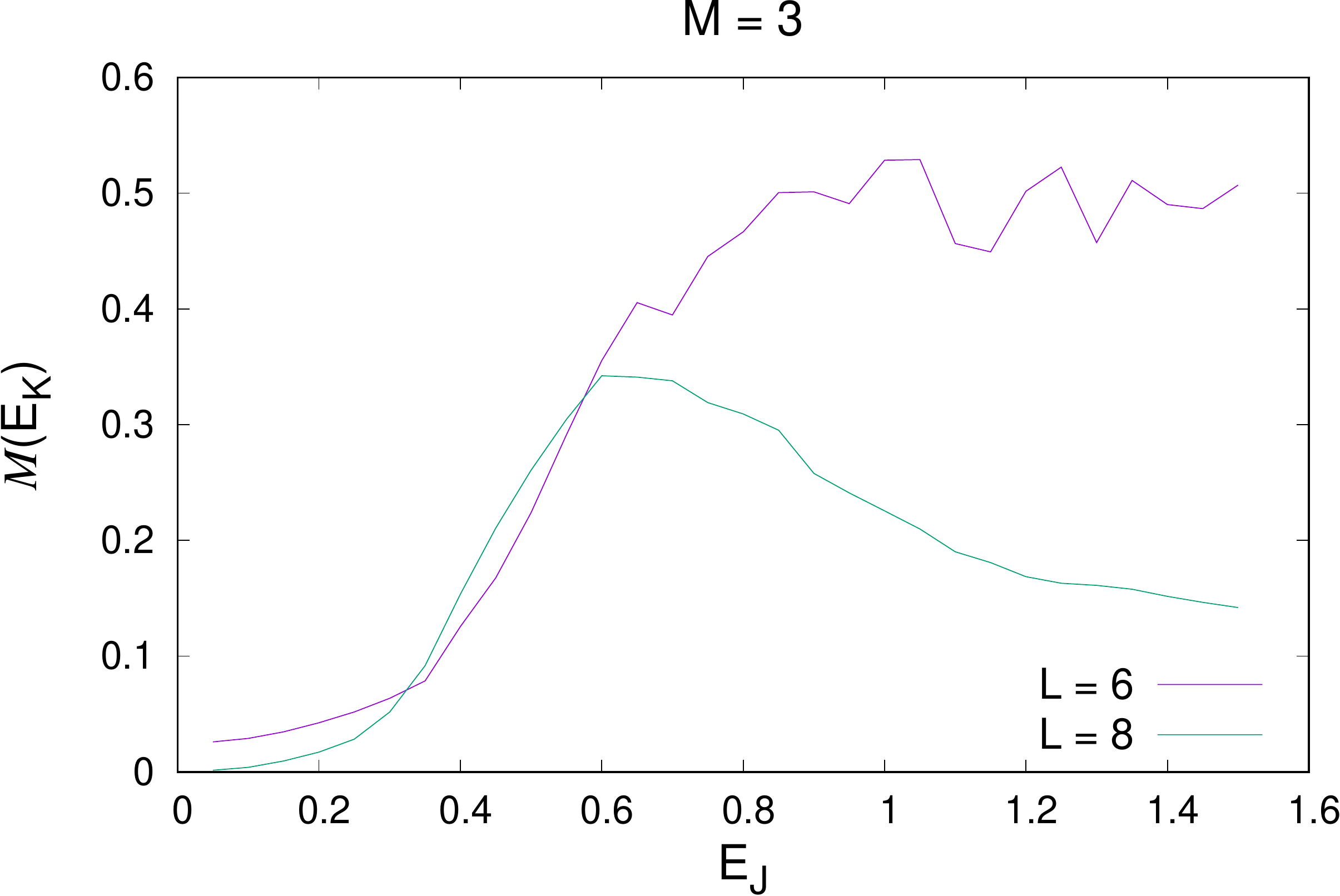}\put(15,65){(d)}\end{overpic}&
     \begin{overpic}[width=60mm]{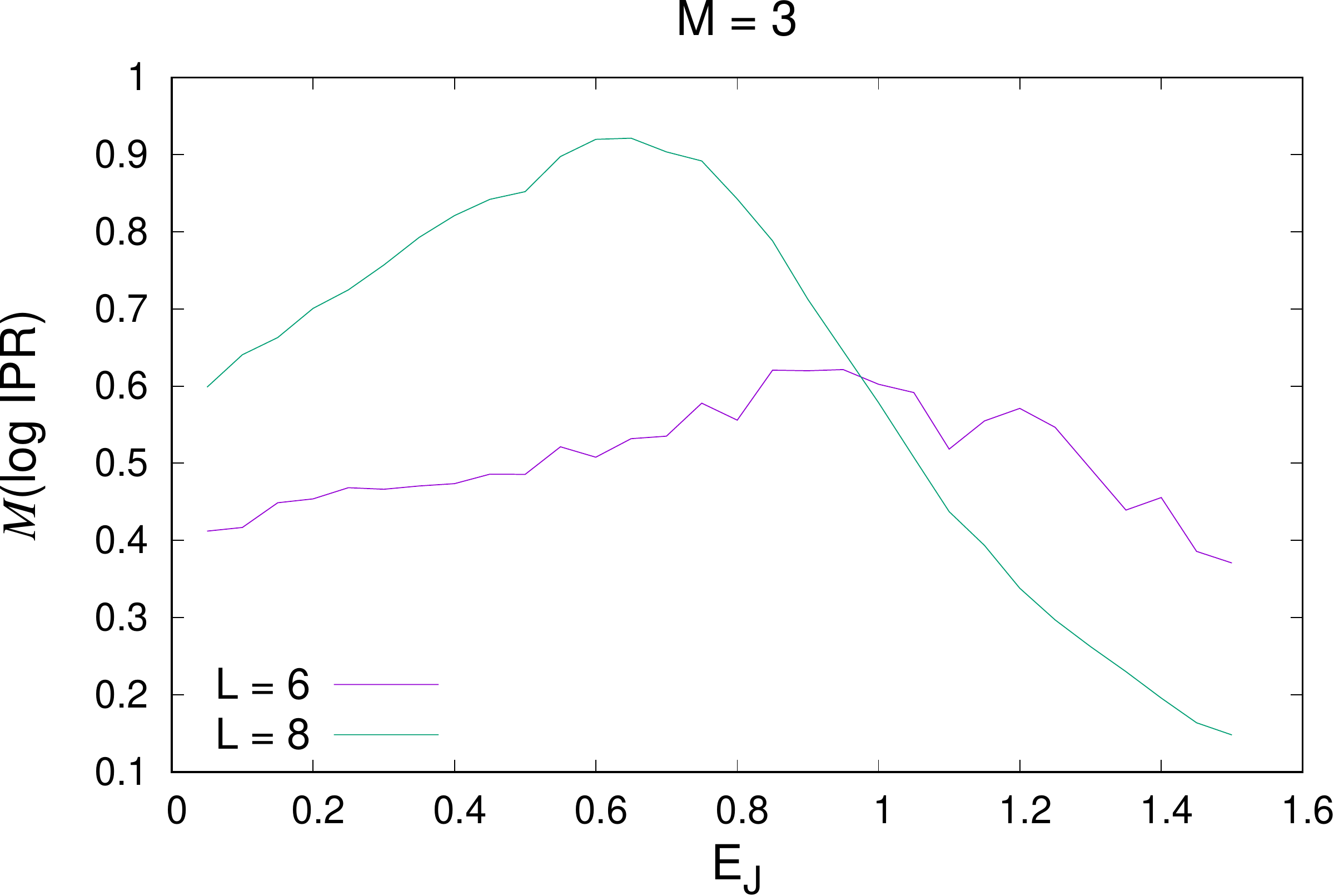}\put(15,65){(e)}\end{overpic}&
          \begin{overpic}[width=60mm]{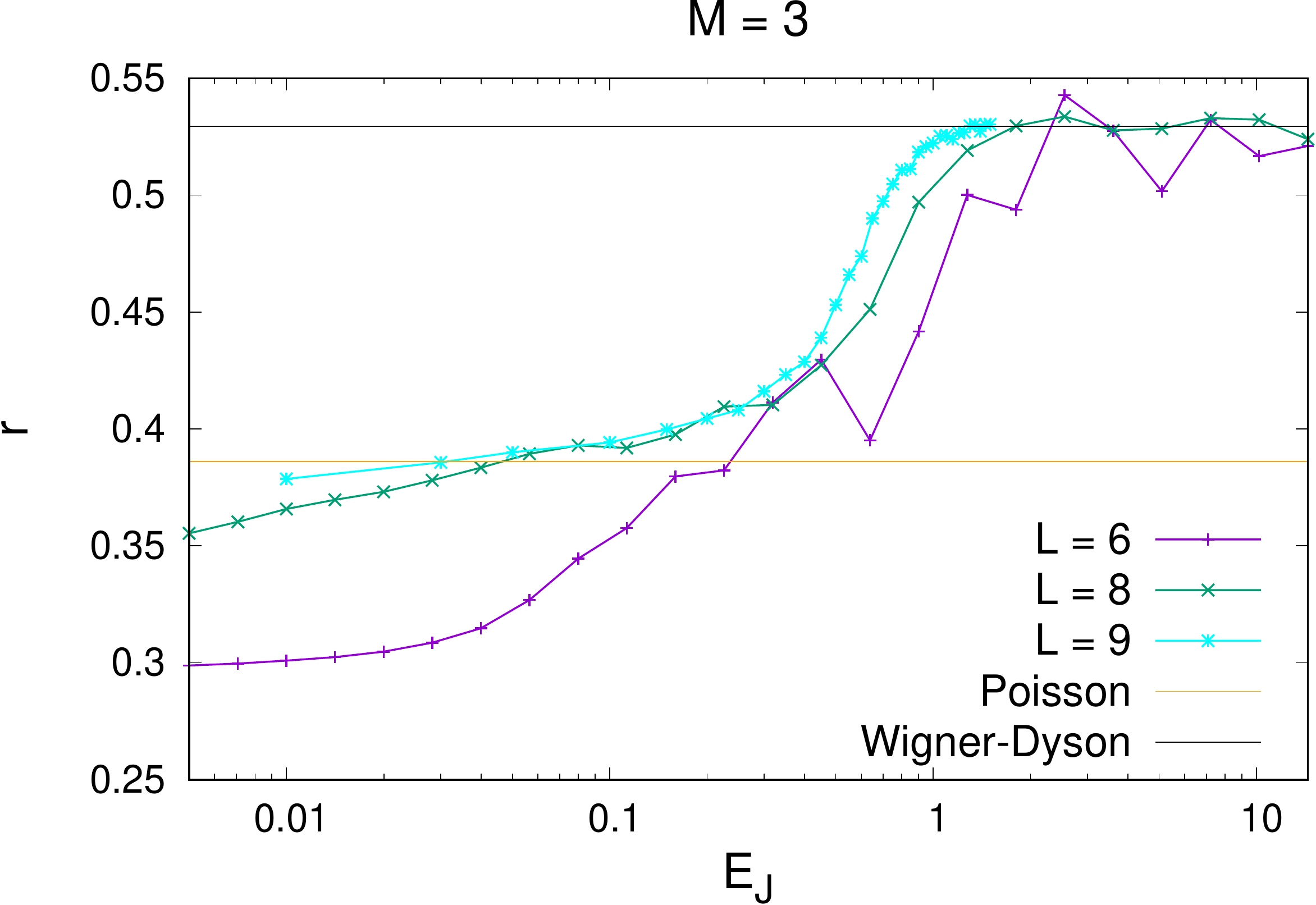}\put(15,65){(f)}\end{overpic}\\
    \end{tabular}
  \end{center}
\caption{Chain model. (Panel a) $\mathcal{M}(E_K)$ versus $E_J$ for different sizes and truncation $M = 2$. (Panel b) The corresponding $\mathcal{M}(\log{\rm IPR})$ versus $E_J$. (Panel c) The corresponding $r$ versus $E_J$. (Panel d-f) The same for truncation $M = 3$.}
\label{r_vs_EJ_M2:fig}
\end{figure*}
\begin{figure}
  \begin{center}
    \begin{tabular}{c}
      \begin{overpic}[width=80mm]{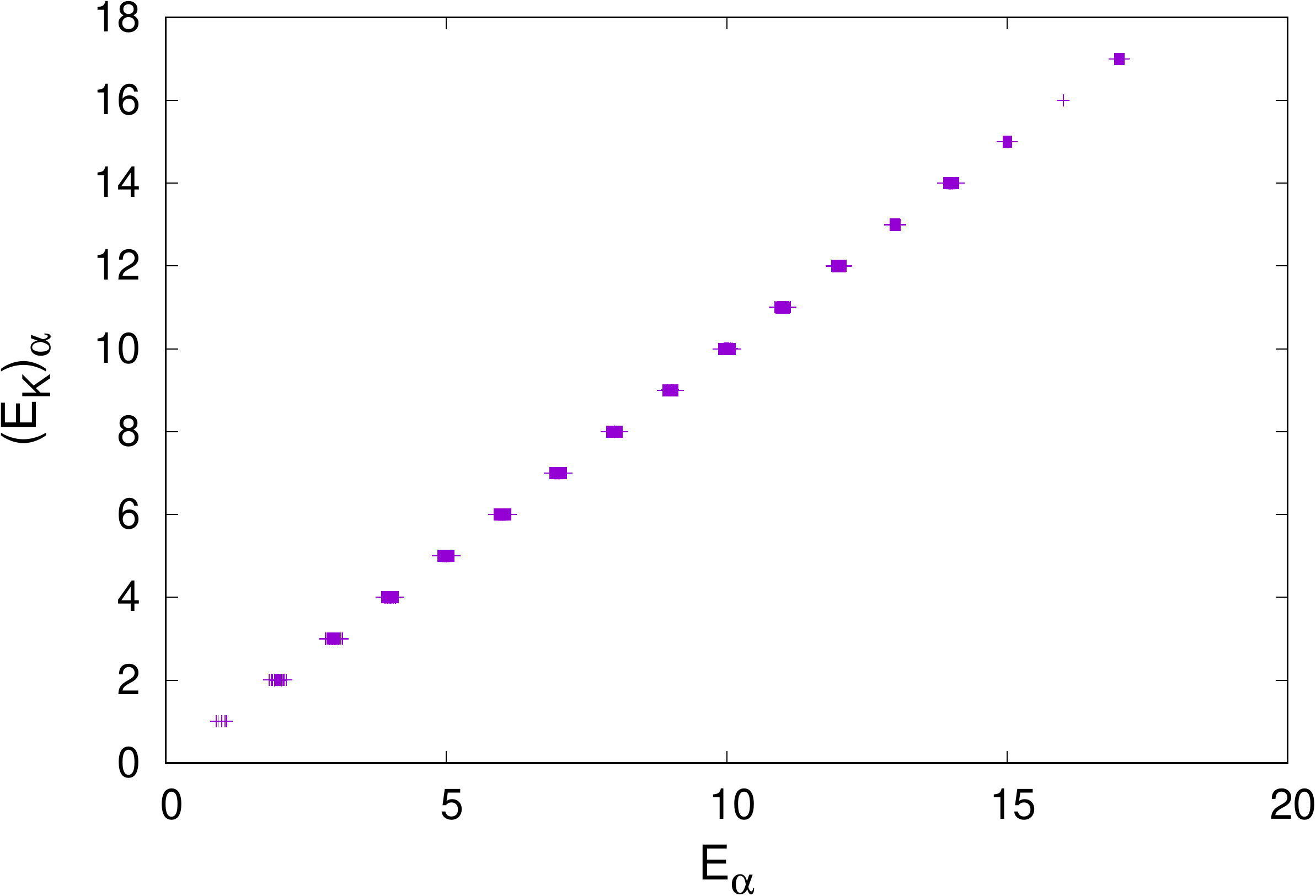}\put(90,32){}\end{overpic}
    \end{tabular}
  \end{center}
\caption{Example of $(E_K)_\alpha$ versus $E_\alpha$ for $E_J=0.05$, $L=10$, $M=2$.}
\label{EK:fig}
\end{figure}
%
\subsection{Ladder}
{The situation for the ladder is strictly similar to the chain, and we discuss it for completeness in Appendix~\ref{App:lad}. Here we focus our attention on the fact that the application of the finite magnetic flux makes the system more ergodic, as witnessed by the probes of quantum chaos and ETH introduced above. Moreover, more ergodicity is strictly related to a larger delocalization in the basis of the charge eigenstates. In order to see it, we fix $L$ and $M$ and compare the cases of vanishing and finite flux. }
\begin{figure}[h!]
  \begin{center}
    \begin{tabular}{c}
    \begin{overpic}[width=80mm]{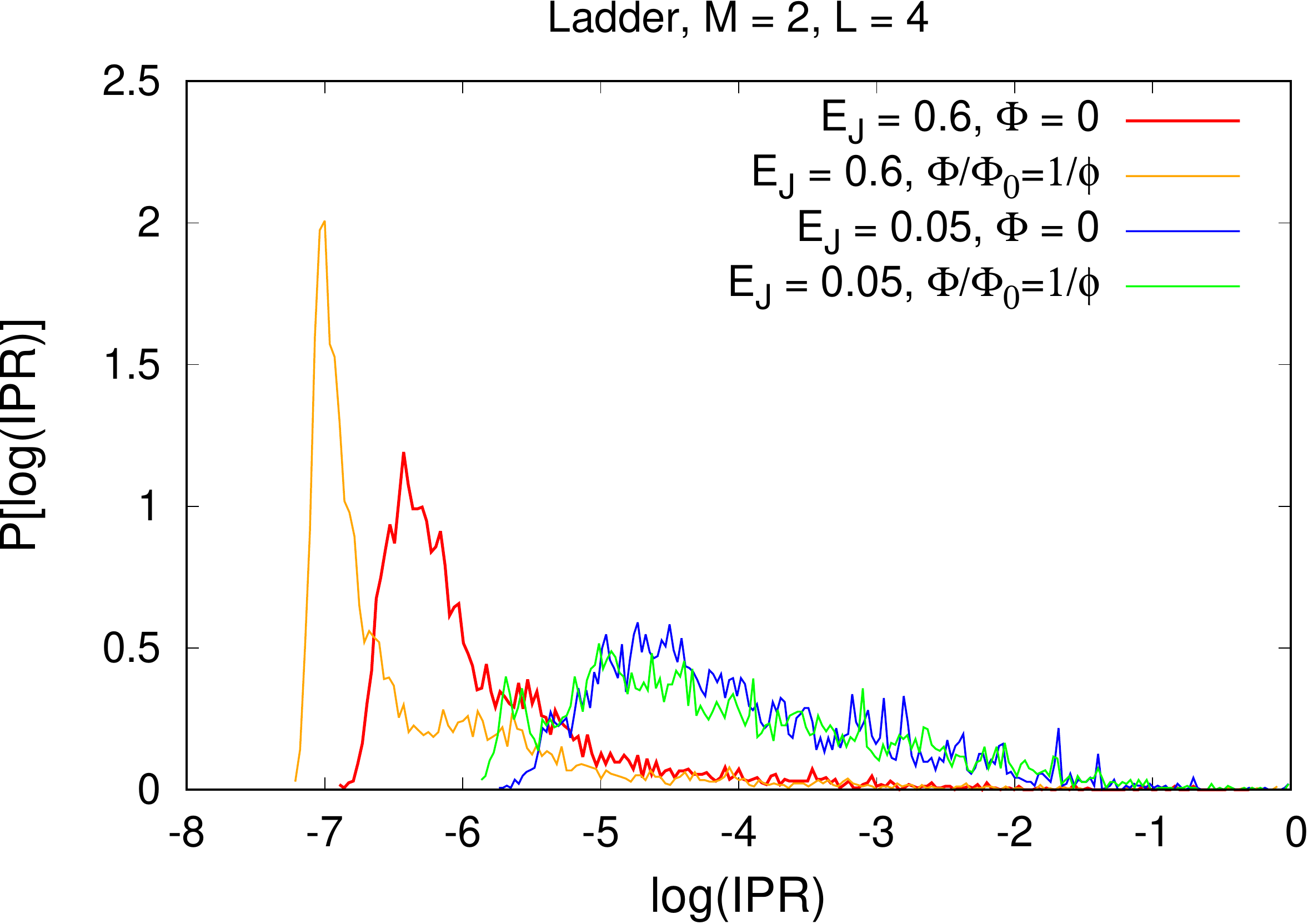}\put(17,70){(a)}\end{overpic}\\ \vspace{1mm}
    \\
    \begin{overpic}[width=80mm]{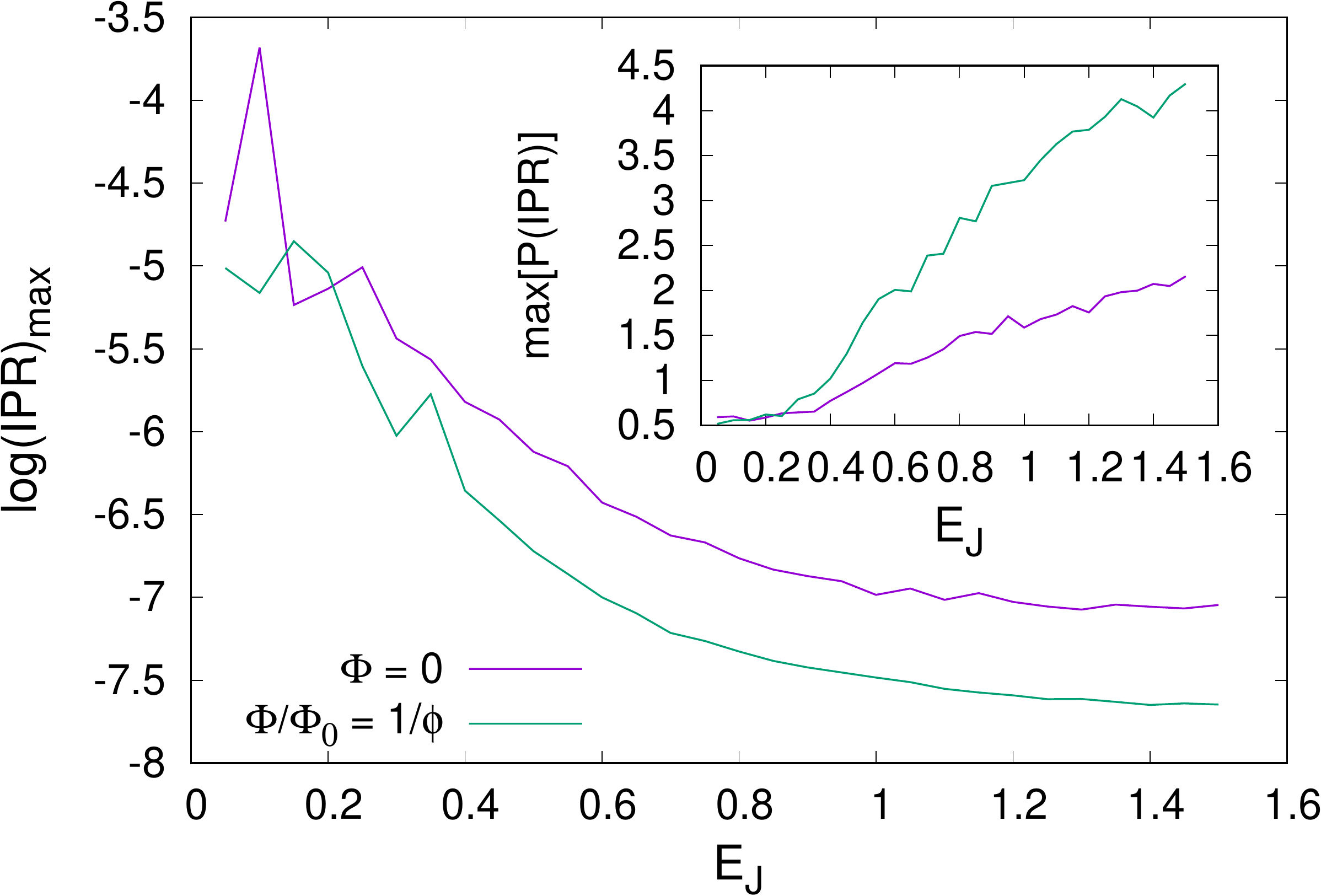}\put(17,70){(b)}\end{overpic}\\
    \end{tabular}
  \end{center}
	\caption{Ladder model. (Panel a) Examples of distributions of the $\log{\rm IPR}_\alpha$ with and without magnetic flux. Given a value of $E_J$, {a finite} magnetic flux shifts the distribution towards the left, marking delocalization of the eigenstates in the charge-configuration basis. (Panel b -- main) Value of $\log{\rm IPR}_\alpha$ where the distribution has a maximum, versus $E_J$, with {vanishing and finite flux}. (Panel b -- inset) Maximum of the $\log{\rm IPR}_\alpha$ distribution versus $E_J$, with {vanishing and finite flux}. Numerical parameters $L=4$, $M=2$, distributions made as histograms with $n_b=200$ boxes.}
	\label{fig:IPRdist1}
\end{figure}
\begin{figure}[h!]
  \begin{center}
    \begin{tabular}{c}
    \begin{overpic}[width=80mm]{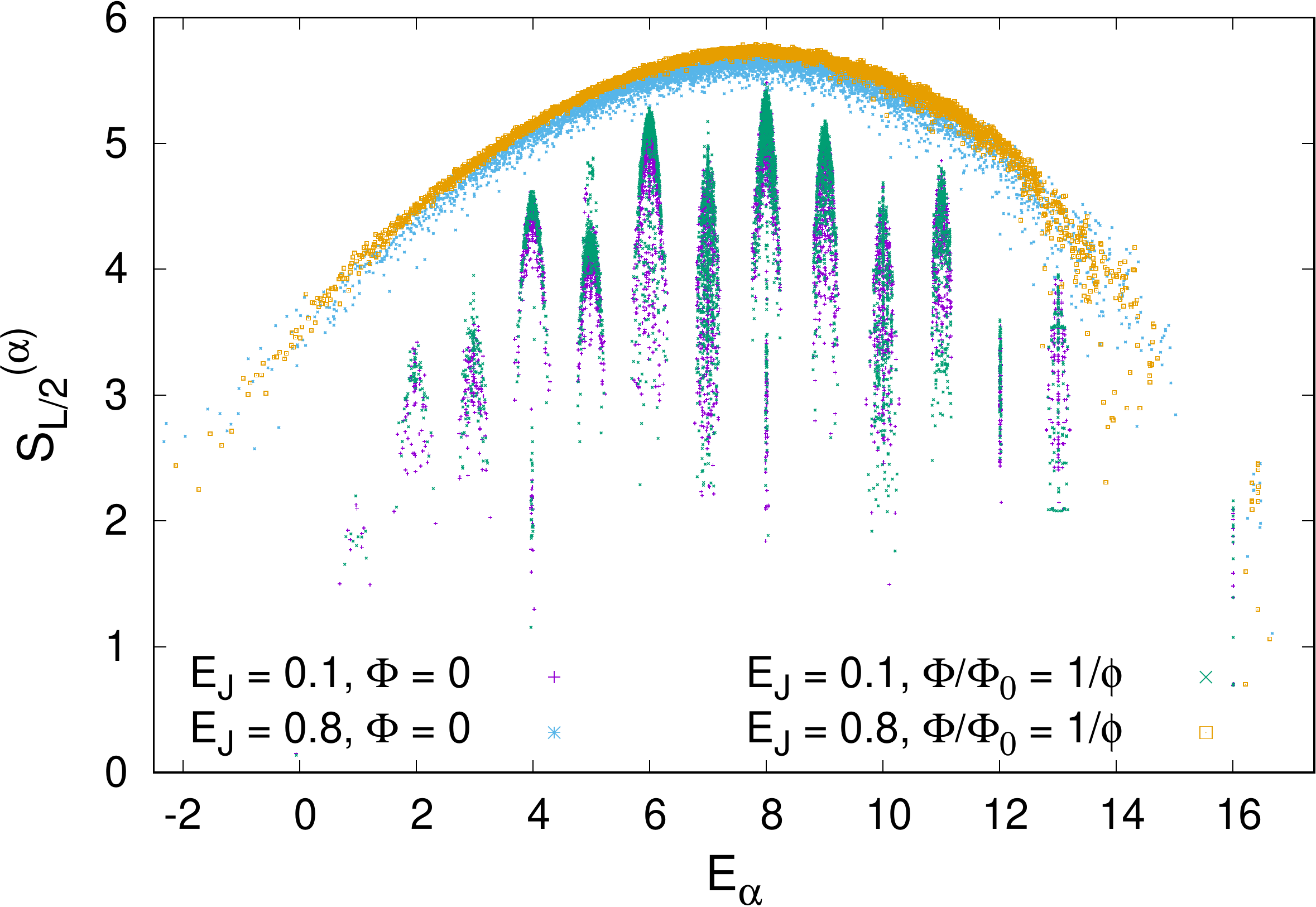}\put(14,72){(a)}\end{overpic}\\ \vspace{1mm}
    \\
    \begin{overpic}[width=85mm]{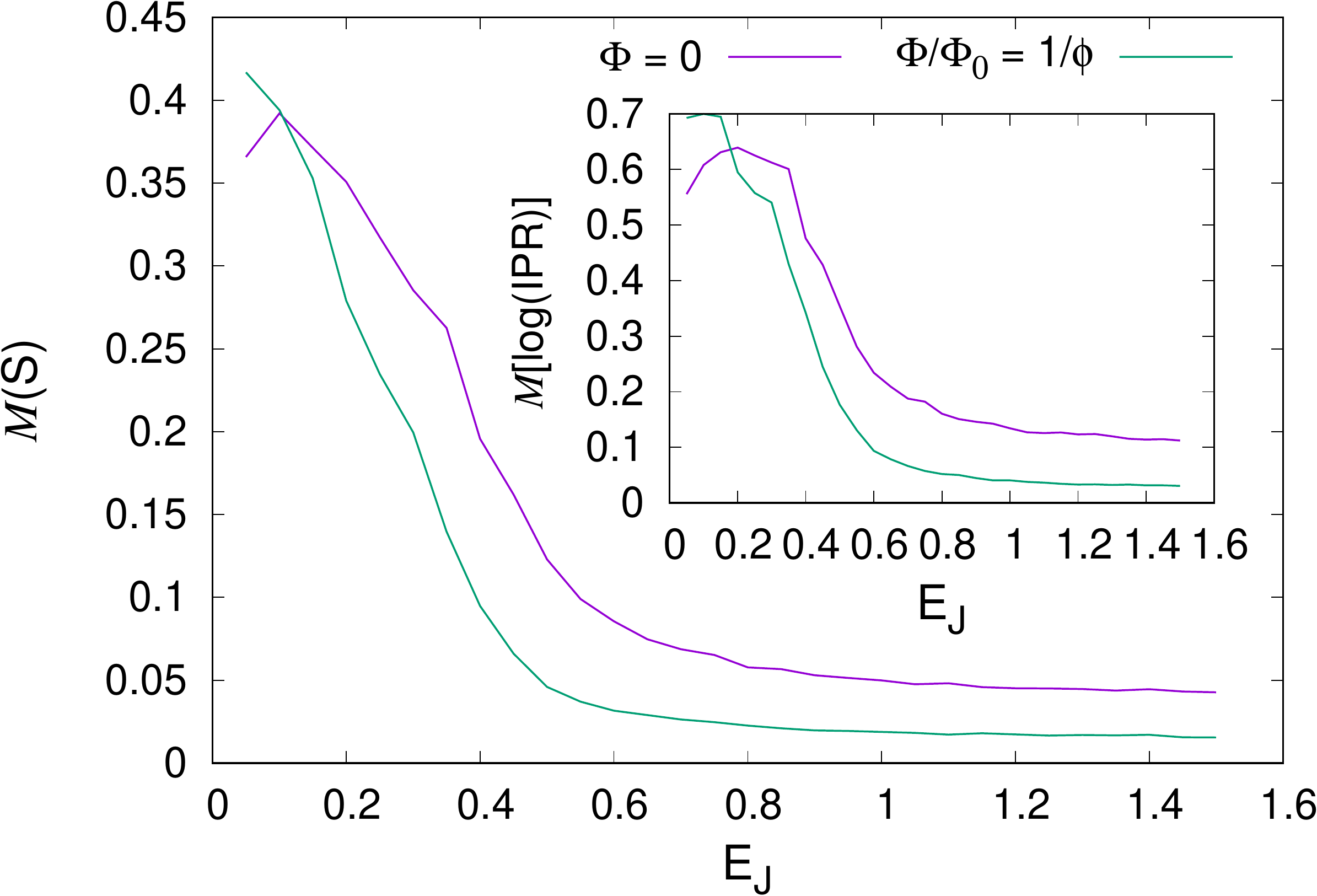}\put(17,70){(b)}\end{overpic}
    \end{tabular}
  \end{center}
 \caption{(Panel a) Examples of $S_{L/2}^{(\alpha)}$ versus $E_\alpha$, in presence of a {finite} magnetic flux the values in the bulk are larger (delocalization) and the curves are smoother (more ergodicity). (Panel b -- main) $\mathcal{M}(S)$ versus $E_J$ with and without flux. (Panel b -- inset) $\mathcal{M}(\log{\rm IPR})$ versus $E_J$ with {vanishing and finite flux}. For $E_J>0.1$ the values {of the $\mathcal{M}$} with the {finite} flux are smaller, marking more ergodicity {(more ETH in the case of $S_{L/2}^{(\alpha)}$ and more quantum chaos in the case of $\log{\rm IPR}_\alpha$)}. Numerical parameters $L=4$, $M=2$.}
  \label{fig:IPRdist}
\end{figure}

{In order to see delocalization, we consider} the distributions of $\log({\rm IPR}_\alpha)$ in Fig.~\ref{fig:IPRdist1}(a). As we can see, {with a finite} magnetic flux, the distribution for a given $E_J$ shifts towards the left, so the ${\rm IPR}_\alpha$ are smaller and then the eigenstates are more delocalized. For $E_J=0.05$ the distribution stays flat and the left shoulder moves left, for $E_J=0.6$ the maximum of the distribution moves left and becomes higher.
 
 We can see how this phenomenon depends on $E_J$ by considering the maximum of the distribution $\max P$ and the corresponding value of $\log({\rm IPR}_\alpha)$, $\log{\rm IPR}_{\rm max}$. We plot $\log{\rm IPR}_{\rm max}$ versus $E_J$ in Fig.~\ref{fig:IPRdist1}(b -- main figure) for $L=4$, $M=2$, and we see that in presence of a magnetic flux the value becomes systematically smaller. The situation is fuzzier for small $E_J$, where the distribution is flat and fluctuations due to the finite number of eigenstates and boxes in the histogram have a more significant effect. In Fig.~\ref{fig:IPRdist}(b -- inset) we plot $\max P$ versus $E_J$. We see that for $E_J$ larger than $\sim 0.2$ the maximum is larger for the case with flux. In presence of the flux, the states are more delocalized and also more similar to each other, so the distributions are more peaked. The value $E_J\sim 0.2$ corresponds to the maximum of $\mathcal{M}(E_K)$ for $L=4$, $M=2$ in presence of flux [Fig.~\ref{fig:tiparb2}(b)] and marks therefore the crossover between regular-like and ergodic behavior for these parameters.
 
{More delocalization in the charge eigenstate basis corresponds to more ergodicity. We can see it probing eigenstate thermalization through the smoothness of 
  $S_{L/2}^{(\alpha)}$ versus $E_\alpha$. We plot $S_{L/2}^{(\alpha)}$ versus $E_\alpha$} for different $E_J$, with {vanishing or finite} flux, in Fig.~\ref{fig:IPRdist}(a). We see by simple inspection that  the entanglement-entropy curves with the flux look smoother. This means that the system is more ETH: The entanglement entropy has smaller fluctuation around the thermal value. The delocalization so comes together with more ergodicity.
 
 These qualitative statements can be quantitatively probed. 
 
 We consider the quantifier $\mathcal{M}(S)$ [see Eq.~\eqref{momo:eqn}] in order to evaluate the smoothness of the $S_{L/2}^{(\alpha)}$ versus $E_\alpha$ curve. We see that $\mathcal{M}(S)$ is smaller in the case when there is magnetic flux [Fig.~\ref{fig:IPRdist}(b -- main)] when $E_J> 0.1$. So, in this parameter range the {finite} magnetic flux induces not only delocalization, but also a more ergodic behavior. This finding is confirmed by looking at $\mathcal{M}(\log{\rm IPR})$ (see Sec.~\ref{ergolo:sec}): A {finite flux gives rise to a} smaller value, {that} means more quantum chaos [Fig.~\ref{fig:IPRdist}(b -- inset)].
 
 A {somewhat} similar delocalizing effect of the magnetic flux occurs also in the Anderson localization problem~\cite{rammer,rama}, which is a disordered single-particle problem with space localization. Remarkably, we see something similar in our case, which is clean and many-body, and the localization phenomenon occurs in the basis of the charge eigenstates. 
\section{Conclusions}\label{conc:sec}
{In conclusion, we have studied weak ergodicity breaking in a Josephson-junction array. We have considered two geometries, a one-dimensional chain and a ladder. For $E_J$ up to order {$E_C$}, we have found a set of nonthermal low-entanglement eigenstates, which lie far from the spectral edges, and persist beyond the perturbative regime, {where the spectrum is globally organized in multiplets} (at least for the sizes we have access to). These eigenstates appear in small {groups}, and are essentially the superposition of few factorized charge-density wave states. Their existence depends on the divisors of the length $L$ of the array: For chains with length multiple of 2 we find doublets of nonthermal eigenstates with entanglement entropy $\log 2$, for lengths multiple of 3, sextuplets of eigenstates with entanglement entropy $\log 6$. They are a countable infinity also at finite size, just like the factorized states whose finite dressed superposition provides these eigenstates. This presence of a minority of nonthermal, robust eigenstates is reminiscent of the many-body quantum scars phenomenon.}

{As in the case of the quantum scars, these nonthermal eigenstates affect the dynamics, when special initial states are chosen. We have considered as initial states simple factorized charge eigenstates with charge density wave order. Defining special operators which probe the existence of the charge-density wave order, we have found its persistence for long times, in a regime of small $E_J$, depending on the geometry of the array, and the pattern of the initial charge density wave (we have checked patterns with period 2, and with period 3). The point is that these initial charge-density-wave states have large overlap with the low-entanglement eigenstates described above, and so the dynamics is strongly affected by them. The persistence of the charge density wave is a probe of the absence of thermalization, and we have studied this phenomenon using exact diagonalization, Krylov technique and tDMRG method.}

{After studying these special eigenstates, we have moved our attention to the full Hamiltonian spectrum, in order to understand if the system overall behaves or not in a thermal way. For that purpose, we have considered global probes of ergodicity, quantities integrated over the spectrum or the bulk of the eigenstates. Specifically, we have studied the average level spacing ratio $r$, and the smoothness quantifiers $\mathcal{M}(E_K)$ and $\mathcal{M}(\log{\rm IPR})$. We have found that the system behaves in a regular-like way at small Josephson coupling, but this parameter range shrinks in favor of an ergodic behavior for increasing system size. This suggests an increasingly ergodic behavior for increasing system size, but no extrapolation to the thermodynamic limit is possible}

{In the case of the ladder geometry, we have studied the effect of {a finite} magnetic flux on the global ergodicity properties.} We have seen that in presence of the flux the eigenstates become systematically more delocalized in the charge-eigenstates basis, as the IPR results show. This delocalization occurs in strict association with more ergodicity, as we see from $\mathcal{M}(S)$ and $\mathcal{M}(\log{\rm IPR})$. 
A similar {delocalizing} effect of the magnetic flux was known in the Anderson space localization problem, which is disordered and non interacting. Our results show that the same occurs in an interacting homogeneous many-body system. 

Perspectives of future research include from one side the analytical interpretation of the persistence of the nonthermal eigenstates beyond the perturbative regime, from the other the experimental observation of the persistence in time of the charge-density wave order. {In this context, it will be important to understand if the nonthermal eigenstates affect at small $E_J$ the behavior of the AC conductivity.} Another exciting possibility would be to analytically understand the delocalizing effect of the magnetic flux along the lines used in the Anderson-localization problem, related to the destruction of the constructive interference of time-reversal paired paths.
\acknowledgments{R.~F. acknowledges financial support from the Google Quantum Research Award. We acknowledge useful discussions with M.~Esposito. A.~R. thanks the Max-Planck Institut f\"ur Physik Komplexer Systeme and P.~Lucignano for the access to the computation facilities where the numerical analysis for this project was performed.}
\appendix
\section{{Energy fluctuation of the charge configurations}}\label{fluc:app}
Here we estimate the energy fluctuation of the charge configurations $\ket{\{q_j\}}$. This is a measure of the broadness over $E_\alpha$ of the overlap distribution $|\braket{\phi_\alpha|\{q_j\}}|^2$. The energy fluctuation is evaluated as follows
\begin{equation}\label{rabarbaro:eqn}
  \Delta E = \sqrt{\braket{\{q_j\}|\hat{H}^2|\{q_j\}}-\braket{\{q_j\}|\hat{H}|\{q_j\}}^2}\,.
\end{equation}
Using Eq.~\eqref{brano:eqn}, we see that $\hat{H}\ket{\{q_j\}}=E_{\{q_j\}}^{(0)}\ket{\{q_j\}}$, being $\ket{\{q_j\}}$ an eigenstate of the charging part of the Hamiltonian. Moreover, we see that $\hat{V}$ is the sum of many nearest-neighbour terms such that -- considering for instance the term acting on the sites $i$, $j$ -- 
\begin{align} \label{ele:eqn}
  \hat{V}_{i\,j}\ket{\ldots\,,q_i\,,q_j,\,\ldots}&\propto\ket{\ldots\,,q_i-1\,,q_j+1,\,\ldots}\nonumber\\
    &+\ket{\ldots\,,q_i+1\,,q_j-1,\,\ldots}\,,
\end{align}
so $\braket{\{q_j\}|\hat{V}|\{q_j\}}=0$. Applying all these relations in Eq.~\eqref{rabarbaro:eqn}, we find
\begin{equation}
  \Delta E = E_J\sqrt{\braket{\{q_j\}|\hat{V}^2|\{q_j\}}}\,.
\end{equation}
Writing the sum over nearest-neighbour terms $\hat{V}=\sum_{<i,\,j>}\hat{V}_{i\,j}$ and using the invariance under horizontal translations, we get
\begin{equation}
  \Delta E = E_J\sqrt{L\braket{\{q_j\}|\sum_{<i,\,1>}\sum_{<l,\,q>}\hat{V}_{i\,1}\hat{V}_{l\,q}|\{q_j\}}}\,.
\end{equation}
Using Eq.~\eqref{ele:eqn}, it is easy to get convinced that the matrix element is order 1. So, we have shown that $\Delta E$ is order $E_J\sqrt{L}$.
\section{Effect of the truncation} \label{effect:app}
Applying the truncation implies an error in the results, of course. In order to see how big it is, we consider the chain model and compare in Fig.~\ref{comparison_M:fig}(a,b) time traces of $\mathcal{I}_{(1)}(t)$ [Eq.~\eqref{iotta:eqn}] versus $E_J t$. We initialize $\ket{\psi_2(q)}$ with $q=1$, and take different values of $M$ ($M=2,3$). We see that changing $M$ the period of the Rabi oscillations slightly changes. Nevertheless, this effect is almost impossible to see for $E_J=0.1$ [Fig.~\ref{comparison_M:fig}(a)], so our approximation is better in the regime $E_J\ll 1$. Our focus is nevertheless on the amplitude of the oscillations (probed by $\mathcal{I}_{(2)}(t)$), which appears not to be affected by the truncation. This is confirmed by results with the tDMRG; As we can see in the example of Fig.~\ref{comparison_M:fig}(c) ($q=4$, $M=5,6$), the dynamics of $\mathcal{I}_{(2)}(t)$ is essentially not affected by the truncation $M$. As a rule of thumb, as soon as $M\geq q + 1$, the dynamics of $\mathcal{I}_{(2)}(t)$ is not affected by the truncation.
%
\begin{figure*}
  \begin{center}
    \begin{tabular}{ccc}
      \hspace{0cm}\begin{overpic}[width=60mm]{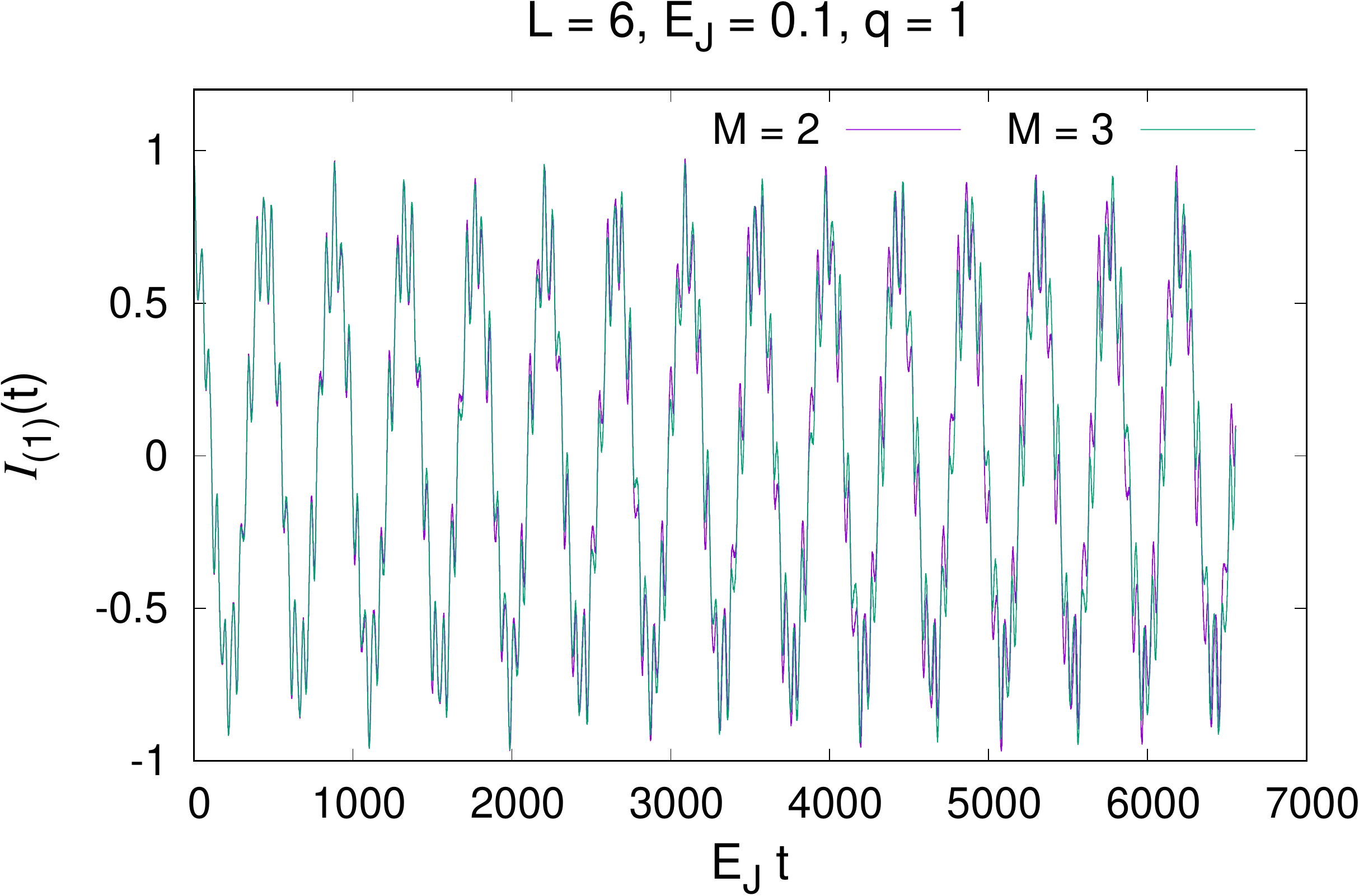}\put(15,68){(a)}\end{overpic}&
      \begin{overpic}[width=60mm]{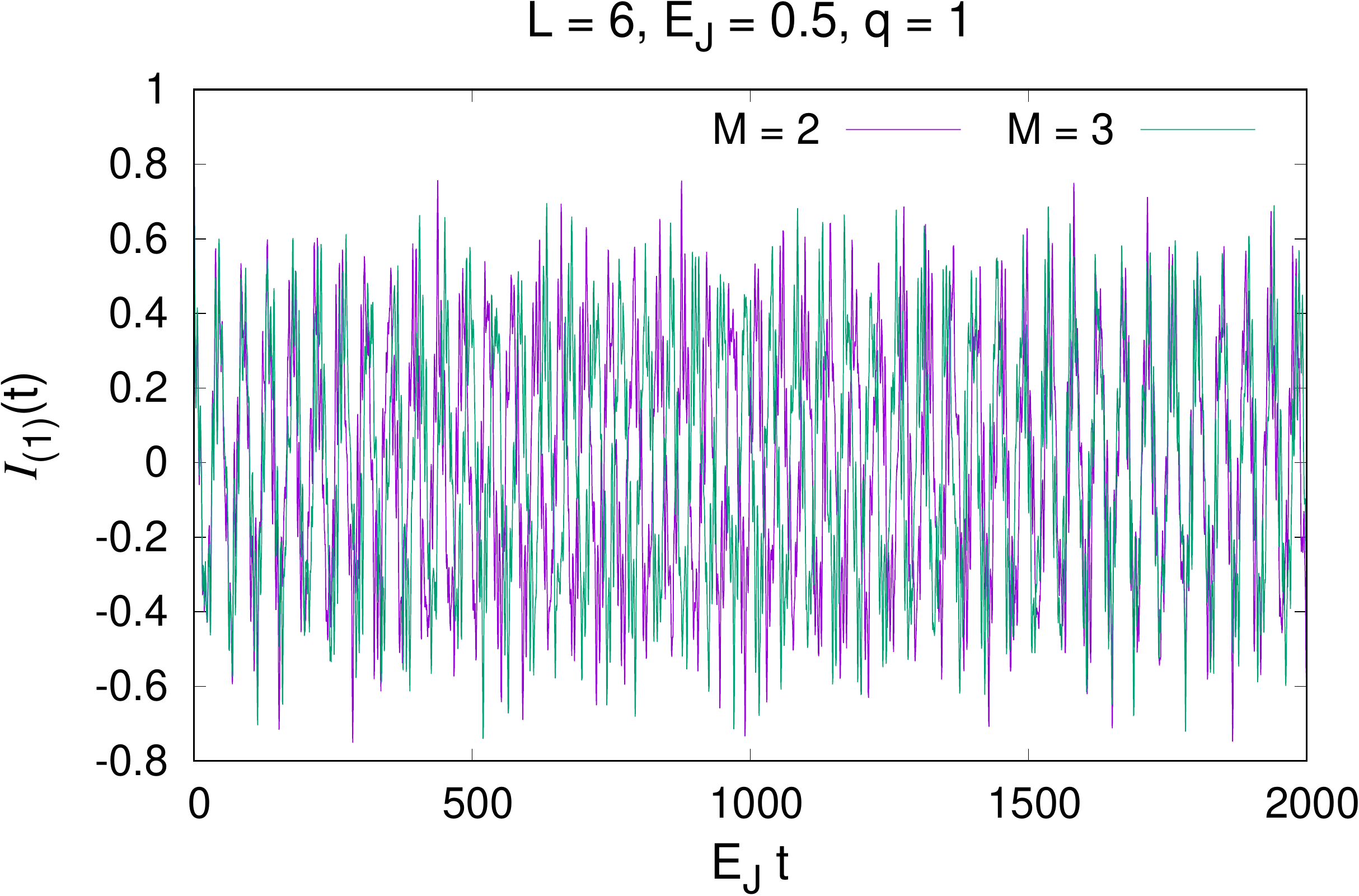}\put(15,68){(b)}\end{overpic}&
      \begin{overpic}[width=60mm]{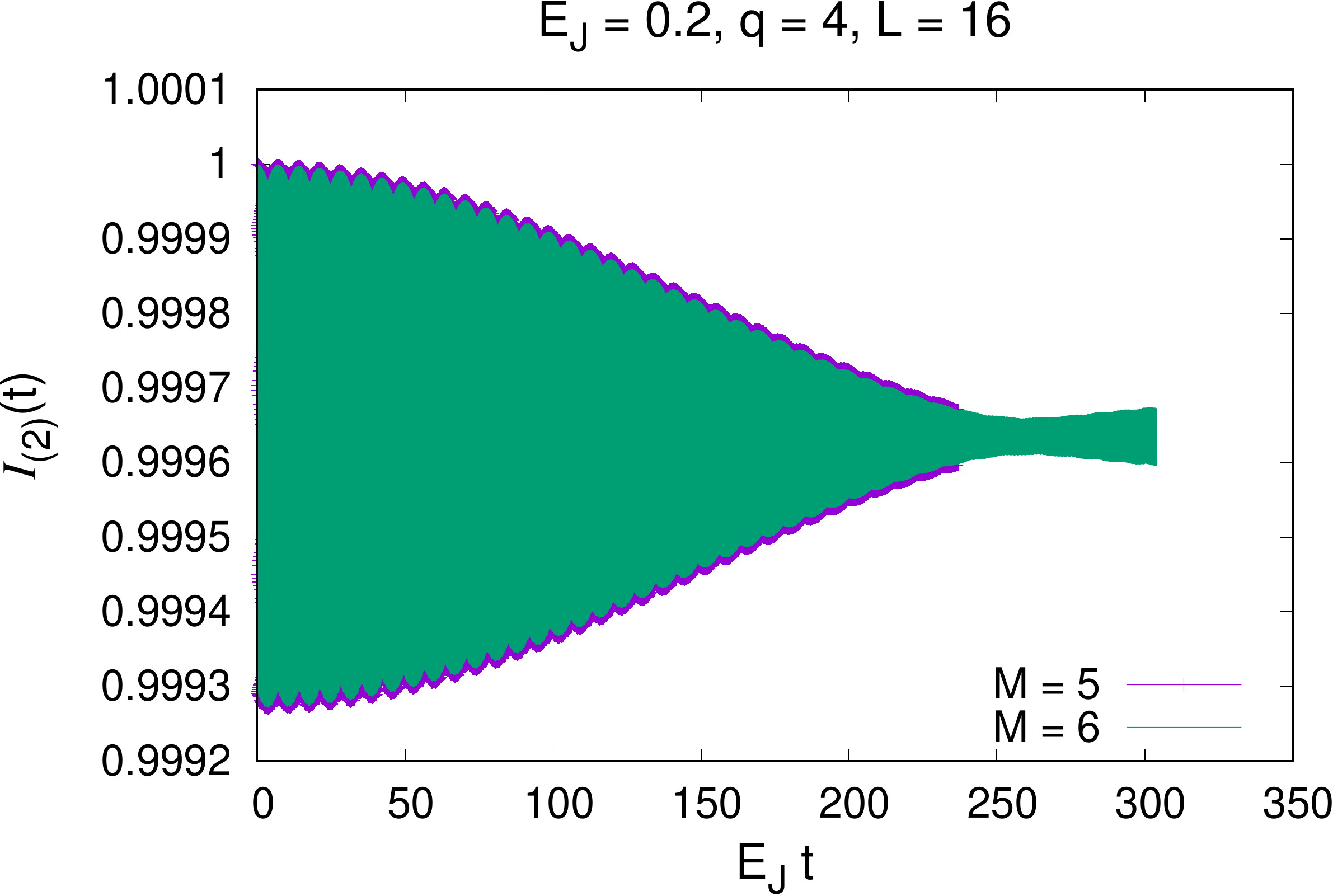}\put(18,68){(c)}\end{overpic}
    \end{tabular}
  \end{center}
\caption{Chain model. Panels (a,b) Comparison of the $\mathcal{I}_{(1)}(t)$ versus $E_J t$ evolution for different values of the truncation $M$. Dynamics performed with exact diagonalization, initial state $\ket{\psi_2(q)}$. (Panel c) Comparison of the $\mathcal{I}_{(2)}(t)$ versus $E_J t$ evolution for different values of the truncation $M$. Dynamics performed with tDMRG method, initial state $\ket{\psi_+(q)}$.}
\label{comparison_M:fig}
\end{figure*}
\section{Ergodicity probes for the ladder system}\label{App:lad}
In Fig.~\ref{fig:tipar02} we show $\mathcal{M}(\log{\rm IPR})$ (panels a,d) and $\mathcal{M}(E_K)$ (panels b,e), and the average level spacing ratio~\cite{PhysRevB.82.174411} $r$ versus $E_J$ (panels c,f) in the case of the ladder with {vanishing} flux ($\Phi=0$). As in the case of the chain, the system tends towards more ergodicity for increasing system size. The value of $r$ increases overall towards the Wigner-Dyson value, and the behavior of the $\mathcal{M}$ shows a small-$E_J$ regular-like region, that shrinks for increasing system size. This can be seen by looking at the intersections of the curves of $\mathcal{M}(\log{\rm IPR})$, or at the moving left and getting smaller with increasing system size of the maximum in $\mathcal{M}(E_K)$. 

We show the situation for {finite flux} $\Phi/\Phi_0=1/\phi$ in Fig.~\ref{fig:tiparb2}. We consider exactly the same quantities as we do in Fig.~\ref{fig:tipar02} for $\Phi=0$. The behavior with $L$ is qualitatively the same as in the $\Phi=0$ case, but -- considering corresponding values of $L$ and $M$ -- the case with $\Phi/\Phi_0=1/\phi$ is more ergodic. This can be seen for instance from the fact that $\mathcal{M}(\log{\rm IPR})$ is systematically smaller than the corresponding $\Phi=0$ value.
\begin{figure*}
  \begin{center}
    \begin{tabular}{ccc}
\\
\vspace{3mm}\\
     \hspace{-0.5cm} \begin{overpic}[width=60mm]{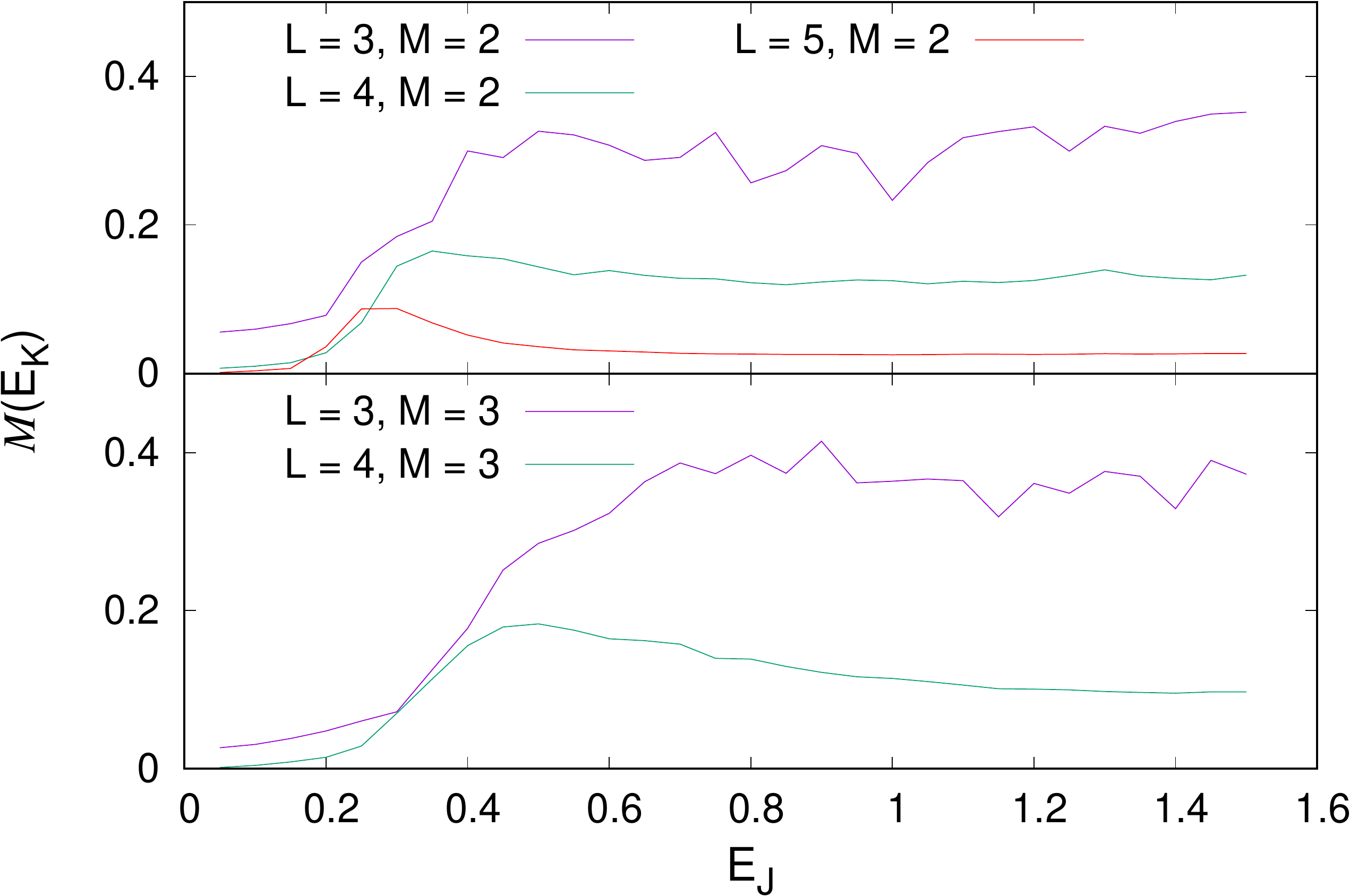}\put(14,70){(a)}\end{overpic}&%
	     \begin{overpic}[width=60mm]{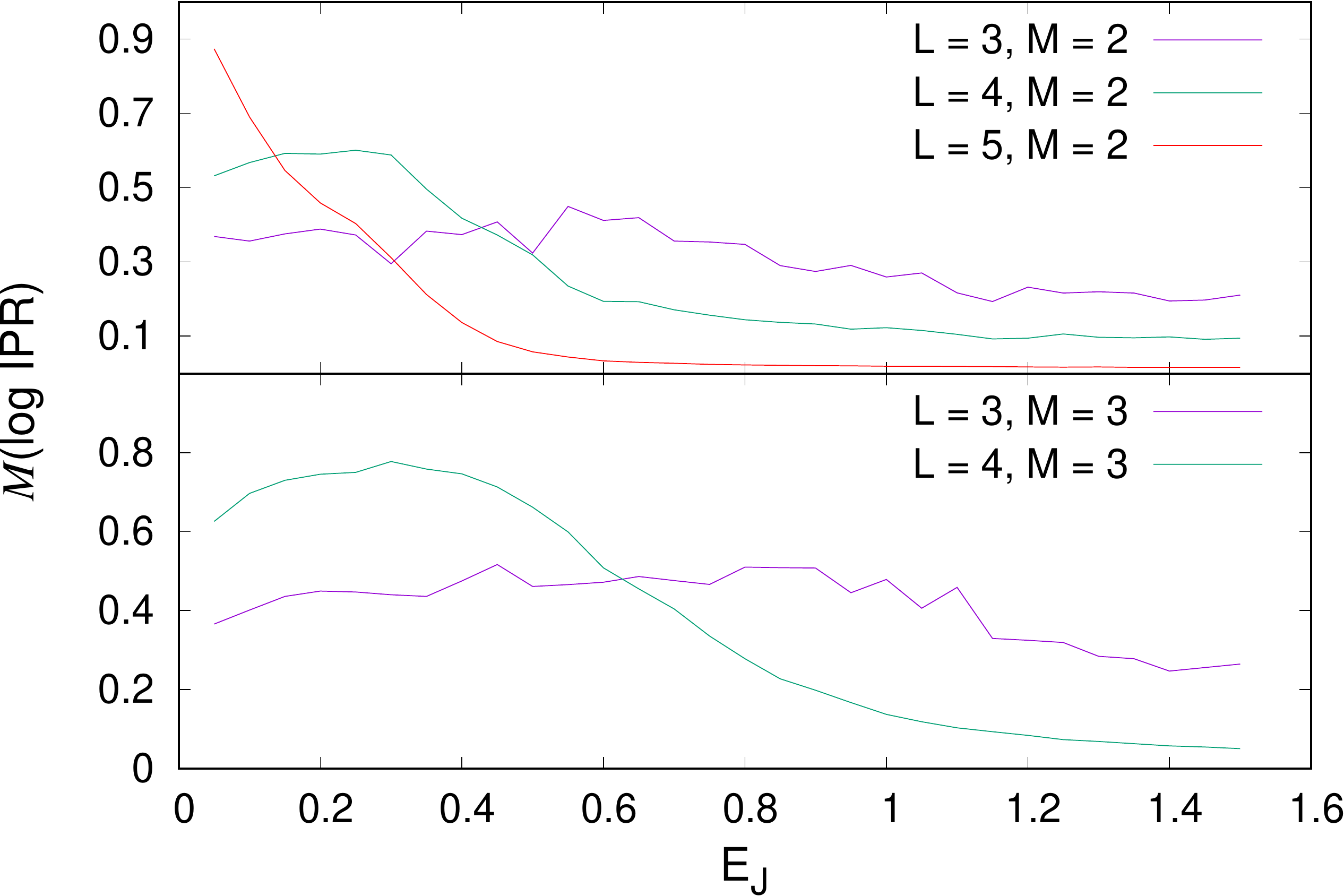}\put(14,70){(b)}\end{overpic}&
             \begin{overpic}[width=60mm]{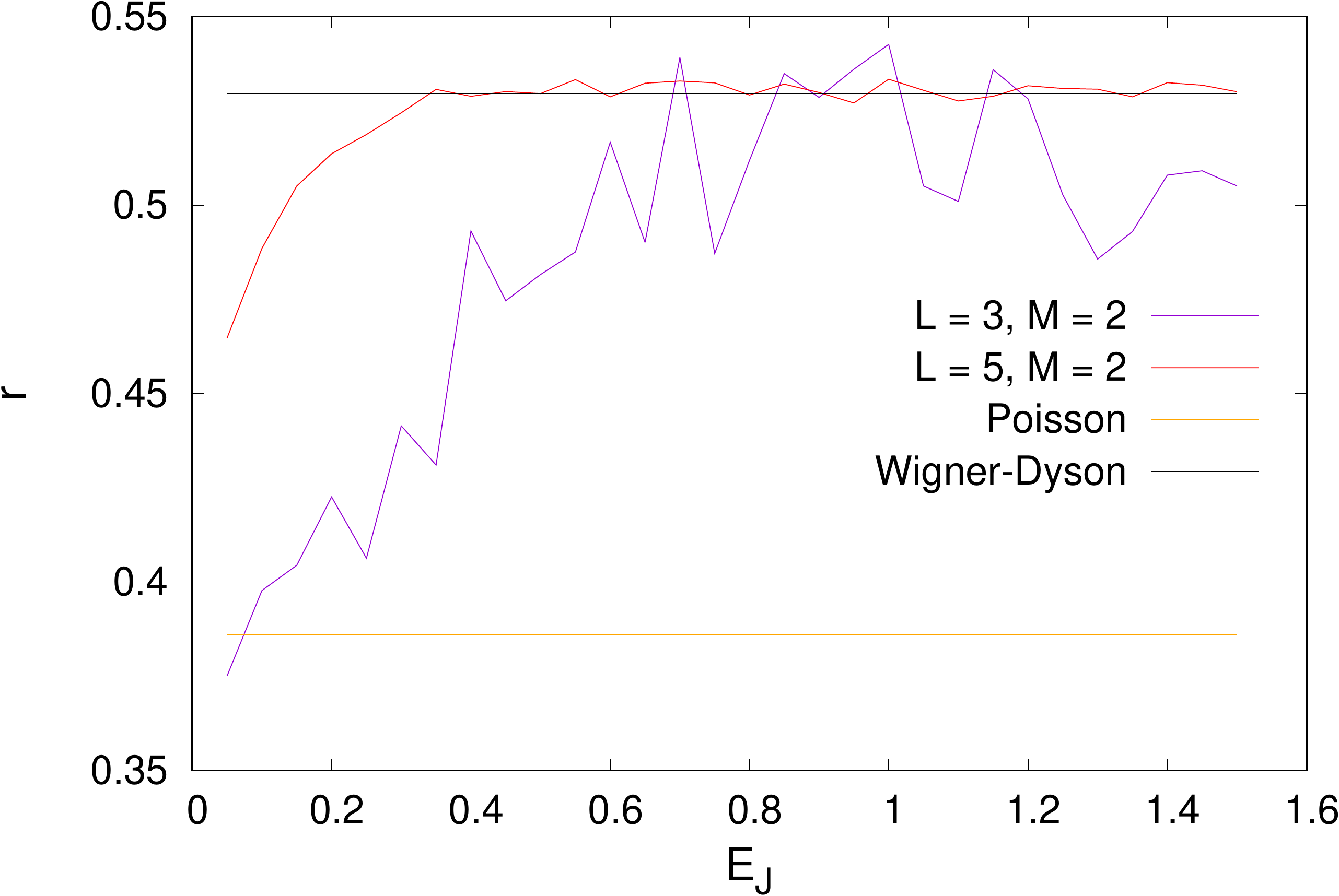}\put(14,70){(c)}\end{overpic}\\
%
    \end{tabular}
  \end{center}
 \caption{Ladder model with vanishing flux $\Phi=0$. (Panel a) $\mathcal{M}(E_K)$ versus $E_J$ and (panel b) $\mathcal{M}(\log{\rm IPR})$ versus $E_J$. (Panel c) Average level spacing ratio $r$ versus $E_J$.}
  \label{fig:tipar02}
\end{figure*}
\begin{figure*}[h!]
  \begin{center} 
    \begin{tabular}{ccc}
     \hspace{-0.5cm} \begin{overpic}[width=60mm]{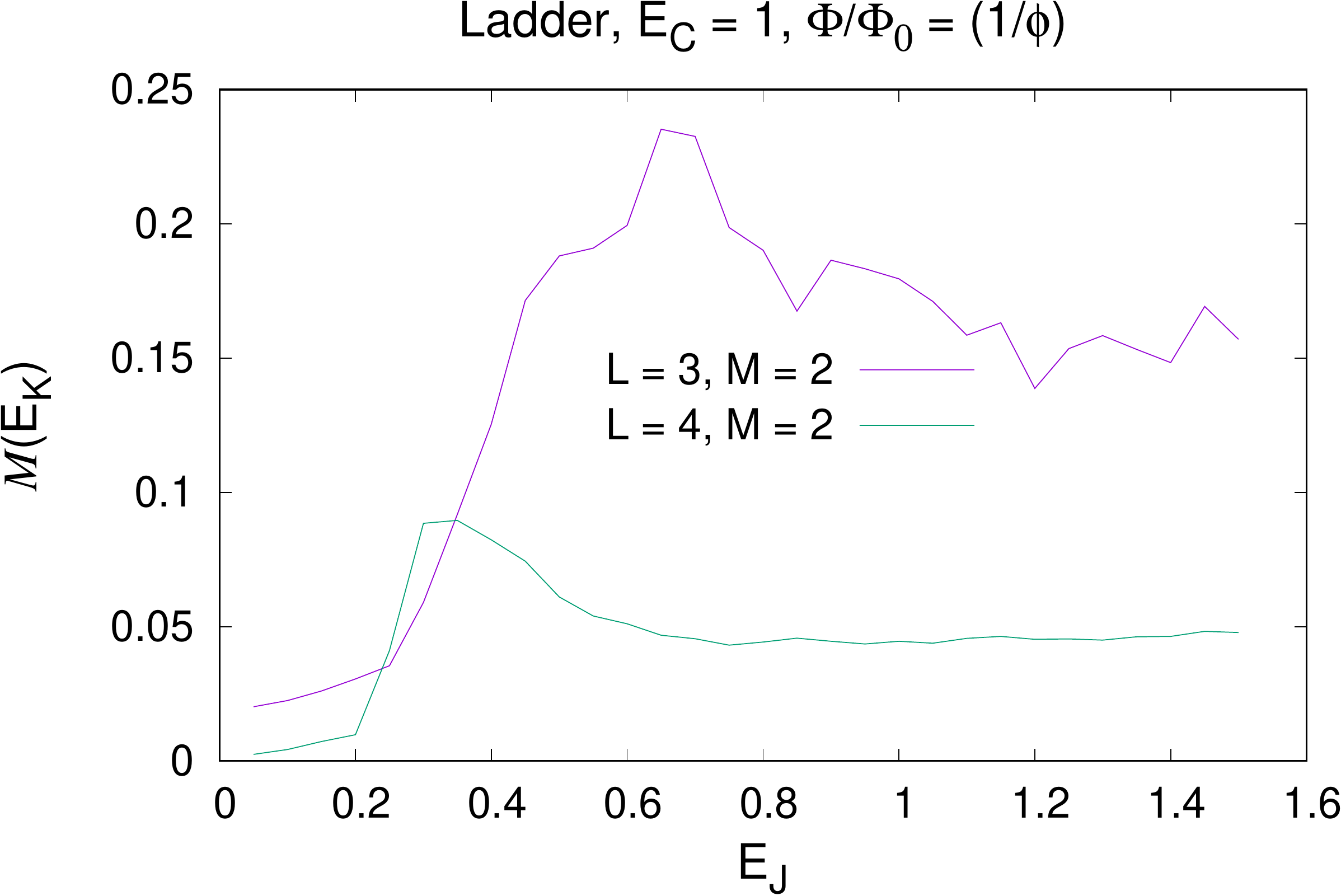}\put(17,64){(a)}\end{overpic}&%
	     \begin{overpic}[width=60mm]{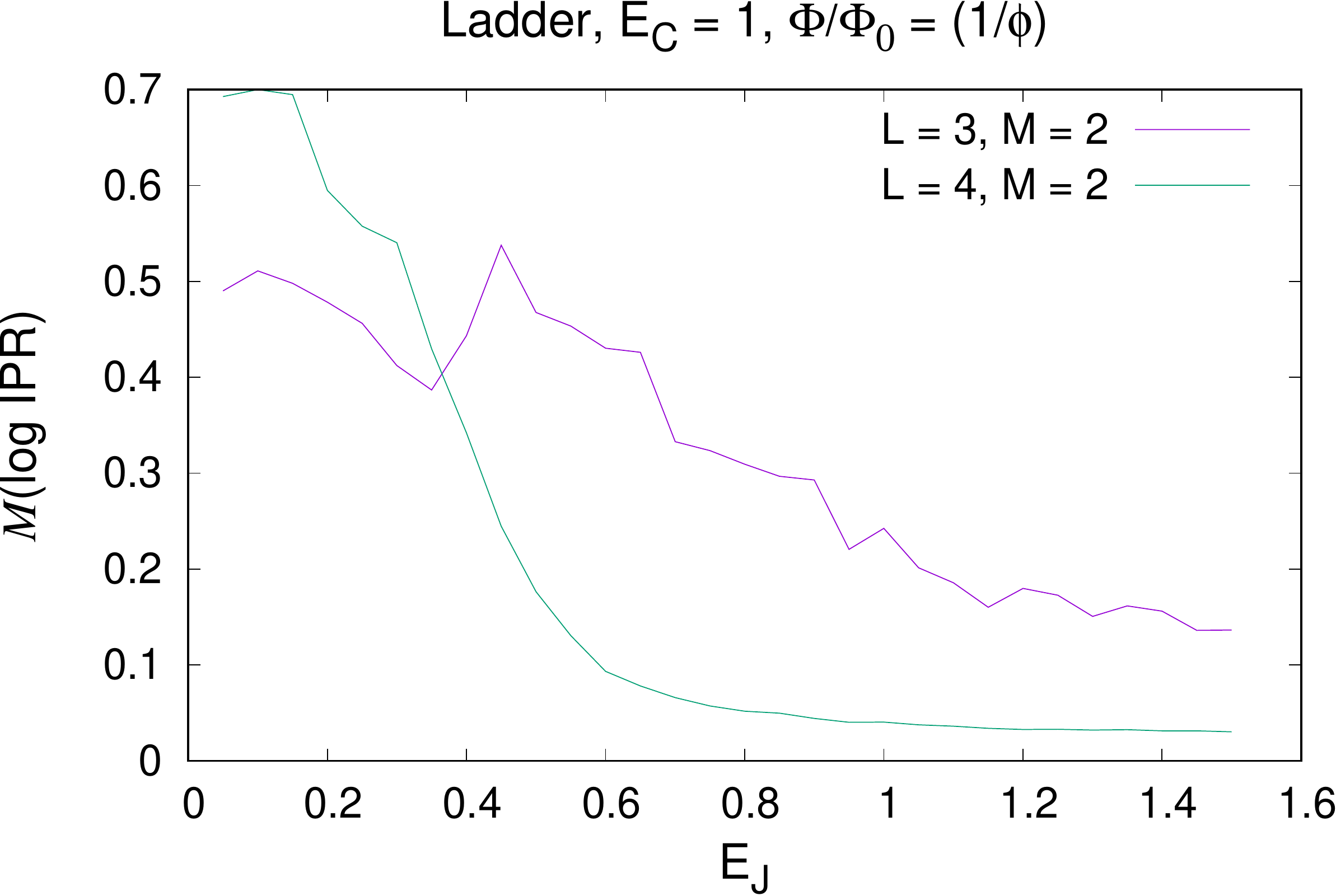}\put(17,64){(b)}\end{overpic}&
            \begin{overpic}[width=60mm]{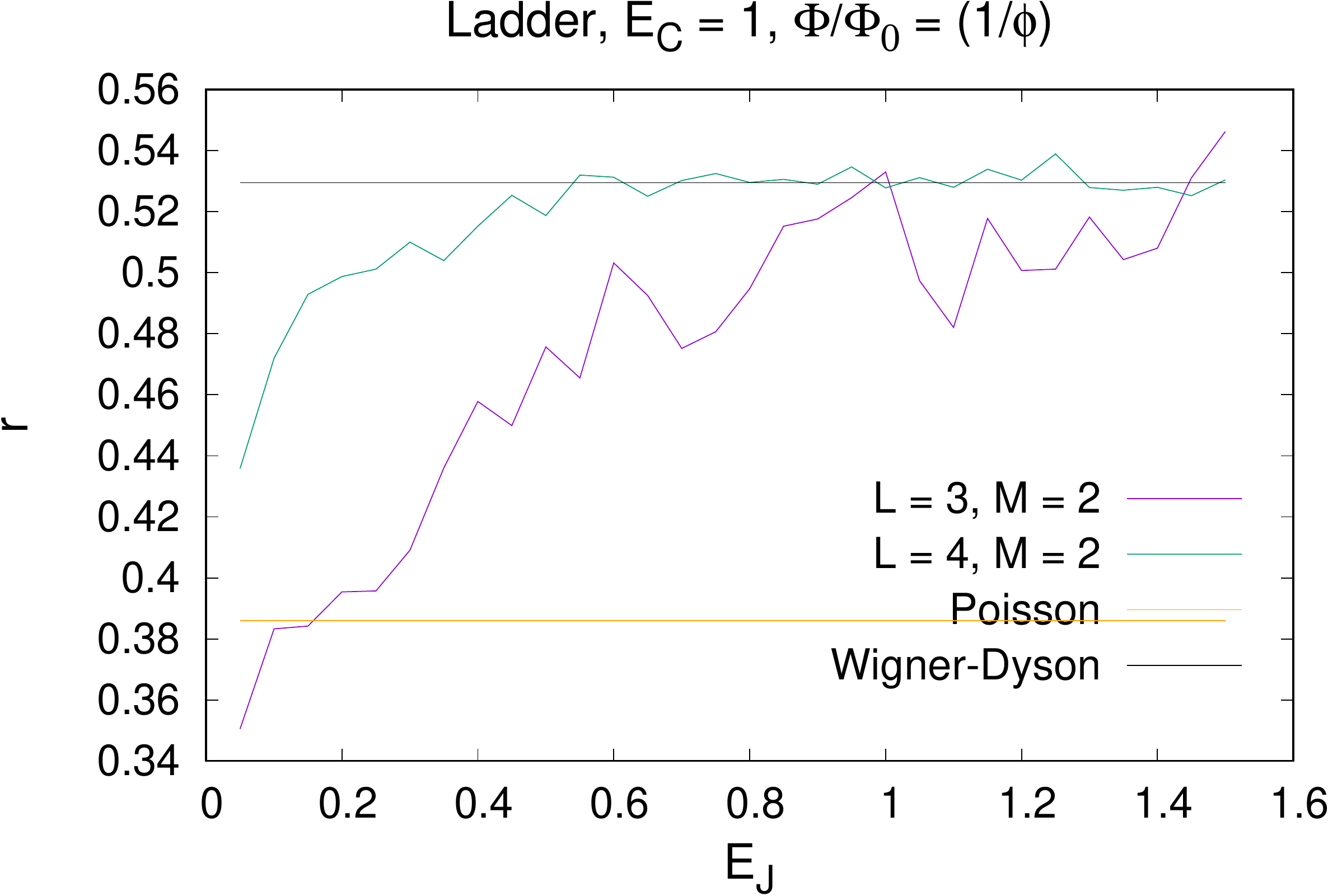}\put(17,64){(c)}\end{overpic}\\
     \hspace{-0.5cm} \begin{overpic}[width=60mm]{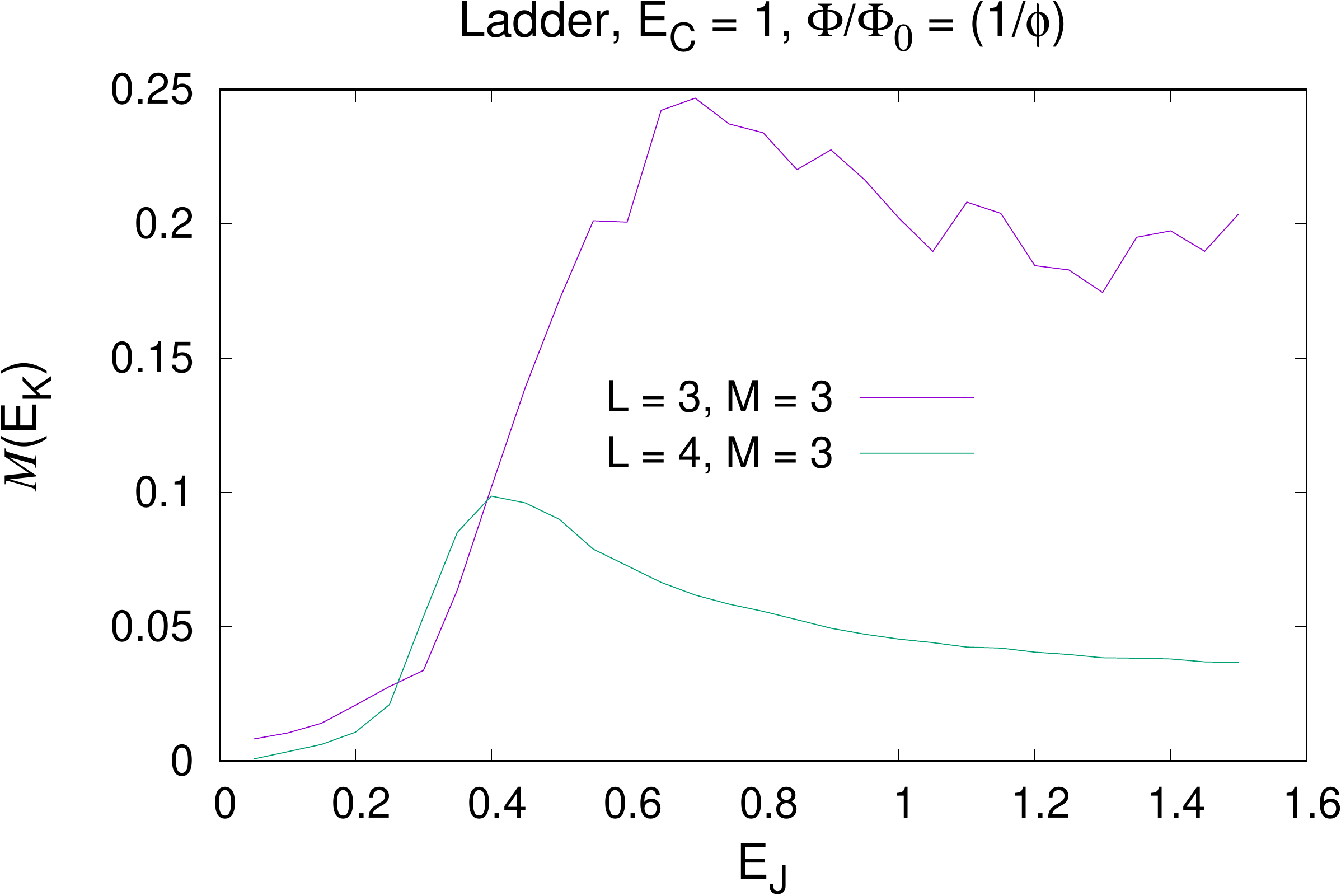}\put(17,64){(d)}\end{overpic}&
              \begin{overpic}[width=60mm]{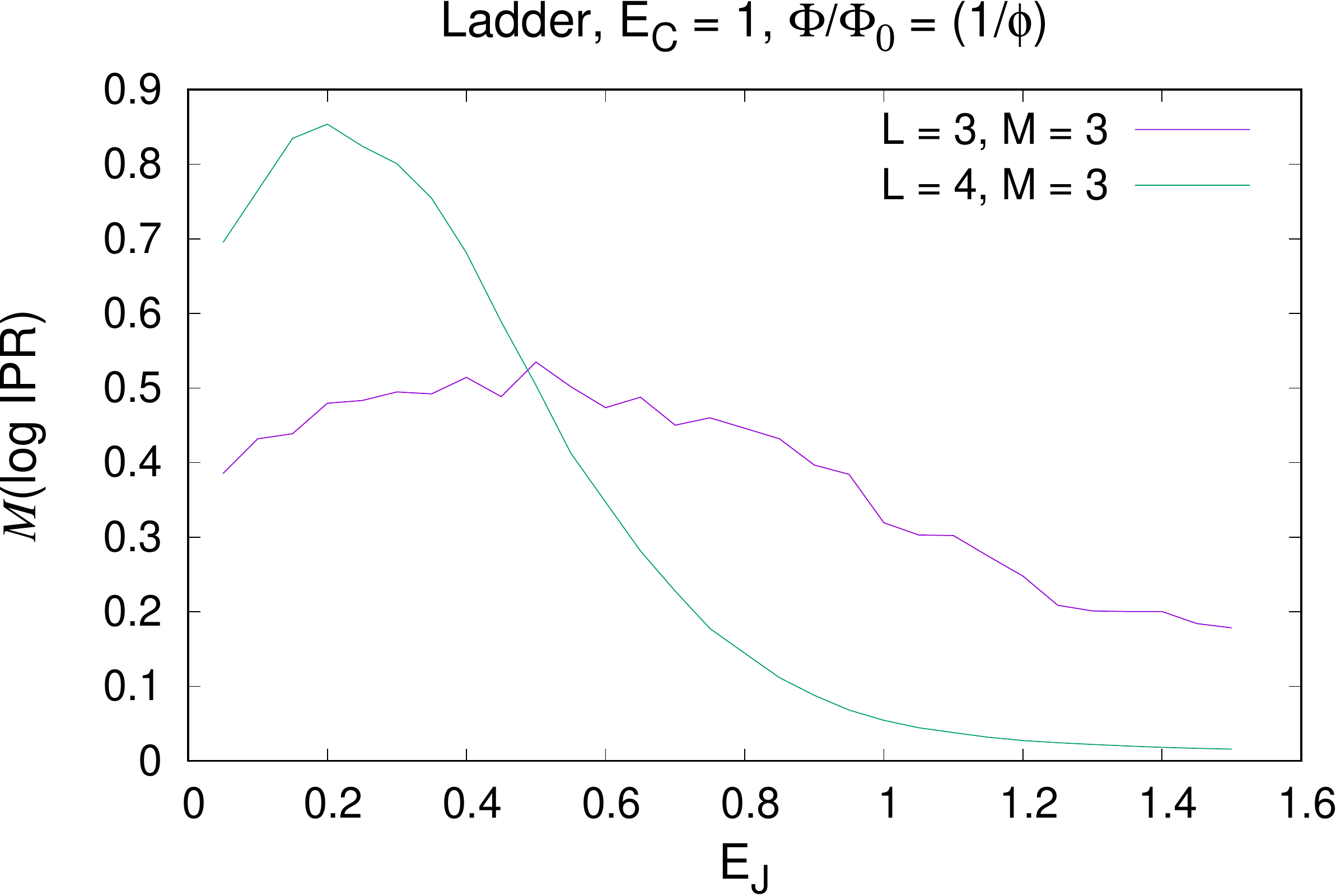}\put(17,64){(e)}\end{overpic}&
             \begin{overpic}[width=60mm]{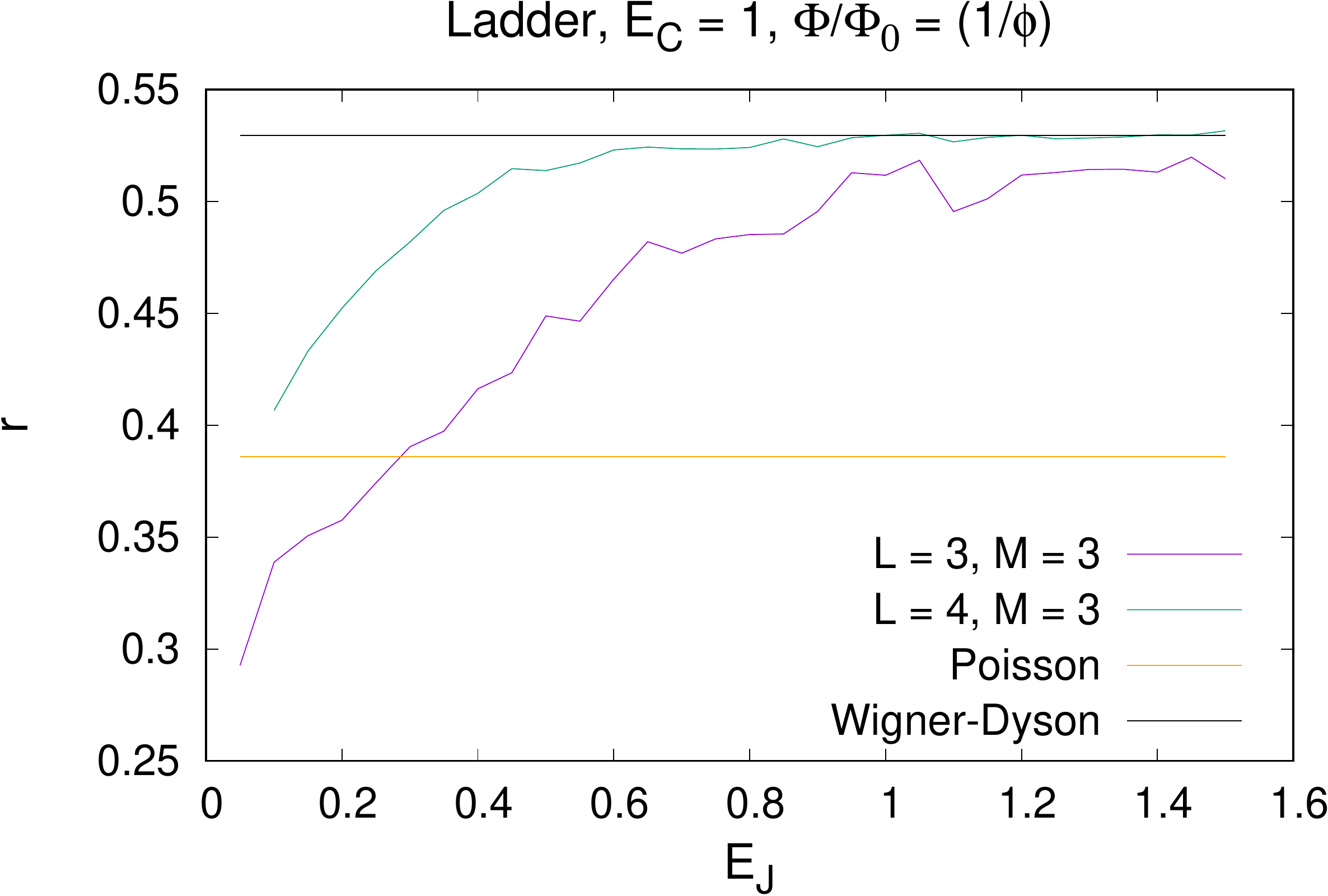}\put(17,64){(f)}\end{overpic}\\
    \end{tabular}
  \end{center}
	\caption{Ladder model with {finite} flux $\Phi/\Phi_0=1/\phi$. (Panel a,d) $\mathcal{M}(E_K)$ versus $E_J$ and (panel b,e) $\mathcal{M}(\log{\rm IPR})$ versus $E_J$. (Panel c,f) Average level spacing ratio $r$ versus $E_J$.}
  \label{fig:tiparb2}
\end{figure*}
\bibliography{junction}

\begin{thebibliography}{67}%
\makeatletter
\providecommand \@ifxundefined [1]{%
 \@ifx{#1\undefined}
}%
\providecommand \@ifnum [1]{%
 \ifnum #1\expandafter \@firstoftwo
 \else \expandafter \@secondoftwo
 \fi
}%
\providecommand \@ifx [1]{%
 \ifx #1\expandafter \@firstoftwo
 \else \expandafter \@secondoftwo
 \fi
}%
\providecommand \natexlab [1]{#1}%
\providecommand \enquote  [1]{``#1''}%
\providecommand \bibnamefont  [1]{#1}%
\providecommand \bibfnamefont [1]{#1}%
\providecommand \citenamefont [1]{#1}%
\providecommand \href@noop [0]{\@secondoftwo}%
\providecommand \href [0]{\begingroup \@sanitize@url \@href}%
\providecommand \@href[1]{\@@startlink{#1}\@@href}%
\providecommand \@@href[1]{\endgroup#1\@@endlink}%
\providecommand \@sanitize@url [0]{\catcode `\\12\catcode `\$12\catcode
  `\&12\catcode `\#12\catcode `\^12\catcode `\_12\catcode `\%12\relax}%
\providecommand \@@startlink[1]{}%
\providecommand \@@endlink[0]{}%
\providecommand \url  [0]{\begingroup\@sanitize@url \@url }%
\providecommand \@url [1]{\endgroup\@href {#1}{\urlprefix }}%
\providecommand \urlprefix  [0]{URL }%
\providecommand \Eprint [0]{\href }%
\providecommand \doibase [0]{https://doi.org/}%
\providecommand \selectlanguage [0]{\@gobble}%
\providecommand \bibinfo  [0]{\@secondoftwo}%
\providecommand \bibfield  [0]{\@secondoftwo}%
\providecommand \translation [1]{[#1]}%
\providecommand \BibitemOpen [0]{}%
\providecommand \bibitemStop [0]{}%
\providecommand \bibitemNoStop [0]{.\EOS\space}%
\providecommand \EOS [0]{\spacefactor3000\relax}%
\providecommand \BibitemShut  [1]{\csname bibitem#1\endcsname}%
\let\auto@bib@innerbib\@empty
\bibitem [{\citenamefont {Polkovnikov}\ \emph {et~al.}(2011)\citenamefont
  {Polkovnikov}, \citenamefont {Sengupta}, \citenamefont {Silva},\ and\
  \citenamefont {Vengalattore}}]{silva_rmp}%
  \BibitemOpen
  \bibfield  {author} {\bibinfo {author} {\bibfnamefont {A.}~\bibnamefont
  {Polkovnikov}}, \bibinfo {author} {\bibfnamefont {K.}~\bibnamefont
  {Sengupta}}, \bibinfo {author} {\bibfnamefont {A.}~\bibnamefont {Silva}},\
  and\ \bibinfo {author} {\bibfnamefont {M.}~\bibnamefont {Vengalattore}},\
  }\bibfield  {title} {\bibinfo {title} {Colloquium: Nonequilibrium dynamics of
  closed interacting quantum systems},\ }\href
  {https://doi.org/10.1103/RevModPhys.83.863} {\bibfield  {journal} {\bibinfo
  {journal} {Rev. Mod. Phys.}\ }\textbf {\bibinfo {volume} {83}},\ \bibinfo
  {pages} {863} (\bibinfo {year} {2011})}\BibitemShut {NoStop}%
\bibitem [{\citenamefont {D’Alessio}\ \emph {et~al.}(2016)\citenamefont
  {D’Alessio}, \citenamefont {Kafri}, \citenamefont {Polkovnikov},\ and\
  \citenamefont {Rigol}}]{kafri_2016}%
  \BibitemOpen
  \bibfield  {author} {\bibinfo {author} {\bibfnamefont {L.}~\bibnamefont
  {D’Alessio}}, \bibinfo {author} {\bibfnamefont {Y.}~\bibnamefont {Kafri}},
  \bibinfo {author} {\bibfnamefont {A.}~\bibnamefont {Polkovnikov}},\ and\
  \bibinfo {author} {\bibfnamefont {M.}~\bibnamefont {Rigol}},\ }\bibfield
  {title} {\bibinfo {title} {From quantum chaos and eigenstate thermalization
  to statistical mechanics and thermodynamics},\ }\href
  {https://doi.org/10.1080/00018732.2016.1198134} {\bibfield  {journal}
  {\bibinfo  {journal} {Advances in Physics}\ }\textbf {\bibinfo {volume}
  {65}},\ \bibinfo {pages} {239–362} (\bibinfo {year} {2016})}\BibitemShut
  {NoStop}%
\bibitem [{\citenamefont {Rigol}\ \emph {et~al.}(2008)\citenamefont {Rigol},
  \citenamefont {Dunjko},\ and\ \citenamefont {Olshanii}}]{Rigol_Nat}%
  \BibitemOpen
  \bibfield  {author} {\bibinfo {author} {\bibfnamefont {M.}~\bibnamefont
  {Rigol}}, \bibinfo {author} {\bibfnamefont {V.}~\bibnamefont {Dunjko}},\ and\
  \bibinfo {author} {\bibfnamefont {M.}~\bibnamefont {Olshanii}},\ }\bibfield
  {title} {\bibinfo {title} {Thermalization and its mechanism for generic
  isolated quantum systems},\ }\href {https://doi.org/10.1038/nature06838}
  {\bibfield  {journal} {\bibinfo  {journal} {Nature}\ }\textbf {\bibinfo
  {volume} {452}},\ \bibinfo {pages} {854} (\bibinfo {year}
  {2008})}\BibitemShut {NoStop}%
\bibitem [{\citenamefont {Prosen}(1998)}]{Prosen_PRL98}%
  \BibitemOpen
  \bibfield  {author} {\bibinfo {author} {\bibfnamefont {T.~c.~v.}\
  \bibnamefont {Prosen}},\ }\bibfield  {title} {\bibinfo {title} {Time
  evolution of a quantum many-body system: Transition from integrability to
  ergodicity in the thermodynamic limit},\ }\href
  {https://doi.org/10.1103/PhysRevLett.80.1808} {\bibfield  {journal} {\bibinfo
   {journal} {Phys. Rev. Lett.}\ }\textbf {\bibinfo {volume} {80}},\ \bibinfo
  {pages} {1808} (\bibinfo {year} {1998})}\BibitemShut {NoStop}%
\bibitem [{\citenamefont {Srednicki}(1994)}]{Sred_PRE94}%
  \BibitemOpen
  \bibfield  {author} {\bibinfo {author} {\bibfnamefont {M.}~\bibnamefont
  {Srednicki}},\ }\bibfield  {title} {\bibinfo {title} {Chaos and quantum
  thermalization},\ }\href {https://doi.org/10.1103/PhysRevE.50.888} {\bibfield
   {journal} {\bibinfo  {journal} {Phys. Rev. E}\ }\textbf {\bibinfo {volume}
  {50}},\ \bibinfo {pages} {888} (\bibinfo {year} {1994})}\BibitemShut
  {NoStop}%
\bibitem [{\citenamefont {Deutsch}(1991)}]{Deutsch_PRA91}%
  \BibitemOpen
  \bibfield  {author} {\bibinfo {author} {\bibfnamefont {J.~M.}\ \bibnamefont
  {Deutsch}},\ }\bibfield  {title} {\bibinfo {title} {Quantum statistical
  mechanics in a closed system},\ }\href
  {https://doi.org/10.1103/PhysRevA.43.2046} {\bibfield  {journal} {\bibinfo
  {journal} {Phys. Rev. A}\ }\textbf {\bibinfo {volume} {43}},\ \bibinfo
  {pages} {2046} (\bibinfo {year} {1991})}\BibitemShut {NoStop}%
\bibitem [{\citenamefont {Abanin}\ \emph {et~al.}(2019)\citenamefont {Abanin},
  \citenamefont {Altman}, \citenamefont {Bloch},\ and\ \citenamefont
  {Serbyn}}]{abanin_rmp}%
  \BibitemOpen
  \bibfield  {author} {\bibinfo {author} {\bibfnamefont {D.~A.}\ \bibnamefont
  {Abanin}}, \bibinfo {author} {\bibfnamefont {E.}~\bibnamefont {Altman}},
  \bibinfo {author} {\bibfnamefont {I.}~\bibnamefont {Bloch}},\ and\ \bibinfo
  {author} {\bibfnamefont {M.}~\bibnamefont {Serbyn}},\ }\bibfield  {title}
  {\bibinfo {title} {Colloquium: Many-body localization, thermalization, and
  entanglement},\ }\href {https://doi.org/10.1103/RevModPhys.91.021001}
  {\bibfield  {journal} {\bibinfo  {journal} {Rev. Mod. Phys.}\ }\textbf
  {\bibinfo {volume} {91}},\ \bibinfo {pages} {021001} (\bibinfo {year}
  {2019})}\BibitemShut {NoStop}%
\bibitem [{\citenamefont {Kohlert}\ \emph {et~al.}(2019)\citenamefont
  {Kohlert}, \citenamefont {Scherg}, \citenamefont {Li}, \citenamefont
  {Lüschen}, \citenamefont {Das~Sarma}, \citenamefont {Bloch},\ and\
  \citenamefont {Aidelsburger}}]{Bloch2019}%
  \BibitemOpen
  \bibfield  {author} {\bibinfo {author} {\bibfnamefont {T.}~\bibnamefont
  {Kohlert}}, \bibinfo {author} {\bibfnamefont {S.}~\bibnamefont {Scherg}},
  \bibinfo {author} {\bibfnamefont {X.}~\bibnamefont {Li}}, \bibinfo {author}
  {\bibfnamefont {H.~P.}\ \bibnamefont {Lüschen}}, \bibinfo {author}
  {\bibfnamefont {S.}~\bibnamefont {Das~Sarma}}, \bibinfo {author}
  {\bibfnamefont {I.}~\bibnamefont {Bloch}},\ and\ \bibinfo {author}
  {\bibfnamefont {M.}~\bibnamefont {Aidelsburger}},\ }\bibfield  {title}
  {\bibinfo {title} {Observation of many-body localization in a one-dimensional
  system with a single-particle mobility edge},\ }\bibfield  {journal}
  {\bibinfo  {journal} {Physical Review Letters}\ }\textbf {\bibinfo {volume}
  {122}},\ \href {https://doi.org/10.1103/physrevlett.122.170403}
  {10.1103/physrevlett.122.170403} (\bibinfo {year} {2019})\BibitemShut
  {NoStop}%
\bibitem [{\citenamefont {Karpov}\ \emph {et~al.}(2020)\citenamefont {Karpov},
  \citenamefont {Verdel}, \citenamefont {Huang}, \citenamefont {Schmitt},\ and\
  \citenamefont {Heyl}}]{karpov2020disorderfree}%
  \BibitemOpen
  \bibfield  {author} {\bibinfo {author} {\bibfnamefont {P.}~\bibnamefont
  {Karpov}}, \bibinfo {author} {\bibfnamefont {R.}~\bibnamefont {Verdel}},
  \bibinfo {author} {\bibfnamefont {Y.~P.}\ \bibnamefont {Huang}}, \bibinfo
  {author} {\bibfnamefont {M.}~\bibnamefont {Schmitt}},\ and\ \bibinfo {author}
  {\bibfnamefont {M.}~\bibnamefont {Heyl}},\ }\href@noop {} {\bibinfo {title}
  {Disorder-free localization in an interacting two-dimensional lattice gauge
  theory}} (\bibinfo {year} {2020}),\ \Eprint
  {https://arxiv.org/abs/2003.04901} {arXiv:2003.04901 [cond-mat.str-el]}
  \BibitemShut {NoStop}%
\bibitem [{\citenamefont {Smith}\ \emph {et~al.}(2018)\citenamefont {Smith},
  \citenamefont {Knolle}, \citenamefont {Moessner},\ and\ \citenamefont
  {Kovrizhin}}]{Adam_prb}%
  \BibitemOpen
  \bibfield  {author} {\bibinfo {author} {\bibfnamefont {A.}~\bibnamefont
  {Smith}}, \bibinfo {author} {\bibfnamefont {J.}~\bibnamefont {Knolle}},
  \bibinfo {author} {\bibfnamefont {R.}~\bibnamefont {Moessner}},\ and\
  \bibinfo {author} {\bibfnamefont {D.~L.}\ \bibnamefont {Kovrizhin}},\
  }\bibfield  {title} {\bibinfo {title} {Dynamical localization in
  ${\ensuremath{\mathbb{z}}}_{2}$ lattice gauge theories},\ }\href
  {https://doi.org/10.1103/PhysRevB.97.245137} {\bibfield  {journal} {\bibinfo
  {journal} {Phys. Rev. B}\ }\textbf {\bibinfo {volume} {97}},\ \bibinfo
  {pages} {245137} (\bibinfo {year} {2018})}\BibitemShut {NoStop}%
\bibitem [{\citenamefont {Smith}\ \emph {et~al.}(2019)\citenamefont {Smith},
  \citenamefont {Knolle}, \citenamefont {Moessner},\ and\ \citenamefont
  {Kovrizhin}}]{Adam_prl19}%
  \BibitemOpen
  \bibfield  {author} {\bibinfo {author} {\bibfnamefont {A.}~\bibnamefont
  {Smith}}, \bibinfo {author} {\bibfnamefont {J.}~\bibnamefont {Knolle}},
  \bibinfo {author} {\bibfnamefont {R.}~\bibnamefont {Moessner}},\ and\
  \bibinfo {author} {\bibfnamefont {D.~L.}\ \bibnamefont {Kovrizhin}},\
  }\bibfield  {title} {\bibinfo {title} {Logarithmic spreading of
  out-of-time-ordered correlators without many-body localization},\ }\href
  {https://doi.org/10.1103/PhysRevLett.123.086602} {\bibfield  {journal}
  {\bibinfo  {journal} {Phys. Rev. Lett.}\ }\textbf {\bibinfo {volume} {123}},\
  \bibinfo {pages} {086602} (\bibinfo {year} {2019})}\BibitemShut {NoStop}%
\bibitem [{\citenamefont {Smith}\ \emph {et~al.}(2017)\citenamefont {Smith},
  \citenamefont {Knolle}, \citenamefont {Moessner},\ and\ \citenamefont
  {Kovrizhin}}]{Adam_prl171}%
  \BibitemOpen
  \bibfield  {author} {\bibinfo {author} {\bibfnamefont {A.}~\bibnamefont
  {Smith}}, \bibinfo {author} {\bibfnamefont {J.}~\bibnamefont {Knolle}},
  \bibinfo {author} {\bibfnamefont {R.}~\bibnamefont {Moessner}},\ and\
  \bibinfo {author} {\bibfnamefont {D.~L.}\ \bibnamefont {Kovrizhin}},\
  }\bibfield  {title} {\bibinfo {title} {Absence of ergodicity without quenched
  disorder: From quantum disentangled liquids to many-body localization},\
  }\href {https://doi.org/10.1103/PhysRevLett.119.176601} {\bibfield  {journal}
  {\bibinfo  {journal} {Phys. Rev. Lett.}\ }\textbf {\bibinfo {volume} {119}},\
  \bibinfo {pages} {176601} (\bibinfo {year} {2017})}\BibitemShut {NoStop}%
\bibitem [{\citenamefont {Pino}\ \emph {et~al.}(2016)\citenamefont {Pino},
  \citenamefont {Ioffe},\ and\ \citenamefont {Altshuler}}]{altshuler}%
  \BibitemOpen
  \bibfield  {author} {\bibinfo {author} {\bibfnamefont {M.}~\bibnamefont
  {Pino}}, \bibinfo {author} {\bibfnamefont {L.~B.}\ \bibnamefont {Ioffe}},\
  and\ \bibinfo {author} {\bibfnamefont {B.~L.}\ \bibnamefont {Altshuler}},\
  }\bibfield  {title} {\bibinfo {title} {Nonergodic metallic and insulating
  phases of josephson junction chains},\ }\href@noop {} {\bibfield  {journal}
  {\bibinfo  {journal} {PNAS}\ }\textbf {\bibinfo {volume} {113}},\ \bibinfo
  {pages} {536} (\bibinfo {year} {2016})}\BibitemShut {NoStop}%
\bibitem [{\citenamefont {Khemani}\ \emph {et~al.}(2019)\citenamefont
  {Khemani}, \citenamefont {Laumann},\ and\ \citenamefont
  {Chandran}}]{PhysRevB.99.161101}%
  \BibitemOpen
  \bibfield  {author} {\bibinfo {author} {\bibfnamefont {V.}~\bibnamefont
  {Khemani}}, \bibinfo {author} {\bibfnamefont {C.~R.}\ \bibnamefont
  {Laumann}},\ and\ \bibinfo {author} {\bibfnamefont {A.}~\bibnamefont
  {Chandran}},\ }\bibfield  {title} {\bibinfo {title} {Signatures of
  integrability in the dynamics of rydberg-blockaded chains},\ }\href
  {https://doi.org/10.1103/PhysRevB.99.161101} {\bibfield  {journal} {\bibinfo
  {journal} {Phys. Rev. B}\ }\textbf {\bibinfo {volume} {99}},\ \bibinfo
  {pages} {161101} (\bibinfo {year} {2019})}\BibitemShut {NoStop}%
\bibitem [{\citenamefont {Turner}\ \emph {et~al.}(2018)\citenamefont {Turner},
  \citenamefont {Michailidis}, \citenamefont {Abanin}, \citenamefont {Serbyn},\
  and\ \citenamefont {Papi\ifmmode~\acute{c}\else
  \'{c}\fi{}}}]{PhysRevB.98.155134}%
  \BibitemOpen
  \bibfield  {author} {\bibinfo {author} {\bibfnamefont {C.~J.}\ \bibnamefont
  {Turner}}, \bibinfo {author} {\bibfnamefont {A.~A.}\ \bibnamefont
  {Michailidis}}, \bibinfo {author} {\bibfnamefont {D.~A.}\ \bibnamefont
  {Abanin}}, \bibinfo {author} {\bibfnamefont {M.}~\bibnamefont {Serbyn}},\
  and\ \bibinfo {author} {\bibfnamefont {Z.}~\bibnamefont
  {Papi\ifmmode~\acute{c}\else \'{c}\fi{}}},\ }\bibfield  {title} {\bibinfo
  {title} {Quantum scarred eigenstates in a rydberg atom chain: Entanglement,
  breakdown of thermalization, and stability to perturbations},\ }\href
  {https://doi.org/10.1103/PhysRevB.98.155134} {\bibfield  {journal} {\bibinfo
  {journal} {Phys. Rev. B}\ }\textbf {\bibinfo {volume} {98}},\ \bibinfo
  {pages} {155134} (\bibinfo {year} {2018})}\BibitemShut {NoStop}%
\bibitem [{\citenamefont {Serbyn}\ \emph {et~al.}(2021)\citenamefont {Serbyn},
  \citenamefont {Abanin},\ and\ \citenamefont {Papić}}]{2021_serbyn_nat}%
  \BibitemOpen
  \bibfield  {author} {\bibinfo {author} {\bibfnamefont {M.}~\bibnamefont
  {Serbyn}}, \bibinfo {author} {\bibfnamefont {D.~A.}\ \bibnamefont {Abanin}},\
  and\ \bibinfo {author} {\bibfnamefont {Z.}~\bibnamefont {Papić}},\
  }\bibfield  {title} {\bibinfo {title} {Quantum many-body scars and weak
  breaking of ergodicity},\ }\href {https://doi.org/10.1038/s41567-021-01230-2}
  {\bibfield  {journal} {\bibinfo  {journal} {Nature Physics}\ }\textbf
  {\bibinfo {volume} {17}},\ \bibinfo {pages} {675–685} (\bibinfo {year}
  {2021})}\BibitemShut {NoStop}%
\bibitem [{\citenamefont {Lichtenberg}\ and\ \citenamefont
  {Lieberman}(1992)}]{lichtenberg1983regular}%
  \BibitemOpen
  \bibfield  {author} {\bibinfo {author} {\bibfnamefont {A.}~\bibnamefont
  {Lichtenberg}}\ and\ \bibinfo {author} {\bibfnamefont {M.}~\bibnamefont
  {Lieberman}},\ }\href@noop {} {\emph {\bibinfo {title} {Regular and Chaotic
  Motion}}}\ (\bibinfo  {publisher} {Springer},\ \bibinfo {year}
  {1992})\BibitemShut {NoStop}%
\bibitem [{\citenamefont {Berry}(1978)}]{Berry_regirr78:proceeding}%
  \BibitemOpen
  \bibfield  {author} {\bibinfo {author} {\bibfnamefont {M.~V.}\ \bibnamefont
  {Berry}},\ }\bibfield  {title} {\bibinfo {title} {{Regular and Irregular
  Motion}},\ }in\ \href@noop {} {\emph {\bibinfo {booktitle} {{Topics in
  Nonlinear Mechanics}}}},\ Vol.~\bibinfo {volume} {46},\ \bibinfo {editor}
  {edited by\ \bibinfo {editor} {\bibfnamefont {S.}~\bibnamefont {Jorna}}}\
  (\bibinfo  {publisher} {Am.Inst.Ph.},\ \bibinfo {year} {1978})\ pp.\ \bibinfo
  {pages} {16--120}\BibitemShut {NoStop}%
\bibitem [{\citenamefont {Berry}(1983{\natexlab{a}})}]{Heller_Les_Houches}%
  \BibitemOpen
  \bibfield  {author} {\bibinfo {author} {\bibfnamefont {M.~V.}\ \bibnamefont
  {Berry}},\ }\bibfield  {title} {\bibinfo {title} {Wavepacket dynamics and
  quantum chaology},\ }in\ \href@noop {} {\emph {\bibinfo {booktitle} {Chaotic
  Behaviour of Deterministic Systems}}},\ \bibinfo {series and number} {Les
  Houches, Session XXXVI, 1981},\ \bibinfo {editor} {edited by\ \bibinfo
  {editor} {\bibfnamefont {R.~S.~G.}\ \bibnamefont {Ioos}}, \bibinfo {editor}
  {\bibfnamefont {R.~H.~G.}\ \bibnamefont {Hellemani}},\ and\ \bibinfo {editor}
  {\bibfnamefont {R.}~\bibnamefont {Stora}}}\ (\bibinfo  {publisher}
  {North-Holland, Amsterdam},\ \bibinfo {year} {1983})\BibitemShut {NoStop}%
\bibitem [{\citenamefont {Turner}\ \emph {et~al.}(2017)\citenamefont {Turner},
  \citenamefont {Michailidis},\ and\ \citenamefont {A.A.}}]{scars_turner}%
  \BibitemOpen
  \bibfield  {author} {\bibinfo {author} {\bibfnamefont {C.}~\bibnamefont
  {Turner}}, \bibinfo {author} {\bibnamefont {Michailidis}},\ and\ \bibinfo
  {author} {\bibfnamefont {D.}~\bibnamefont {A.A.}, \bibfnamefont {Abanin}},\
  }\bibfield  {title} {\bibinfo {title} {Weak ergodicity breaking from quantum
  many-body scars},\ }\href@noop {} {\bibfield  {journal} {\bibinfo  {journal}
  {Nature}\ }\textbf {\bibinfo {volume} {14}},\ \bibinfo {pages} {745}
  (\bibinfo {year} {2017})}\BibitemShut {NoStop}%
\bibitem [{\citenamefont {Houck}\ \emph {et~al.}(2012)\citenamefont {Houck},
  \citenamefont {T{\"u}reci},\ and\ \citenamefont {Koch}}]{houck}%
  \BibitemOpen
  \bibfield  {author} {\bibinfo {author} {\bibfnamefont {A.~A.}\ \bibnamefont
  {Houck}}, \bibinfo {author} {\bibfnamefont {H.~E.}\ \bibnamefont
  {T{\"u}reci}},\ and\ \bibinfo {author} {\bibfnamefont {J.}~\bibnamefont
  {Koch}},\ }\bibfield  {title} {\bibinfo {title} {On-chip quantum simulation
  with superconducting circuits},\ }\href@noop {} {\bibfield  {journal}
  {\bibinfo  {journal} {Nat. Phys.}\ }\textbf {\bibinfo {volume} {8}},\
  \bibinfo {pages} {292} (\bibinfo {year} {2012})}\BibitemShut {NoStop}%
\bibitem [{\citenamefont {Fazio}\ and\ \citenamefont {van~der
  Zant}(2001)}]{FAZIO2001235}%
  \BibitemOpen
  \bibfield  {author} {\bibinfo {author} {\bibfnamefont {R.}~\bibnamefont
  {Fazio}}\ and\ \bibinfo {author} {\bibfnamefont {H.}~\bibnamefont {van~der
  Zant}},\ }\bibfield  {title} {\bibinfo {title} {Quantum phase transitions and
  vortex dynamics in superconducting networks},\ }\href
  {https://doi.org/https://doi.org/10.1016/S0370-1573(01)00022-9} {\bibfield
  {journal} {\bibinfo  {journal} {Physics Reports}\ }\textbf {\bibinfo {volume}
  {355}},\ \bibinfo {pages} {235 } (\bibinfo {year} {2001})}\BibitemShut
  {NoStop}%
\bibitem [{\citenamefont {Cedergren}\ \emph {et~al.}(2017)\citenamefont
  {Cedergren}, \citenamefont {Ackroyd}, \citenamefont {Kafanov}, \citenamefont
  {Vogt}, \citenamefont {Shnirman},\ and\ \citenamefont
  {Duty}}]{PhysRevLett.119.167701}%
  \BibitemOpen
  \bibfield  {author} {\bibinfo {author} {\bibfnamefont {K.}~\bibnamefont
  {Cedergren}}, \bibinfo {author} {\bibfnamefont {R.}~\bibnamefont {Ackroyd}},
  \bibinfo {author} {\bibfnamefont {S.}~\bibnamefont {Kafanov}}, \bibinfo
  {author} {\bibfnamefont {N.}~\bibnamefont {Vogt}}, \bibinfo {author}
  {\bibfnamefont {A.}~\bibnamefont {Shnirman}},\ and\ \bibinfo {author}
  {\bibfnamefont {T.}~\bibnamefont {Duty}},\ }\bibfield  {title} {\bibinfo
  {title} {Insulating josephson junction chains as pinned luttinger liquids},\
  }\href {https://doi.org/10.1103/PhysRevLett.119.167701} {\bibfield  {journal}
  {\bibinfo  {journal} {Phys. Rev. Lett.}\ }\textbf {\bibinfo {volume} {119}},\
  \bibinfo {pages} {167701} (\bibinfo {year} {2017})}\BibitemShut {NoStop}%
\bibitem [{\citenamefont {Russomanno}\ \emph {et~al.}(2020)\citenamefont
  {Russomanno}, \citenamefont {Fava},\ and\ \citenamefont
  {Fazio}}]{russomanno2020nonergodic}%
  \BibitemOpen
  \bibfield  {author} {\bibinfo {author} {\bibfnamefont {A.}~\bibnamefont
  {Russomanno}}, \bibinfo {author} {\bibfnamefont {M.}~\bibnamefont {Fava}},\
  and\ \bibinfo {author} {\bibfnamefont {R.}~\bibnamefont {Fazio}},\ }\bibfield
   {title} {\bibinfo {title} {Nonergodic behavior of the clean bose-hubbard
  chain},\ }\bibfield  {journal} {\bibinfo  {journal} {Physical Review B}\
  }\textbf {\bibinfo {volume} {102}},\ \href
  {https://doi.org/10.1103/physrevb.102.144302} {10.1103/physrevb.102.144302}
  (\bibinfo {year} {2020})\BibitemShut {NoStop}%
\bibitem [{\citenamefont {Kollath}\ \emph {et~al.}(2007)\citenamefont
  {Kollath}, \citenamefont {L\"auchli},\ and\ \citenamefont
  {Altman}}]{kollath}%
  \BibitemOpen
  \bibfield  {author} {\bibinfo {author} {\bibfnamefont {C.}~\bibnamefont
  {Kollath}}, \bibinfo {author} {\bibfnamefont {A.~M.}\ \bibnamefont
  {L\"auchli}},\ and\ \bibinfo {author} {\bibfnamefont {E.}~\bibnamefont
  {Altman}},\ }\bibfield  {title} {\bibinfo {title} {Quench dynamics and
  nonequilibrium phase diagram of the bose-hubbard model},\ }\href
  {https://doi.org/10.1103/PhysRevLett.98.180601} {\bibfield  {journal}
  {\bibinfo  {journal} {Phys. Rev. Lett.}\ }\textbf {\bibinfo {volume} {98}},\
  \bibinfo {pages} {180601} (\bibinfo {year} {2007})}\BibitemShut {NoStop}%
\bibitem [{\citenamefont {Kollath}\ \emph {et~al.}(2010)\citenamefont
  {Kollath}, \citenamefont {Roux}, \citenamefont {Biroli},\ and\ \citenamefont
  {Läuchli}}]{kollath1}%
  \BibitemOpen
  \bibfield  {author} {\bibinfo {author} {\bibfnamefont {C.}~\bibnamefont
  {Kollath}}, \bibinfo {author} {\bibfnamefont {G.}~\bibnamefont {Roux}},
  \bibinfo {author} {\bibfnamefont {G.}~\bibnamefont {Biroli}},\ and\ \bibinfo
  {author} {\bibfnamefont {A.~M.}\ \bibnamefont {Läuchli}},\ }\bibfield
  {title} {\bibinfo {title} {Statistical properties of the spectrum of the
  extended bose{\textendash}hubbard model},\ }\href
  {https://doi.org/10.1088/1742-5468/2010/08/p08011} {\bibfield  {journal}
  {\bibinfo  {journal} {Journal of Statistical Mechanics: Theory and
  Experiment}\ }\textbf {\bibinfo {volume} {2010}},\ \bibinfo {pages} {P08011}
  (\bibinfo {year} {2010})}\BibitemShut {NoStop}%
\bibitem [{\citenamefont {Sorg}\ \emph {et~al.}(2014)\citenamefont {Sorg},
  \citenamefont {Vidmar}, \citenamefont {Pollet},\ and\ \citenamefont
  {Heidrich-Meisner}}]{PhysRevA.90.033606}%
  \BibitemOpen
  \bibfield  {author} {\bibinfo {author} {\bibfnamefont {S.}~\bibnamefont
  {Sorg}}, \bibinfo {author} {\bibfnamefont {L.}~\bibnamefont {Vidmar}},
  \bibinfo {author} {\bibfnamefont {L.}~\bibnamefont {Pollet}},\ and\ \bibinfo
  {author} {\bibfnamefont {F.}~\bibnamefont {Heidrich-Meisner}},\ }\bibfield
  {title} {\bibinfo {title} {Relaxation and thermalization in the
  one-dimensional bose-hubbard model: A case study for the interaction quantum
  quench from the atomic limit},\ }\href
  {https://doi.org/10.1103/PhysRevA.90.033606} {\bibfield  {journal} {\bibinfo
  {journal} {Phys. Rev. A}\ }\textbf {\bibinfo {volume} {90}},\ \bibinfo
  {pages} {033606} (\bibinfo {year} {2014})}\BibitemShut {NoStop}%
\bibitem [{\citenamefont {Biroli}\ \emph {et~al.}(2010)\citenamefont {Biroli},
  \citenamefont {Kollath},\ and\ \citenamefont
  {L{\"a}uchli}}]{biroli2010effect}%
  \BibitemOpen
  \bibfield  {author} {\bibinfo {author} {\bibfnamefont {G.}~\bibnamefont
  {Biroli}}, \bibinfo {author} {\bibfnamefont {C.}~\bibnamefont {Kollath}},\
  and\ \bibinfo {author} {\bibfnamefont {A.~M.}\ \bibnamefont {L{\"a}uchli}},\
  }\bibfield  {title} {\bibinfo {title} {Effect of rare fluctuations on the
  thermalization of isolated quantum systems},\ }\href@noop {} {\bibfield
  {journal} {\bibinfo  {journal} {Physical review letters}\ }\textbf {\bibinfo
  {volume} {105}},\ \bibinfo {pages} {250401} (\bibinfo {year}
  {2010})}\BibitemShut {NoStop}%
\bibitem [{\citenamefont {Carleo}\ \emph {et~al.}(2012)\citenamefont {Carleo},
  \citenamefont {Becca}, \citenamefont {Schir{\`o}},\ and\ \citenamefont
  {Fabrizio}}]{Carleo}%
  \BibitemOpen
  \bibfield  {author} {\bibinfo {author} {\bibfnamefont {G.}~\bibnamefont
  {Carleo}}, \bibinfo {author} {\bibfnamefont {F.}~\bibnamefont {Becca}},
  \bibinfo {author} {\bibfnamefont {M.}~\bibnamefont {Schir{\`o}}},\ and\
  \bibinfo {author} {\bibfnamefont {M.}~\bibnamefont {Fabrizio}},\ }\bibfield
  {title} {\bibinfo {title} {{Localization and Glassy Dynamics Of Many-Body
  Quantum Systems}},\ }\href@noop {} {\bibfield  {journal} {\bibinfo  {journal}
  {Scientific Reports}\ }\textbf {\bibinfo {volume} {2}},\ \bibinfo {pages}
  {243} (\bibinfo {year} {2012})}\BibitemShut {NoStop}%
\bibitem [{\citenamefont {Russomanno}\ \emph {et~al.}(2021)\citenamefont
  {Russomanno}, \citenamefont {Fava},\ and\ \citenamefont
  {Heyl}}]{michele2021}%
  \BibitemOpen
  \bibfield  {author} {\bibinfo {author} {\bibfnamefont {A.}~\bibnamefont
  {Russomanno}}, \bibinfo {author} {\bibfnamefont {M.}~\bibnamefont {Fava}},\
  and\ \bibinfo {author} {\bibfnamefont {M.}~\bibnamefont {Heyl}},\ }\bibfield
  {title} {\bibinfo {title} {Quantum chaos and ensemble inequivalence of
  quantum long-range ising chains},\ }\bibfield  {journal} {\bibinfo  {journal}
  {Physical Review B}\ }\textbf {\bibinfo {volume} {104}},\ \href
  {https://doi.org/10.1103/physrevb.104.094309} {10.1103/physrevb.104.094309}
  (\bibinfo {year} {2021})\BibitemShut {NoStop}%
\bibitem [{\citenamefont {Baldovin}\ \emph {et~al.}(2021)\citenamefont
  {Baldovin}, \citenamefont {Gradenigo},\ and\ \citenamefont
  {Vulpiani}}]{gradenigo}%
  \BibitemOpen
  \bibfield  {author} {\bibinfo {author} {\bibfnamefont {M.}~\bibnamefont
  {Baldovin}}, \bibinfo {author} {\bibfnamefont {G.}~\bibnamefont
  {Gradenigo}},\ and\ \bibinfo {author} {\bibfnamefont {A.}~\bibnamefont
  {Vulpiani}},\ }\href {https://doi.org/10.48550/ARXIV.2106.06609} {\bibinfo
  {title} {Statistical features of high-dimensional hamiltonian systems}}
  (\bibinfo {year} {2021})\BibitemShut {NoStop}%
\bibitem [{\citenamefont {Pal}\ and\ \citenamefont
  {Huse}(2010)}]{PhysRevB.82.174411}%
  \BibitemOpen
  \bibfield  {author} {\bibinfo {author} {\bibfnamefont {A.}~\bibnamefont
  {Pal}}\ and\ \bibinfo {author} {\bibfnamefont {D.~A.}\ \bibnamefont {Huse}},\
  }\bibfield  {title} {\bibinfo {title} {Many-body localization phase
  transition},\ }\href {https://doi.org/10.1103/PhysRevB.82.174411} {\bibfield
  {journal} {\bibinfo  {journal} {Phys. Rev. B}\ }\textbf {\bibinfo {volume}
  {82}},\ \bibinfo {pages} {174411} (\bibinfo {year} {2010})}\BibitemShut
  {NoStop}%
\bibitem [{\citenamefont {Edwards}\ and\ \citenamefont
  {Thouless}(1972)}]{thouless}%
  \BibitemOpen
  \bibfield  {author} {\bibinfo {author} {\bibfnamefont {J.~T.}\ \bibnamefont
  {Edwards}}\ and\ \bibinfo {author} {\bibfnamefont {D.~J.}\ \bibnamefont
  {Thouless}},\ }\href@noop {} {\bibfield  {journal} {\bibinfo  {journal} {J.
  Phys. C}\ }\textbf {\bibinfo {volume} {5}},\ \bibinfo {pages} {807} (\bibinfo
  {year} {1972})}\BibitemShut {NoStop}%
\bibitem [{\citenamefont {Torres-Herrera}\ \emph {et~al.}(2015)\citenamefont
  {Torres-Herrera}, \citenamefont {Kollmar},\ and\ \citenamefont
  {Santos}}]{Torres_Herrera_2015}%
  \BibitemOpen
  \bibfield  {author} {\bibinfo {author} {\bibfnamefont {E.~J.}\ \bibnamefont
  {Torres-Herrera}}, \bibinfo {author} {\bibfnamefont {D.}~\bibnamefont
  {Kollmar}},\ and\ \bibinfo {author} {\bibfnamefont {L.~F.}\ \bibnamefont
  {Santos}},\ }\bibfield  {title} {\bibinfo {title} {Relaxation and
  thermalization of isolated many-body quantum systems},\ }\href
  {https://doi.org/10.1088/0031-8949/2015/t165/014018} {\bibfield  {journal}
  {\bibinfo  {journal} {Physica Scripta}\ }\textbf {\bibinfo {volume} {T165}},\
  \bibinfo {pages} {014018} (\bibinfo {year} {2015})}\BibitemShut {NoStop}%
\bibitem [{\citenamefont {Santos}\ \emph {et~al.}(2012)\citenamefont {Santos},
  \citenamefont {Borgonovi},\ and\ \citenamefont {Izrailev}}]{Santos_2012}%
  \BibitemOpen
  \bibfield  {author} {\bibinfo {author} {\bibfnamefont {L.~F.}\ \bibnamefont
  {Santos}}, \bibinfo {author} {\bibfnamefont {F.}~\bibnamefont {Borgonovi}},\
  and\ \bibinfo {author} {\bibfnamefont {F.~M.}\ \bibnamefont {Izrailev}},\
  }\bibfield  {title} {\bibinfo {title} {Onset of chaos and relaxation in
  isolated systems of interacting spins: Energy shell approach},\ }\bibfield
  {journal} {\bibinfo  {journal} {Physical Review E}\ }\textbf {\bibinfo
  {volume} {85}},\ \href {https://doi.org/10.1103/physreve.85.036209}
  {10.1103/physreve.85.036209} (\bibinfo {year} {2012})\BibitemShut {NoStop}%
\bibitem [{\citenamefont {Santos}\ and\ \citenamefont
  {Rigol}(2010)}]{Santos_2010}%
  \BibitemOpen
  \bibfield  {author} {\bibinfo {author} {\bibfnamefont {L.~F.}\ \bibnamefont
  {Santos}}\ and\ \bibinfo {author} {\bibfnamefont {M.}~\bibnamefont {Rigol}},\
  }\bibfield  {title} {\bibinfo {title} {Localization and the effects of
  symmetries in the thermalization properties of one-dimensional quantum
  systems},\ }\bibfield  {journal} {\bibinfo  {journal} {Physical Review E}\
  }\textbf {\bibinfo {volume} {82}},\ \href
  {https://doi.org/10.1103/physreve.82.031130} {10.1103/physreve.82.031130}
  (\bibinfo {year} {2010})\BibitemShut {NoStop}%
\bibitem [{\citenamefont {Gubin}\ and\ \citenamefont
  {Santos}(2012)}]{Gubin_2012}%
  \BibitemOpen
  \bibfield  {author} {\bibinfo {author} {\bibfnamefont {A.}~\bibnamefont
  {Gubin}}\ and\ \bibinfo {author} {\bibfnamefont {L.~F.}\ \bibnamefont
  {Santos}},\ }\bibfield  {title} {\bibinfo {title} {Quantum chaos: An
  introduction via chains of interacting spins 1/2},\ }\href
  {https://doi.org/10.1119/1.3671068} {\bibfield  {journal} {\bibinfo
  {journal} {American Journal of Physics}\ }\textbf {\bibinfo {volume} {80}},\
  \bibinfo {pages} {246} (\bibinfo {year} {2012})}\BibitemShut {NoStop}%
\bibitem [{\citenamefont {Haake}(2006)}]{Haake}%
  \BibitemOpen
  \bibfield  {author} {\bibinfo {author} {\bibfnamefont {F.}~\bibnamefont
  {Haake}},\ }\href@noop {} {\emph {\bibinfo {title} {Quantum Signatures of
  Chaos}}}\ (\bibinfo  {publisher} {Springer-Verlag New York, Inc.},\ \bibinfo
  {address} {Secaucus, NJ, USA},\ \bibinfo {year} {2006})\ Chap.~\bibinfo
  {chapter} {7}, pp.\ \bibinfo {pages} {263--274}\BibitemShut {NoStop}%
\bibitem [{\citenamefont {Rammer}(2004)}]{rammer}%
  \BibitemOpen
  \bibfield  {author} {\bibinfo {author} {\bibfnamefont {J.}~\bibnamefont
  {Rammer}},\ }\href@noop {} {\emph {\bibinfo {title} {Quantum Transport
  Theory}}}\ (\bibinfo  {publisher} {Avalon, Reading, MA},\ \bibinfo {year}
  {2004})\BibitemShut {NoStop}%
\bibitem [{\citenamefont {Lee}\ and\ \citenamefont
  {Ramakrishnan}(1985)}]{rama}%
  \BibitemOpen
  \bibfield  {author} {\bibinfo {author} {\bibfnamefont {P.~A.}\ \bibnamefont
  {Lee}}\ and\ \bibinfo {author} {\bibfnamefont {T.~V.}\ \bibnamefont
  {Ramakrishnan}},\ }\bibfield  {title} {\bibinfo {title} {Disordered
  electronic systems},\ }\href@noop {} {\bibfield  {journal} {\bibinfo
  {journal} {Rev. Mod. Phys.}\ }\textbf {\bibinfo {volume} {57}},\ \bibinfo
  {pages} {287} (\bibinfo {year} {1985})}\BibitemShut {NoStop}%
\bibitem [{\citenamefont {Suthar}\ \emph {et~al.}(2020)\citenamefont {Suthar},
  \citenamefont {Sierant},\ and\ \citenamefont
  {Zakrzewski}}]{PhysRevB.101.134203}%
  \BibitemOpen
  \bibfield  {author} {\bibinfo {author} {\bibfnamefont {K.}~\bibnamefont
  {Suthar}}, \bibinfo {author} {\bibfnamefont {P.}~\bibnamefont {Sierant}},\
  and\ \bibinfo {author} {\bibfnamefont {J.}~\bibnamefont {Zakrzewski}},\
  }\bibfield  {title} {\bibinfo {title} {Many-body localization with synthetic
  gauge fields in disordered hubbard chains},\ }\href
  {https://doi.org/10.1103/PhysRevB.101.134203} {\bibfield  {journal} {\bibinfo
   {journal} {Phys. Rev. B}\ }\textbf {\bibinfo {volume} {101}},\ \bibinfo
  {pages} {134203} (\bibinfo {year} {2020})}\BibitemShut {NoStop}%
\bibitem [{\citenamefont {Cole}\ \emph {et~al.}(2014)\citenamefont {Cole},
  \citenamefont {Lepp{\"a}kangas},\ and\ \citenamefont {Marthaler}}]{Cole}%
  \BibitemOpen
  \bibfield  {author} {\bibinfo {author} {\bibfnamefont {J.~H.}\ \bibnamefont
  {Cole}}, \bibinfo {author} {\bibfnamefont {J.}~\bibnamefont
  {Lepp{\"a}kangas}},\ and\ \bibinfo {author} {\bibfnamefont {M.}~\bibnamefont
  {Marthaler}},\ }\bibfield  {title} {\bibinfo {title} {Correlated transport
  through junction arrays in the small josephson energy limit: incoherent
  cooper-pairs and hot electrons},\ }\href@noop {} {\bibfield  {journal}
  {\bibinfo  {journal} {New Journal of Physics}\ }\textbf {\bibinfo {volume}
  {16}},\ \bibinfo {pages} {063019} (\bibinfo {year} {2014})}\BibitemShut
  {NoStop}%
\bibitem [{\citenamefont {Hu}\ and\ \citenamefont
  {O'Connell}(1994)}]{PhysRevB.49.16773}%
  \BibitemOpen
  \bibfield  {author} {\bibinfo {author} {\bibfnamefont {G.~Y.}\ \bibnamefont
  {Hu}}\ and\ \bibinfo {author} {\bibfnamefont {R.~F.}\ \bibnamefont
  {O'Connell}},\ }\bibfield  {title} {\bibinfo {title} {Exact solution for the
  charge soliton in a one-dimensional array of small tunnel junctions},\ }\href
  {https://doi.org/10.1103/PhysRevB.49.16773} {\bibfield  {journal} {\bibinfo
  {journal} {Phys. Rev. B}\ }\textbf {\bibinfo {volume} {49}},\ \bibinfo
  {pages} {16773} (\bibinfo {year} {1994})}\BibitemShut {NoStop}%
\bibitem [{\citenamefont {Leggett}(1987)}]{Leggett}%
  \BibitemOpen
  \bibfield  {author} {\bibinfo {author} {\bibfnamefont {A.~J.}\ \bibnamefont
  {Leggett}},\ }\bibfield  {title} {\bibinfo {title} {{Quantum Mechanics at the
  Macroscopic Level}},\ }in\ \href@noop {} {\emph {\bibinfo {booktitle} {{Le
  hasard et la mati{\`e}re/Chance and matter}}}},\ \bibinfo {series and number}
  {Les Houches, Session XLVI, 1986},\ \bibinfo {editor} {edited by\ \bibinfo
  {editor} {\bibnamefont {{J. Souletie and J. Vannimenus and R. Stora}}}}\
  (\bibinfo  {publisher} {Elsevier Science},\ \bibinfo {year}
  {1987})\BibitemShut {NoStop}%
\bibitem [{\citenamefont {Clarke}\ and\ \citenamefont
  {Wilhelm}(2008)}]{clarke}%
  \BibitemOpen
  \bibfield  {author} {\bibinfo {author} {\bibfnamefont {J.}~\bibnamefont
  {Clarke}}\ and\ \bibinfo {author} {\bibfnamefont {F.~K.}\ \bibnamefont
  {Wilhelm}},\ }\bibfield  {title} {\bibinfo {title} {Superconducting quantum
  bits},\ }\href@noop {} {\bibfield  {journal} {\bibinfo  {journal} {Nature}\
  }\textbf {\bibinfo {volume} {453}},\ \bibinfo {pages} {1031} (\bibinfo {year}
  {2008})}\BibitemShut {NoStop}%
\bibitem [{\citenamefont {{Exploiting the commutation relation
  $\left[\hat{q}_i,\nep^{\pm i\hat{\theta}_j}\right]=\pm\delta_{i,j}\nep^{\pm
  i\hat{\theta}_j}$}}()}]{note_Q}%
  \BibitemOpen
  \bibfield  {author} {\bibinfo {author} {\bibnamefont {{Exploiting the
  commutation relation $\left[\hat{q}_i,\nep^{\pm
  i\hat{\theta}_j}\right]=\pm\delta_{i,j}\nep^{\pm i\hat{\theta}_j}$}}},\
  }\href@noop {} {\bibinfo  {journal} {{we easily find that
  $\left[\hat{Q},\hat{H}\right]=0$}}\ }\BibitemShut {NoStop}%
\bibitem [{\citenamefont {{Up to immaterial constants}}()}]{notabile}%
  \BibitemOpen
\bibfield  {journal} {  }\bibfield  {author} {\bibinfo {author} {\bibnamefont
  {{Up to immaterial constants}}},\ }\href@noop {} {\ }\bibinfo {note} {{t}he
  precise form of the Hamiltonian Eq.~\eqref{Ham0:eqn} in the charge basis is
  \begin{align}
  \hat{H}&=\frac{1}{2}E_C\sum_{i=1}^L\sum_{n_i=-\infty}^{\infty}{n}_i^2\ket{q_i}\bra{q_i}\nonumber\\
  &\hspace{-0.3cm}-\frac{E_J}{2}\sum_{i=1}^{L}\sum_{q_i,q_{i+1}=-\infty}^{\infty}\hspace{-0.5cm}\left(\ket{q_i+1}\bra{q_i}\otimes\ket{q_{i+1}}\bra{q_{i+1}+1}+{\rm
  H.c.}\right)\nonumber\,. \end{align}}\BibitemShut {NoStop}%
\bibitem [{\citenamefont {Tinkham}(1996)}]{tinkham}%
  \BibitemOpen
  \bibfield  {author} {\bibinfo {author} {\bibfnamefont {M.}~\bibnamefont
  {Tinkham}},\ }\href@noop {} {\emph {\bibinfo {title} {Introduction to
  Superconductivity}}},\ \bibinfo {edition} {2nd}\ ed.\ (\bibinfo  {publisher}
  {Mc Graw-Hill, New York},\ \bibinfo {year} {1996})\BibitemShut {NoStop}%
\bibitem [{\citenamefont {Hofstadter}(1976)}]{PhysRevB.14.2239}%
  \BibitemOpen
  \bibfield  {author} {\bibinfo {author} {\bibfnamefont {D.~R.}\ \bibnamefont
  {Hofstadter}},\ }\bibfield  {title} {\bibinfo {title} {Energy levels and wave
  functions of bloch electrons in rational and irrational magnetic fields},\
  }\href {https://doi.org/10.1103/PhysRevB.14.2239} {\bibfield  {journal}
  {\bibinfo  {journal} {Phys. Rev. B}\ }\textbf {\bibinfo {volume} {14}},\
  \bibinfo {pages} {2239} (\bibinfo {year} {1976})}\BibitemShut {NoStop}%
\bibitem [{\citenamefont {Sidje}(1998)}]{EXPOKIT}%
  \BibitemOpen
  \bibfield  {author} {\bibinfo {author} {\bibfnamefont {R.~B.}\ \bibnamefont
  {Sidje}},\ }\bibfield  {title} {\bibinfo {title} {{\sc Expokit.} {A} software
  package for computing matrix exponentials},\ }\href@noop {} {\bibfield
  {journal} {\bibinfo  {journal} {ACM Trans. Math. Softw.}\ }\textbf {\bibinfo
  {volume} {24}},\ \bibinfo {pages} {130} (\bibinfo {year} {1998})}\BibitemShut
  {NoStop}%
\bibitem [{\citenamefont {Schollwock}(2011)}]{Schollwock_rev}%
  \BibitemOpen
  \bibfield  {author} {\bibinfo {author} {\bibfnamefont {U.}~\bibnamefont
  {Schollwock}},\ }\href@noop {} {\bibfield  {journal} {\bibinfo  {journal}
  {Ann. Phys.}\ }\textbf {\bibinfo {volume} {326}},\ \bibinfo {pages} {96}
  (\bibinfo {year} {2011})}\BibitemShut {NoStop}%
\bibitem [{\citenamefont {Fishman}\ \emph {et~al.}(2020)\citenamefont
  {Fishman}, \citenamefont {White},\ and\ \citenamefont
  {Stoudenmire}}]{ITensor}%
  \BibitemOpen
  \bibfield  {author} {\bibinfo {author} {\bibfnamefont {M.}~\bibnamefont
  {Fishman}}, \bibinfo {author} {\bibfnamefont {S.~R.}\ \bibnamefont {White}},\
  and\ \bibinfo {author} {\bibfnamefont {E.~M.}\ \bibnamefont {Stoudenmire}},\
  }\bibfield  {title} {\bibinfo {title} {The \mbox{ITensor} software library
  for tensor network calculations},\ }\href@noop {} {\  (\bibinfo {year}
  {2020})},\ \Eprint {https://arxiv.org/abs/2007.14822} {arXiv:2007.14822}
  \BibitemShut {NoStop}%
\bibitem [{\citenamefont {{In the tDMRG we impose no constraint on the bond
  dimension}}()}]{notaD}%
  \BibitemOpen
  \bibfield  {author} {\bibinfo {author} {\bibnamefont {{In the tDMRG we impose
  no constraint on the bond dimension}}},\ }\href@noop {} {\bibinfo  {journal}
  {{and set in the SVD eigenvalues a cutoff smaller than $10^{-10}$. So, the
  evolution is essentially exact, until it is numerically no more feasible,
  being the eigenvalues retained in the SVD too many. In all tDMRG calculations
  we take open boundary conditions}}\ }\BibitemShut {NoStop}%
\bibitem [{\citenamefont {{In the ladder we draw vertical lines in
  Fig.~\ref{schema:fig}(b) perpendicularly to the legs of the
  ladder}}()}]{note_s}%
  \BibitemOpen
\bibfield  {journal} {  }\bibfield  {author} {\bibinfo {author} {\bibnamefont
  {{In the ladder we draw vertical lines in Fig.~\ref{schema:fig}(b)
  perpendicularly to the legs of the ladder}}},\ }\href@noop {} {\bibinfo
  {journal} {{so that the two resulting subsystems $A$ and $B$ have the same
  number of sites. There being periodic boundary conditions in the horizontal
  direction we need to draw two vertical lines in order to make this partition,
  and this operation is possible only for a number of sites multiple of 4. In
  the case of the linear chain, the number of sites must for the same reason be
  multiple of 2.}}\ }\BibitemShut {NoStop}%
\bibitem [{\citenamefont {Luitz}(2016)}]{PhysRevB.93.134201}%
  \BibitemOpen
\bibfield  {journal} {  }\bibfield  {author} {\bibinfo {author} {\bibfnamefont
  {D.~J.}\ \bibnamefont {Luitz}},\ }\bibfield  {title} {\bibinfo {title} {Long
  tail distributions near the many-body localization transition},\ }\href
  {https://doi.org/10.1103/PhysRevB.93.134201} {\bibfield  {journal} {\bibinfo
  {journal} {Phys. Rev. B}\ }\textbf {\bibinfo {volume} {93}},\ \bibinfo
  {pages} {134201} (\bibinfo {year} {2016})}\BibitemShut {NoStop}%
\bibitem [{\citenamefont {Huang}(2019)}]{Huang_NPB19}%
  \BibitemOpen
  \bibfield  {author} {\bibinfo {author} {\bibfnamefont {Y.}~\bibnamefont
  {Huang}},\ }\bibfield  {title} {\bibinfo {title} {Universal eigenstate
  entanglement of chaotic local hamiltonians},\ }\href@noop {} {\bibfield
  {journal} {\bibinfo  {journal} {Nuclear Physics B}\ }\textbf {\bibinfo
  {volume} {938}},\ \bibinfo {pages} {594} (\bibinfo {year}
  {2019})}\BibitemShut {NoStop}%
\bibitem [{\citenamefont {Kuwahara}\ and\ \citenamefont {Saito}(2020)}]{saito}%
  \BibitemOpen
  \bibfield  {author} {\bibinfo {author} {\bibfnamefont {T.}~\bibnamefont
  {Kuwahara}}\ and\ \bibinfo {author} {\bibfnamefont {K.}~\bibnamefont
  {Saito}},\ }\bibfield  {title} {\bibinfo {title} {Area law of noncritical
  ground states in 1d long-range interacting systems},\ }\href@noop {}
  {\bibfield  {journal} {\bibinfo  {journal} {Nature Communications}\ }\textbf
  {\bibinfo {volume} {11}},\ \bibinfo {pages} {4478} (\bibinfo {year}
  {2020})}\BibitemShut {NoStop}%
\bibitem [{\citenamefont {Eisert}\ \emph {et~al.}(2010)\citenamefont {Eisert},
  \citenamefont {Cramer},\ and\ \citenamefont {Plenio}}]{RevModPhys.82.277}%
  \BibitemOpen
  \bibfield  {author} {\bibinfo {author} {\bibfnamefont {J.}~\bibnamefont
  {Eisert}}, \bibinfo {author} {\bibfnamefont {M.}~\bibnamefont {Cramer}},\
  and\ \bibinfo {author} {\bibfnamefont {M.~B.}\ \bibnamefont {Plenio}},\
  }\bibfield  {title} {\bibinfo {title} {Colloquium: Area laws for the
  entanglement entropy},\ }\href {https://doi.org/10.1103/RevModPhys.82.277}
  {\bibfield  {journal} {\bibinfo  {journal} {Rev. Mod. Phys.}\ }\textbf
  {\bibinfo {volume} {82}},\ \bibinfo {pages} {277} (\bibinfo {year}
  {2010})}\BibitemShut {NoStop}%
\bibitem [{\citenamefont {Hastings}(2007)}]{Hastings_2007}%
  \BibitemOpen
  \bibfield  {author} {\bibinfo {author} {\bibfnamefont {M.~B.}\ \bibnamefont
  {Hastings}},\ }\bibfield  {title} {\bibinfo {title} {An area law for
  one-dimensional quantum systems},\ }\href
  {https://doi.org/10.1088/1742-5468/2007/08/p08024} {\bibfield  {journal}
  {\bibinfo  {journal} {Journal of Statistical Mechanics: Theory and
  Experiment}\ }\textbf {\bibinfo {volume} {2007}},\ \bibinfo {pages} {P08024}
  (\bibinfo {year} {2007})}\BibitemShut {NoStop}%
\bibitem [{\citenamefont {{To be picky}}()}]{notec}%
  \BibitemOpen
  \bibfield  {author} {\bibinfo {author} {\bibnamefont {{To be picky}}},\
  }\href@noop {} {\bibinfo  {journal} {{the period-3 pattern is not a charge
  density wave (which should have period 2), but we use the same wording in
  order to avoid too much complex writing}}\ }\BibitemShut {NoStop}%
\bibitem [{\citenamefont {{There are other low-entanglement states with
  entanglement entropies between $\log(2)$}}()}]{notaccola}%
  \BibitemOpen
\bibfield  {journal} {  }\bibfield  {author} {\bibinfo {author} {\bibnamefont
  {{There are other low-entanglement states with entanglement entropies
  between $\log(2)$}}},\ }\href@noop {} {\bibinfo  {journal} {{$\log(4)$
  and $\log(6)$, for values $E_J\ll E_C$ or at the edges of the spectrum, but
  they are not special. Indeed, those of them in the inner part of the spectrum
  disappear when one goes beyond the perturbative regime. The ones at the upper
  edge of the spectrum are an artifact due to the finite value of the
  truncation $M$. The ones at the lower edge of the spectrum cannot be
  distinguished from a finite-size effect due to the small microcanonical
  entropy in those energy shell, as we have discussed above.}}\ }\BibitemShut
  {NoStop}%
\bibitem [{\citenamefont {{This perturbative analysis is similar to one already
  performed for the Bose-Hubbard model}}()}]{note_mik}%
  \BibitemOpen
\bibfield  {journal} {  }\bibfield  {author} {\bibinfo {author} {\bibnamefont
  {{This perturbative analysis is similar to one already performed for the
  Bose-Hubbard model}}},\ }\href@noop {} {\bibinfo  {journal}
  {{\cite{russomanno2020nonergodic}}}\ }\BibitemShut {NoStop}%
\bibitem [{\citenamefont {Atas}\ \emph {et~al.}(2013)\citenamefont {Atas},
  \citenamefont {Bogomolny}, \citenamefont {Giraud},\ and\ \citenamefont
  {Roux}}]{PhysRevLett.110.084101}%
  \BibitemOpen
\bibfield  {journal} {  }\bibfield  {author} {\bibinfo {author} {\bibfnamefont
  {Y.~Y.}\ \bibnamefont {Atas}}, \bibinfo {author} {\bibfnamefont
  {E.}~\bibnamefont {Bogomolny}}, \bibinfo {author} {\bibfnamefont
  {O.}~\bibnamefont {Giraud}},\ and\ \bibinfo {author} {\bibfnamefont
  {G.}~\bibnamefont {Roux}},\ }\bibfield  {title} {\bibinfo {title}
  {Distribution of the ratio of consecutive level spacings in random matrix
  ensembles},\ }\href {https://doi.org/10.1103/PhysRevLett.110.084101}
  {\bibfield  {journal} {\bibinfo  {journal} {Phys. Rev. Lett.}\ }\textbf
  {\bibinfo {volume} {110}},\ \bibinfo {pages} {084101} (\bibinfo {year}
  {2013})}\BibitemShut {NoStop}%
\bibitem [{\citenamefont {Berry}(1983{\natexlab{b}})}]{Berry_Les_Houches}%
  \BibitemOpen
  \bibfield  {author} {\bibinfo {author} {\bibfnamefont {M.~V.}\ \bibnamefont
  {Berry}},\ }\bibfield  {title} {\bibinfo {title} {Semiclassical mechanics of
  regular and irregular motion},\ }in\ \href@noop {} {\emph {\bibinfo
  {booktitle} {Chaotic Behaviour of Deterministic Systems}}},\ \bibinfo {series
  and number} {Les Houches, Session XXXVI, 1981},\ \bibinfo {editor} {edited
  by\ \bibinfo {editor} {\bibfnamefont {R.~S.~G.}\ \bibnamefont {Ioos}},
  \bibinfo {editor} {\bibfnamefont {R.~H.~G.}\ \bibnamefont {Hellemani}},\ and\
  \bibinfo {editor} {\bibfnamefont {R.}~\bibnamefont {Stora}}}\ (\bibinfo
  {publisher} {North-Holland, Amsterdam},\ \bibinfo {year} {1983})\ p.\
  \bibinfo {pages} {174–271}\BibitemShut {NoStop}%
\bibitem [{\citenamefont {Berry}\ and\ \citenamefont
  {Tabor}(1977)}]{Berry_PRS77}%
  \BibitemOpen
  \bibfield  {author} {\bibinfo {author} {\bibfnamefont {M.~V.}\ \bibnamefont
  {Berry}}\ and\ \bibinfo {author} {\bibfnamefont {M.}~\bibnamefont {Tabor}},\
  }\bibfield  {title} {\bibinfo {title} {Level clustering in the regular
  spectrum},\ }\href@noop {} {\bibfield  {journal} {\bibinfo  {journal} {Proc.
  Roy. Soc. A}\ }\textbf {\bibinfo {volume} {356}},\ \bibinfo {pages} {375}
  (\bibinfo {year} {1977})}\BibitemShut {NoStop}%
\bibitem [{\citenamefont {Arnol'd}(2013)}]{arnol2013mathematical}%
  \BibitemOpen
  \bibfield  {author} {\bibinfo {author} {\bibfnamefont {V.~I.}\ \bibnamefont
  {Arnol'd}},\ }\href@noop {} {\emph {\bibinfo {title} {Mathematical methods of
  classical mechanics}}},\ Vol.~\bibinfo {volume} {60}\ (\bibinfo  {publisher}
  {Springer Science \& Business Media},\ \bibinfo {year} {2013})\BibitemShut
  {NoStop}%
\bibitem [{\citenamefont {Schulz}\ \emph {et~al.}(2019)\citenamefont {Schulz},
  \citenamefont {Hooley}, \citenamefont {Moessner},\ and\ \citenamefont
  {Pollmann}}]{Schulz_2019}%
  \BibitemOpen
  \bibfield  {author} {\bibinfo {author} {\bibfnamefont {M.}~\bibnamefont
  {Schulz}}, \bibinfo {author} {\bibfnamefont {C.}~\bibnamefont {Hooley}},
  \bibinfo {author} {\bibfnamefont {R.}~\bibnamefont {Moessner}},\ and\
  \bibinfo {author} {\bibfnamefont {F.}~\bibnamefont {Pollmann}},\ }\bibfield
  {title} {\bibinfo {title} {Stark many-body localization},\ }\bibfield
  {journal} {\bibinfo  {journal} {Physical Review Letters}\ }\textbf {\bibinfo
  {volume} {122}},\ \href {https://doi.org/10.1103/physrevlett.122.040606}
  {10.1103/physrevlett.122.040606} (\bibinfo {year} {2019})\BibitemShut
  {NoStop}%
\end{thebibliography}%
\end{document}